\documentclass[prd,showpacs,amsmath,amssymb]{revtex4}

\usepackage{graphicx}
\usepackage{dcolumn}
\usepackage{bm}
\usepackage{psfrag}
\usepackage{subfigure}

\newcommand{\be}{\begin{eqnarray}}
\newcommand{\ee}{\end{eqnarray}}
\newcommand{\bn}{\begin{eqnarray*}}
\newcommand{\en}{\end{eqnarray*}}

\newcommand{\nn}{\nonumber \\}
\newcommand{\nl}{\\}

\renewcommand{\vec}[1]{\mbox{\boldmath$#1$}}
\renewcommand{\d}{\mbox{\rm d}}

\renewcommand{\th}{\ensuremath{\theta}}

\newcommand{\vph}{\ensuremath{\varphi}}
\newcommand{\al}{\ensuremath{\alpha}}
\newcommand{\bt}{\ensuremath{\beta}}
\newcommand{\sg}{\ensuremath{\sigma}}
\newcommand{\gm}{\ensuremath{\gamma}}

\newcommand{\lm}{\ensuremath{\lambda}}
\newcommand{\Lm}{\ensuremath{\Lambda}}

\newcommand{\Dl}{\ensuremath{\Delta}}
\newcommand{\Sg}{\ensuremath{\Sigma}}

\newcommand{\Om}{\ensuremath{\Omega}}

\newcommand{\Ct}{\ensuremath{\hat{t}}}

\newcommand{\Cx}{\ensuremath{\hat{x}}}
\newcommand{\Cy}{\ensuremath{\hat{y}}}
\newcommand{\Cz}{\ensuremath{\hat{z}}}
\newcommand{\Cmu}{\ensuremath{\hat{\mu}}}

\newcommand{\Cj}{\ensuremath{\hat{\jmath}}}

\newcommand{\ze}{\ensuremath{\hat{0}}}

\newcommand{\pvec}{\ensuremath{\vec{p}}}

\newcommand{\kvec}{\ensuremath{\vec{k}}}

\newcommand{\Lvec}{\ensuremath{\vec{L}}}

\newcommand{\rvec}{\ensuremath{\vec{r}}}

\newcommand{\alvec}{\ensuremath{\vec{\al}}}
\newcommand{\sgvec}{\ensuremath{\vec{\sg}}}
\newcommand{\Sgvec}{\ensuremath{\vec{\Sg}}}

\newcommand{\lt}{\ensuremath{\left}}
\newcommand{\rt}{\ensuremath{\right}}

\newcommand{\nabvec}{\ensuremath{\vec{\nabla}}}
\newcommand{\PhiG}{\ensuremath{\Phi_{\rm G}}}

\renewcommand{\k}{\ensuremath{k_0}}
\newcommand{\krq}{\ensuremath{(\k R/q)^2}}
\newcommand{\erf}{\ensuremath{{\rm erf}}}

\renewcommand{\d}{\mbox{\rm d}}

\begin{document}







\title{The Distinction Between Dirac and Majorana Neutrino Wave Packets \\
Due to Gravity and Its Impact on Neutrino Oscillations}


\author{Dinesh Singh$^{a}$}
\altaffiliation[Electronic address:  ]{singhd@uregina.ca}
\author{Nader Mobed$^{a}$}
\altaffiliation[Electronic address:  ]{nader.mobed@uregina.ca}
\author{Giorgio Papini$^{a,b,c}$}
\altaffiliation[Electronic address:  ]{papini@uregina.ca}
\address{$^a$Department of Physics, University of Regina, Regina, Saskatchewan, S4S 0A2, Canada}
\address{$^b$Prairie Particle Physics Institute, Regina, Saskatchewan, S4S 0A2, Canada}
\address{$^c$International Institute for Advanced Scientific Studies, 89019 Vietri sul Mare (SA), Italy}

\date{\today}

\begin{abstract}

We present the possibility that Dirac and Majorana neutrino wave packets can be distinguished
when subject to spin-gravity interaction while propagating through vacuum described by the Lense-Thirring metric.
By adopting the techniques of gravitational phase and time-independent perturbation theory following the
Brillouin-Wigner method, we generate spin-gravity matrix elements from a perturbation Hamiltonian and
show that this distinction is easily reflected in well-defined gravitational corrections to the
neutrino oscillation length for a two-flavour system.
Explicit examples are presented using the Sun and SN1987A as the gravitational sources for
the Lense-Thirring metric.
This approach offers the possibility to determine the absolute neutrino masses by this method and identify
a theoretical upper bound for the absolute neutrino mass difference, where the distinctions between the Dirac
and Majorana cases are evident.
We discuss the relevance of this analysis to the upcoming attempts to measure the properties of low-energy
neutrinos by SNO and other solar neutrino observatories.

\end{abstract}

\pacs{04.90.+e, 14.60.Pq, 04.80.+z, 97.60.Bw}

\maketitle








\section{Introduction}

As arguably the most elusive of the known subatomic particles in existence, neutrinos
nontheless offer great promise for revealing insights at both the microscopic and cosmological level.
Because they are generated solely from the weak interaction and have an extremely small scattering cross section,
to just even observe a neutrino event is a major experimental accomplishment.
Neutrinos have the surprising property of having exclusively negative helicity due to parity violation
of the weak interaction, and that only their left-handed chiral projections interact with matter.
For all these reasons, it was understandable to assume that neutrinos were strictly massless particles like photons which
propagate at the speed of light, and are represented as such in the Standard Model of particle physics.
However, there is clear evidence that the Standard Model is incomplete because of the recent
experimental discovery by Superkamiokande \cite{Kam} and the Sudbury Neutrino Observatory
(SNO) \cite{SNO} demonstrating that neutrinos can transform from one species to another
while propagating through space.
This property, known as neutrino flavour oscillation \cite{gribov}, strongly suggests that
neutrinos have non-zero rest mass, since the standard treatment of the theory for a two flavour
system requires that each flavour has a distinct mass eigenstate.

The current theory of neutrino oscillations is based on solving the Dirac equation in flat
space-time for the mass eigenstates and their associated energy eigenvalues.
One immediate consequence is that we lack sufficient information to determine their absolute masses outright,
since only the mass-squared difference can be determined directly by observation.
Because of this fact, it is not obvious how to obtain the absolute neutrino masses
without resorting to debatable theoretical speculations.
Since an enormous number of neutrinos were produced in the early Universe, having a reliable
figure for the neutrino masses can give cosmology researchers better insight about the nature of dark matter,
with implications for the Universe's overall evolution.
There is also a conceptual difficulty with the current approach to neutrino oscillations, in that neutrinos
are largely produced in supernovae and within the stellar cores of astrophysical sources,
where for the case of a neutron star the curvature of space-time near its surface is close to the
Schwarzschild limit.
Therefore, a more complete approach should be within the framework of a curved space-time background,
as described by Einstein's general theory of relativity.

On a different level entirely, neutrinos can be modelled as the propagation of plane waves, wave packets, or quantum fields.
The obvious advantage for using plane waves comes from the simplicity in performing calculations,
but at the expense of losing possibly significant physical insights.
Indeed, the fact that neutrinos are quantum objects which can be localized in a detector suggests that they
should at least be treated as wave packets \cite{giunti} with some spatial extension subject to the
Heisenberg uncertainty relations.
This point becomes especially relevant when considering the effects of gravitation on neutrino
wave packets, since both are inherently non-local in nature, where space-time curvature
may have a non-trivial influence on their intrinsic properties as they propagate through space.
Conversely, it is theoretically possible to glean insights about the nature of space-time
at the quantum mechanical level that may provide some observational clues towards attaining
a viable quantum gravity theory.
In fact, much research has been done to better understand the interface between quantum mechanics
and classical gravitation, where most recently it was shown \cite{singh1} that the
gravitational field can distinguish between the helicity and chirality of {\em massless} spin-1/2 particles.
This is in contrast to the prevalent understanding \cite{itzykson} that the helicity and chirality operators coincide
for massless spin-1/2 particles in a flat space-time background.

Another more fundamental debate concerns whether neutrinos exist as Dirac particles, where the antineutrinos are
physically unique particles, or Majorana particles, where the neutrinos and antineutrinos are identical.
More precisely, this means that Dirac neutrinos and antineutrinos have a distinct set of quantum numbers,
while the Majorana set share the same quantum numbers.
While lepton number is a conserved quantity for Dirac neutrinos, it is violated for Majorana neutrinos.
Currently the only proposed method to distinguish between the two types comes from neutrinoless double beta
decay \cite{mohapatra,fukugita}, which can only happen for Majorana neutrinos.
Because of the extreme difficulty in making such an observation, it may be a long time before this ambiguity
can become clarified by this approach.
As for neutrino oscillations, current understanding suggests that it is not possible to distinguish between
Dirac and Majorana neutrinos by this method.
This is because both neutrino types share the same left-handed projection of their respective four-spinors \cite{bilenky},
which generate the same physical predictions in this context.
Again, however, it is important to emphasize that this understanding comes about in the absence of a
gravitational field.

In this paper, we consider the possibility of whether classical gravitation, described in terms of the
Lense-Thirring metric \cite{lense} can have an impact on neutrino wave packet propagation, with
consequence for neutrino oscillations.
This is done by treating the gravitational field as a phase contribution to the neutrino wavefunction
leading to an interaction Hamiltonian.
By using the techniques of time-independent perturbation theory, we can determine the perturbation energy
from the matrix elements corresponding to the propagation of neutrino wave packets, which are represented
by a power series expansion in terms of their rest masses.
A preliminary investigation \cite{singh2} into this problem following a semi-analytical approach shows that it is
theoretically possible to extract the absolute neutrino masses from direct observation because the oscillation
length becomes dependent on both the {\em linear} mass difference coupled to the gravitational field,
as well as the mass-squared difference.

Since this first investigation considered the problem exclusively for Dirac neutrinos, it left open the
question of whether Majorana neutrinos would behave differently from their Dirac counterparts under the same conditions.
The purpose of this paper is to address this question by performing the same general analysis as before,
but with much greater attention to detail emphasized throughout.
As a result, we have a set of calculations to present with virtually no approximations employed, in which we show
that the gravitational field can fundamentally distinguish between Dirac and Majorana neutrinos \cite{singh2a}.
We begin in Sec.~\ref{section:grav-phase} with the formalism on neutrinos in curved space-time, the Lense-Thirring metric,
and a description of gravitational phase required for the time-independent perturbation theory calculations
in this paper.
This is followed by Sec.~\ref{section:wave-packets}, which provides both the formalism of neutrino wave packets.
The resulting matrix element calculations are then described by Sec.~\ref{section:grav-matrix-element}
for both Dirac and Majorana neutrinos.
After this, we proceed in Sec.~\ref{section:oscill-length} to determine the mass-dependent perturbation energy
via the Brillouin-Wigner approach normally used in condensed matter theory.
This has the advantage of generating {\em exact} energy corrections for a two-level spin system,
eventually leading to the main formal expressions in this paper.
We then perform some extensive numerical analysis in Sec.~\ref{section:analysis} using both the Sun and the SN1987A
supernova as examples to predict the gravitational corrections on the neutrino oscillation length,
and present a method to extract the absolute neutrino masses by this approach.
Finally, we conclude in Sec.~\ref{section:conclusion} with a discussion about possible future developments
that may follow from this investigation.

\section{Neutrinos in curved space-time with gravitational phase}
\label{section:grav-phase}

\subsection{Covariant Dirac Equation}

For all calculations to follow, the space-time metric has $-2$ signature and we adopt geometric units
$G = c = 1$~\cite{MTW}, such that all physical quantities are described in units of length defined by $M$,
the mass of the gravitational source.
We start with the covariant Dirac equation in curved space-time \cite{singh3} for spin-1/2 particles with mass $m$:
\be
\lt[i \gamma^\mu (x) D_\mu - {m \over \hbar}\rt]\psi (x) & = & 0,
\label{cov-Dirac}
\ee
where the set of space-time dependent gamma matrices $\lt\{\gamma^\mu(x) \rt\}$ satisfy
$\lt\{\gamma^\mu (x), \gamma^\nu (x) \rt\} = 2 \, g^{\mu \nu}(x)$, and the covariant derivative operator
$D_\mu \equiv \partial_\mu + i \Gamma_\mu$ is described in terms of
spin connection $\Gamma_\mu$, such that $D_\mu \, \gamma^\nu(x) = 0$.
By defining a local Minkowski frame at each space-time point on the manifold for a given set of
orthonormal tetrads $\lt\{\vec{e}_{\hat{\mu}} \rt\}$ and basis one-forms $\lt\{\vec{e}^{\hat{\mu}} \rt\}$ labelled by hatted indices
and satisfying $\lt\langle \vec{e}^{\hat{\mu}} , \vec{e}_{\hat{\nu}} \rt\rangle = \delta^{\hat{\mu}}{}_{\hat{\nu}} \,$,
the curved space-time metric is $\vec{g} = \eta_{\hat{\mu}\hat{\nu}} \, \vec{e}^{\hat{\mu}} \otimes \vec{e}^{\hat{\nu}}$,
where $\eta_{\hat{\mu}\hat{\nu}}$ is the Minkowski metric tensor.
The general metric tensor $g_{\mu \nu}$ is then related to $\eta_{\hat{\mu}\hat{\nu}}$
in terms of vierbein projections $\lt\{ e^{\hat{\alpha}}{}_\mu \rt\} \,$,
$\lt\{ e^\mu{}_{\hat{\alpha}} \rt\}$ satisfying
$\vec{e}^{\hat{\alpha}} = e^{\hat{\alpha}}{}_\beta \, \vec{e}^\beta$ and
$\vec{e}_{\hat{\alpha}} = e^\beta{}_{\hat{\alpha}} \, \vec{e}_\beta$, such that
\be
e^{\hat{\alpha}}{}_\mu \, e^\mu{}_{\hat{\beta}} & = &
\delta^{\hat{\alpha}}{}_{\hat{\beta}}, \quad
e^{\mu}{}_{\hat{\alpha}} \, e^{\hat{\alpha}}{}_{\nu} \ = \
\delta^\mu{}_\nu,
\label{vierbein=}
\nl
g_{\mu \nu} & = & \eta_{\hat{\alpha}\hat{\beta}} \,
e^{\hat{\alpha}}{}_\mu \, e^{\hat{\beta}}{}_\nu \, .
\label{gtensor=}
\ee
The spin connection matrix is then
\be
\Gamma_\mu & = & -{1 \over 4} \, \sigma^{\alpha \beta}(x) \,
\Gamma_{\alpha \beta \mu} \ = \
-{1 \over 4} \, \sigma^{\hat{\alpha} \hat{\beta}} \,
\Gamma_{\hat{\alpha} \hat{\beta} \hat{\mu}} \, e^{\hat{\mu}}{}_\mu \, ,
\label{Gammadef=}
\ee
where $\sigma^{\hat{\alpha} \hat{\beta}} =
{i \over 2} [\gamma^{\hat{\alpha}}, \gamma^{\hat{\beta}}]$ are the
Minkowski space-time spin matrices, and $\Gamma_{\hat{\alpha} \hat{\beta} \hat{\mu}}$
are the Ricci rotation coefficients derived from the Cartan equation of differential forms
\be
\d\vec{e}^{\hat{\mu}} + \Gamma^{\hat{\mu}}{}_{\hat{\beta} \hat{\alpha}} \,
\vec{e}^{\hat{\alpha}} \wedge \vec{e}^{\hat{\beta}} & = & 0.
\label{Cartan=}
\ee

\subsection{Dirac Hamiltonian in Lense-Thirring Space-Time}

The Lense-Thirring metric in cartesian co-ordinates $x^\mu = \lt(x^0, x^1, x^2, x^3\rt) = \lt(t,x,y,z\rt)$ is given by
\be
\vec{g} & = & \lt(1 - {2M \over r}\rt) \d t \otimes \d t -
\lt(1 + {2M \over r}\rt) \lt(\d x \otimes \d x + \d y \otimes \d y + \d z \otimes \d z \rt)
\nn
&  &{} + {4 \over 5} \, {M \Omega R^2 \over r^3} \lt[x \lt(\d y \otimes \d t + \d t \otimes \d y \rt)
- y \lt(\d x \otimes \d t + \d t \otimes \d x \rt)\rt],
\label{g-cart=}
\ee
used to describe a weak gravitational field in the vicinity of a rotating star, where
$R$ is the radius of the gravitational source, $\Omega$ is its rotational velocity, and $r = \sqrt{x^2 + y^2 + z^2}$.
Given that $\alvec$ and $\bt$ are the usual four-dimensional Dirac matrices and $\vec{\Sigma}^{\Cj} = \sg^{\ze \Cj}$
is the spin angular momentum operator, the Lense-Thirring spin connection components following (\ref{Gammadef=}) are then
\be
\Gamma_t & = & \lt(1 - {M \over r}\rt)^{-1} \lt(1 + {M \over r}\rt)^{-2}
\lt\{ {M \Om R^2 \over 5 \, r^3} \lt[{3 \, z \over r^2} \lt(1 + {2M \over 3 \, r}\rt)\lt(\Sgvec \cdot \rvec\rt) - \Sgvec^{\Cz} \rt]
- i \, {M \over 2 \, r^3} \lt(1 + {M \over r}\rt) \lt(\alvec \cdot \rvec\rt)\rt\},
\label{Gm0}
\nl
\Gamma_x & = & \lt(1 + {M \over r}\rt)^{-2}
\lt\{ {M \over 2 \, r^3} \lt(\Sgvec \times \rvec\rt)^{\Cx}
- i \, {3 \, M \Om R^2 \over 5 \, r^5} \lt(1 + {2 \, M \over 3 \, r}\rt) \lt(1 - {M \over r}\rt)^{-1} \lt[y \lt(\alvec \cdot \rvec\rt) + x \lt(\alvec \times \rvec \rt)^{\Cz} \rt] \rt\},
\label{Gm1}
\nl
\Gamma_y & = & \lt(1 + {M \over r}\rt)^{-2}
\lt\{ {M \over 2 \, r^3} \lt(\Sgvec \times \rvec\rt)^{\Cy}
+ i \, {3 \, M \Om R^2 \over 5 \, r^5} \lt(1 + {2 \, M \over 3 \, r}\rt) \lt(1 - {M \over r}\rt)^{-1} \lt[x \lt(\alvec \cdot \rvec\rt) - y \lt(\alvec \times \rvec \rt)^{\Cz} \rt] \rt\},
\label{Gm2}
\nl
\Gamma_z & = & \lt(1 + {M \over r}\rt)^{-2}
\lt\{ {M \over 2 \, r^3} \lt(\Sgvec \times \rvec\rt)^{\Cz}
- i \, {3 \, M \Om R^2 \over 5 \, r^5} \lt(1 + {2 \, M \over 3 \, r}\rt) \lt(1 - {M \over r}\rt)^{-1} \lt[z \lt(\alvec \times \rvec \rt)^{\Cz} \rt] \rt\}.
\label{Gm3}
\ee
Therefore, the corresponding Dirac Hamiltonian derived from (\ref{cov-Dirac}) up to leading order in $M/r$ is
\be
H_0 & \approx & \lt(1 - {2M \over r}\rt) \alvec \cdot \pvec
+ m \lt(1 - {M \over r}\rt) \bt + {i \hbar \, {M \over 2 \, r^3}} \lt(\alvec \cdot \rvec \rt)
\nn
& &{}+ {4 \over 5} \, {M \Omega R^2 \over r^3} \, \Lvec^{\Cz} + {1 \over 5} \, {\hbar \, M \Omega R^2 \over r^3}
\lt[{3 \, z \over r^2} \lt(\Sgvec \cdot \rvec\rt) - \Sgvec^{\Cz}\rt],
\label{H0=}
\ee
where $\Lvec^{\Cz} = x \, \pvec^{\Cy} - y \, \pvec^{\Cx}$ is the orbital angular momentum operator in the $z$-direction.
For future reference, (\ref{H0=}) is regarded as the zeroth-order part of the total Hamiltonian with an interaction term, where the latter
is a perturbation due to the gravitational phase contribution, whose details are listed immediately below.

\subsection{Gravitational Phase and the Interaction Hamiltonian}

For this investigation, we want to consider the effect of gravitation on the {\em quantum} propagation of a neutrino wave packet from the gravitational
source to the Earth.
Such a condition implies that, unlike other semiclassical treatments often found in the literature \cite{bhattacharya},
we need to average over all possible trajectories within phase space that are accessible by the neutrino,
as opposed to integrating over its classical trajectory described by the geodesic equation.
Such a consideration requires that we use time-independent perturbation theory, where the interaction Hamiltonian is dependent on the space-time metric.
This can come about following the introduction of the gravitational field via a phase shift in the wavefunction.
The gravitational phase $\PhiG$ for a weak gravitational field is then described by \cite{cai,singh4,papini1,papini2}
\be
\PhiG & \equiv & {1 \over 2} \int_{x_0^\mu}^{x^\mu}   
\d z^\lambda h_{\lm \al} (z) \pvec^\alpha
 - {1 \over 4} \int_{x_0^\mu}^{x^\mu}    
 \d z^\lambda \lt[h_{\lm \al, \bt} (z) - h_{\lm \bt, \al} (z) \rt]
\vec{L}^{\alpha \beta} (z),
\label{PhiG=}
\ee
where $h_{\mu \nu} = g_{\mu \nu} - \eta_{\mu\nu}$ is the metric deviation,
$\pvec^\mu$ and $\vec{L}^{\alpha \beta}$ are the momentum and orbital angular momentum operators of the free particle,
and $z^\mu = \lt(t', x', y', z' \rt)$ is the integration variable over some {\em arbitrary} classical path in space-time.
One of the main properties of the gravitational phase in this form is that a wavefunction described with (\ref{PhiG=}) becomes a
solution of the weak-field covariant Klein-Gordon equation, exact to first order in $h_{\mu \nu}$
and invariant under $h_{\mu \nu} \rightarrow h_{\mu \nu} - \lt(\xi_{\mu,\nu} + \xi_{\nu,\mu}\rt)$, the co-ordinate gauge conditions
which follow from the co-ordinate transformations $x_\mu \rightarrow x_\mu + \xi_\mu$.

When applied to a closed space-time integration path, (\ref{PhiG=}) leads to the covariant generalization of Berry's phase
\cite{cai1} with useful applications for calculations involving particle interferometry \cite{cai2}.
This form of the gravitational phase is comparable with other formulations \cite{DeWitt,Papini3,Anandan,Stodolsky} available in the literature.
It has also been shown \cite{lamb} that use of (\ref{PhiG=}) correctly reproduces the degree of particle deflection predicted by general relativity
in the geometrical optics limit.
When $\Phi_{\rm G}$ is applied to spin-1/2 particles accelerated in storage rings for a metric describing noninertial motion,
it gives rise to all gravitational-inertial effects that have been either observed outright \cite{page,bonse} or are very likely to exist \cite{hehlni,mash}.

\begin{figure}
\psfrag{x}[tc][][2.5][0]{\large ${\rm x}$}
\psfrag{y}[tc][][2.5][0]{\large ${\rm y}$}
\psfrag{z}[tc][][2.5][0]{\large ${\rm z}$}
\psfrag{th}[cc][][2.5][0]{$\theta$}
\psfrag{ph}[cc][][2.5][0]{$\varphi$}
\psfrag{M}[bc][][2.5][0]{$M$}
\psfrag{R}[tc][][2.5][0]{$R$}
\psfrag{Om}[tc][][2.5][0]{$\Omega$}
\psfrag{nu}[tc][][2.5][0]{$\nu$}
\begin{minipage}[t]{0.3 \textwidth}
\centering
\rotatebox{0}{\includegraphics[width = 8cm, height = 6cm, scale = 1]{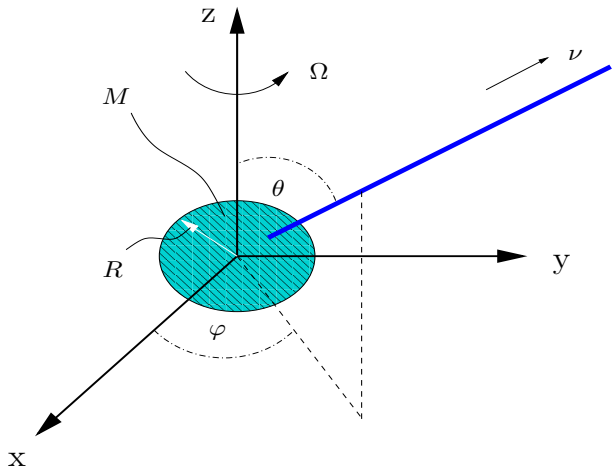}}
\end{minipage}
\vspace{0.5cm}
\caption{\label{fig:orientation} Orientation of the neutrino beam in relation to a rotating gravitational source
of mass $M$, radius $R$, and angular velocity $\Omega$ which describes the Lense-Thirring metric.
As a quantum mechanical particle, the neutrino has the dynamical freedom to occupy any position $\rvec' = \lt(r', \th', \vph'\rt)$
while propagating through space from the source to the detector defined by the co-ordinates $\rvec = \lt(r, \th, \vph \rt)$. }
\end{figure}

For the purpose of this investigation, the gravitational phase can be introduced by means of the transformation
$\psi(x) \rightarrow \exp\lt(i\Phi_{\rm G}/\hbar\rt) \psi(x)$, leading to the total Hamiltonian $H = H_0 + H_{\Phi_{\rm G}}$,
where
\be
H_{\PhiG} & = & \alvec \cdot \lt(\nabvec \PhiG \rt) + \lt(\nabvec_t \PhiG \rt)
\label{H-phiG=}
\ee
is a first-order correction to $H_0$.
Therefore, (\ref{H-phiG=}) is the interaction Hamiltonian sought after, which describes a {\em direct} spin coupling to the gravitational field via $\PhiG$,
where the gravitational phase can be written as
\be
\PhiG(\rvec) & = & \int_{t_0}^{t} \, \d t' \, \lt(\nabvec_t \PhiG\rt) +
\int_{x_0}^{x} \, \d x' \, \lt(\nabvec_x \PhiG\rt) +
\int_{y_0}^{y} \, \d y' \, \lt(\nabvec_y \PhiG\rt) +
\int_{z_0}^{z} \, \d z' \, \lt(\nabvec_z \PhiG\rt).
\label{Berry1a=}
\ee
In terms of the Lense-Thirring metric, the explicit expressions for $\nabvec_\mu \Phi_{\rm G}$ are
\be
\lt(\nabvec_t \PhiG\rt) & = &
\lt[{M \over r'^3} (\rvec \cdot \rvec') - {2M \over r'} \rt] \pvec^{\Ct} -
{M \over r'^3} \, (t - t') \, \rvec' \cdot \pvec
\nn
& &{}- {2 \over 5} \, {M \Omega R^2 \over r'^3} \lt[
\lt(\rvec \times \pvec \rt)^{\Cz} - 2 \lt(\rvec' \times \pvec \rt)^{\Cz}
+ {3 \, z' \over r'^2} \lt(\rvec \times \rvec' \rt) \cdot \pvec \rt],
\label{grad-PhiG-t=}
\nl
\lt(\nabvec_x \PhiG\rt) & = &
\lt[{M \over r'^3} (\rvec \cdot \rvec') - {2M \over r'} \rt] \pvec^{\Cx} -
{M \over r'^3} \, (x - x') \, \rvec' \cdot \vec{p}
\nn
&  &{} + {2 \over 5} \, {M \Omega R^2 \over r'^3} \lt[
\lt[(t - t') \, \pvec^{\Cy} - y \, \pvec^{\Ct} \rt] + {3 \, y' \over r'^2}
\lt\{\rvec' \cdot \lt(\rvec - \rvec' \rt) \pvec^{\Ct}
- (t - t')\lt[x' \, \pvec^{\Cx} + \rvec' \cdot \pvec \rt] \rt\} \rt],
\label{grad-PhiG-x=}
\nl
\lt(\nabvec_y \PhiG\rt) & = &
\lt[{M \over r'^3} (\rvec \cdot \rvec') - {2M \over r'} \rt] \pvec^{\Cy} -
{M \over r'^3} \, (y - y') \, \rvec' \cdot \pvec
\nn
&  &{} - {2 \over 5} \, {M \Omega R^2 \over r'^3} \lt[
\lt[(t - t') \, \pvec^{\Cx} - x \, \pvec^{\Ct} \rt] + {3 \, x' \over r'^2}
\lt\{\rvec' \cdot \lt(\rvec - \rvec' \rt) \pvec^{\Ct}
- (t - t')\lt[y' \, \pvec^{\Cy} + \rvec' \cdot \pvec \rt] \rt\} \rt],
\label{grad-PhiG-y=}
\nl
\lt(\nabvec_z \PhiG\rt) & = &
\lt[{M \over r'^3} \lt( \rvec \cdot \rvec'\rt) -
{2M \over r'} \rt] \pvec^{\Cz} - {M \over r'^3} \lt(z - z' \rt) \rvec' \cdot \pvec,
\label{grad-PhiG-z=}
\ee
where $\rvec = \lt(x, y, z\rt) = \lt(r, \th, \vph\rt)$ is a fixed observation point in space, and
\be
x & = & r \, \sin \th \, \cos \vph, \qquad y \ = \ r \, \sin \th \, \sin \vph, \qquad z \ = \ r \, \cos \th.
\label{observation-pt}
\ee
The orientation of the neutrino beam in space, expressed in terms of spherical polar co-ordinates, is shown in Figure~\ref{fig:orientation}.

%
%
%
While this treatment does not pose any problems for the spatial separation $\rvec - \rvec'$ found in (\ref{grad-PhiG-x=})--(\ref{grad-PhiG-z=}),
the time interval $t - t'$ requires more careful consideration, given that the spatial part of the event $x^\mu = \lt(t,\rvec\rt)$ is not
constrained to satisfy a classical trajectory defined by the Einstein field equations.
However, there is a way to at least approximate $t - t'$ quantum mechanically by imposing Poincar\'{e} invariance on
$\Dl x^\mu \equiv x^\mu - x'^\mu$ for a timelike interval $\Dl \tau$, such that
\be
\lt(\Dl \tau\rt)^2 & = & \lt(\Dl t\rt)^2 - \lt(\Dl \rvec \cdot \Dl \rvec\rt) \ \approx {1 \over \gm^2} \, \lt(\Dl t\rt)^2,
\label{Poincare}
\ee
where $\gm = E/m$, and $E = \sqrt{\lt(\pvec \cdot \pvec\rt) + m^2}$.
It follows that
\be
t - t' & \approx & \lt(1 - {m^2 \over E^2} \rt)^{-1/2} \sqrt{\lt(\rvec - \rvec'\rt) \cdot \lt(\rvec - \rvec'\rt)}.
\label{t-t'=}
\ee
Even with this approximation, however, there is a technical challenge in performing the integration over all spatial angles
contained in (\ref{t-t'=}), so we further approximate this time interval by assuming
\be
t - t' & \approx &
\lt(1 - {m^2 \over E^2} \rt)^{-1/2} {\lt(\rvec - \rvec'\rt) \cdot \lt(\rvec - \rvec'\rt) \over
\lt. \sqrt{\lt\langle \Dl \rvec \cdot \Dl \rvec\rt\rangle} \rt|_{r' = r}},
\label{t-t'-approx=}
\ee
where
\be
\lt\langle \Dl \rvec \cdot \Dl \rvec\rt\rangle & = & \int \lt(\rvec - \rvec'\rt) \cdot \lt(\rvec - \rvec'\rt) \d \Omega \ = \ 4 \pi \lt(r^2 + r'^2\rt)
\label{t-t'-avg=}
\ee
is the angular average over $\d \Om = \sin \th \, \d \th \, \d \vph$.
This construction leads to the expression
\be
t - t' & \approx & {1 \over \sqrt{8 \pi}} \lt(1 - {m^2 \over E^2} \rt)^{-1/2}
\lt[r + {r'^2 \over r} - {2 \, r' \over r} \lt(x \, \sin \th' \, \cos \vph' + y \, \sin \th' \, \sin \vph' + z \, \cos \th' \rt) \rt],
\label{t-t'-subst=}
\ee
which we substitute into (\ref{grad-PhiG-t=})--(\ref{grad-PhiG-z=}) for further evaluation.
It immediately follows that (\ref{t-t'-subst=}) identically vanishes in the limit as $x'^\mu \rightarrow x^\mu$.

\section{Neutrino wave packets and spin-gravity interactions}
\label{section:wave-packets}
%

There is some debate \cite{giunti} over whether neutrinos exist strictly as plane waves with precisely defined momentum eigenstates, or as a linear superposition of
plane waves with some momentum spread.
While the computational simplicity of the plane wave approach over using wave packets is obvious, the problem with such an assumption is that the neutrino is not
localizable in space, according to the Heisenberg uncertainty principle.
Although the choice to use wave packets is motivated more by the need to see the effects of curved space-time on some localized region where the neutrino can be identified,
we acknowledge that it makes intuitive sense to treat neutrinos like any other quantum particle according to standard quantum mechanics.
Therefore, we assume that the neutrino wavefunction is a Gaussian wave packet composed of a linear superposition of plane waves in momentum space, described by
\be
| \psi \rangle & = &
{1  \over (2\pi)^{3/2}} \int \d^3 k \, \xi(\vec{k}) \,
e^{i \vec{k} \cdot \vec{r}} |U(\vec{k})\rangle,
\label{psi=}
\ee
where $|U(\vec{k})\rangle$ is a free-particle spinor with momentum $\pvec = \hbar \, \kvec$, and the normalized Gaussian function of width $\sg_{\rm p}$ and
mean momentum $\langle \pvec \rangle = \hbar \, \kvec_0$ is
\be
\xi(\kvec) & = & {1 \over ({\sqrt{2\pi} \, \sigma_{\rm p}})^{3/2}} \, \exp\lt[- {(\kvec - \kvec_0)^2 \over 4 \, \sigma_{\rm p}^2} \rt].
\label{Gaussian=}
\ee
%

In order to make use of time-independent perturbation theory consistently, we need to assume that $\sg_{\rm p}$
is constant.
Of course, we expect that a realistic neutrino wave packet will spread with time, but assume in this model that the spreading is so rapid that
it will not adversely affect the main results to follow.
Therefore, we assume implicitly that the perturbation energy emerges over a very short time scale and remains frozen almost
immediately after the neutrinos start propagating through space, when the gravitational potential rapidly decays to zero.

The matrix element for a transition involving neutrino wave packets due to spin-gravity coupling is then formally described by
\be
\langle \psi(\rvec) |H_{\PhiG}| \psi(\rvec) \rangle & = &
{1 \over (2\pi)^3} \int \d^3 r' \ \d^3 k \ \d^3 k' \ \xi(\kvec) \, \xi(\kvec') \, \exp\lt[i\lt(\kvec - \kvec'\rt)\cdot \rvec'\rt]
\langle U(\kvec') | H_{\PhiG}\lt(\rvec, \rvec'\rt) | U(\kvec) \rangle,
\label{matrix-element}
\ee
where we integrate over all phase space, excluding the region occupied by the gravitational source.
It is from the choice of $|U(\vec{k})\rangle$ in (\ref{matrix-element}) which determines the nature of the matrix element due to spin-gravity coupling,
and hence the degree of mass-dependent energy difference predicted for the Dirac and Majorana neutrino.
This comparison between Dirac and Majorana wave packet constructions and their impact on (\ref{matrix-element}) immediately follows below.

\subsection{Dirac Neutrinos}

Assuming the Dirac representation \cite{itzykson} for $\alvec$ and $\bt$, the corresponding free-particle Dirac spinor for use in (\ref{psi=}) is
\be
|U(\vec{k})\rangle_{\pm}^{\rm Dirac} & = & \sqrt{E + m \over 2E}
\lt(
\begin{array}{c}
1 \\
{\hbar \, \sgvec \cdot \kvec \over {E + m}}
\end{array} \rt) \otimes | \pm \rangle, 
\label{Uspinor=}
\ee
where
\be
| + \rangle & = & \lt(
\begin{array}{c}
\cos \lt(\th/2\rt) \\
\sin \lt(\th/2\rt) e^{i \vph}
\end{array} \rt), \qquad
| - \rangle \ = \ \lt(
\begin{array}{c}
-\sin \lt(\th/2\rt) \\
\cos \lt(\th/2\rt) e^{i \vph}
\end{array} \rt)
\label{2-comp+/-}
\ee
are the two-component spinors \cite{bransden} with positive and negative helicity, respectively,
whose spin quantization axis is defined in terms of (\ref{observation-pt}) along the neutrino beam's radial direction.
Then the interaction Hamiltonian (\ref{H-phiG=}) in the Dirac representation is
\be
H_{\PhiG} & = & \lt(\sgvec \cdot \nabvec \PhiG\rt) \otimes
\lt(
\begin{array}{cc}
0 & 1 \\
1 & 0
\end{array} \rt)
+ \lt(\nabvec_t \PhiG\rt) \otimes
\lt(
\begin{array}{cc}
1 & 0 \\
0 & 1
\end{array} \rt).
\label{H-phiG-Dirac=}
\ee

It is straightforward to show that the transition from $| U(\pvec) \rangle$ to  $| U(\pvec') \rangle$ to evaluate (\ref{matrix-element}) is
described by
\be
\lefteqn{
\langle U(\pvec') | H_{\PhiG} | U(\pvec) \rangle^{\rm Dirac} \ = \ }
\nn \nn \nn
&& \sqrt{E + m \over 2E} \sqrt{E' + m \over 2E'} \lt[\lt(\nabvec_t \PhiG \rt)_{\rm S}
\lt(1 + {\lt(\pvec \cdot \pvec'\rt) \over (E + m)(E' + m)} \rt)
+ \lt({\lt(\lt(\nabvec \PhiG \rt)_{\rm S} \cdot \pvec\rt) \over E + m} + {\lt(\lt(\nabvec \PhiG \rt)_{\rm S} \cdot \pvec'\rt) \over E' + m}\rt) \rt.
\nn \nn \nn
&&{} - \lt. i  \, \sgvec \cdot \lt[ {\lt(\nabvec_t \PhiG \rt)_{\rm S} \lt(\pvec \times \pvec'\rt) \over (E + m)(E' + m)}
- \lt({\lt(\lt(\nabvec \PhiG\rt)_{\rm S} \times \pvec\rt) \over E + m} - {\lt(\lt(\nabvec \PhiG\rt)_{\rm S} \times \pvec'\rt) \over E' + m}\rt) \rt] \rt],
\label{transition-Dirac}
\ee
where
\be
\lt(\nabvec_\mu \PhiG \rt)_{\rm S} & = & {1 \over 2} \lt[\lt(\nabvec_\mu \PhiG \rt) (\pvec) + \lt(\nabvec_\mu \PhiG \rt) (\pvec')\rt]
\label{grad-PhiG-symm}
\ee
is the gradient of $\PhiG$ symmetrized with respect to $\pvec$ and $\pvec'$ for the purpose of integration over all intermediate momentum states.
While this symmetrization of $\nabvec_\mu \PhiG$ has no bearing on the final evaluation of (\ref{matrix-element}), it does serve
to help identify in (\ref{transition-Dirac}) the symmetric and antisymmetric contributions of the final matrix element
under the interchange of $\pvec$ and $\pvec'$.
We note immediately that the symmetric part of (\ref{transition-Dirac}) is independent of the initial spin polarization of the neutrino, while
the antisymmetric part is coupled to the spin operator.
This highlights an important feature about the wave packet treatment, in that a spread of momentum states in the neutrino allows for the possibility to
register a non-trivial contribution of the Dirac matrix element due to spin flip.
In contrast, it is impossible for a plane-wave neutrino to see the same effect, since the spin flip part of
(\ref{transition-Dirac}) identically vanishes as $\pvec \rightarrow \pvec'$.
When integrating over phase space, it follows from (\ref{transition-Dirac}) that the Dirac matrix element is
\be
\lefteqn{\langle \psi(\rvec) |H_{\PhiG}| \psi(\rvec) \rangle^{\rm Dirac} \ = \ 
{1 \over (2\pi)^3} \int
\d^3 r' \ \d^3 k \ \d^3 k' \ \ \xi(\kvec) \, \xi(\kvec') \, \sqrt{E + m \over 2E} \sqrt{E' + m \over 2E'}}
\nn \nn \nn
&&{} \times \lt\{\cos\lt[(\kvec - \kvec') \cdot \rvec'\rt]
\lt[\lt(\nabvec_t \PhiG \rt)_{\rm S}
\lt(1 + {\hbar^2 \lt(\kvec \cdot \kvec'\rt) \over (E + m)(E' + m)} \rt)
+ \hbar \lt({\lt(\lt(\nabvec \PhiG \rt)_{\rm S} \cdot \kvec\rt) \over E + m} + {\lt(\lt(\nabvec \PhiG \rt)_{\rm S} \cdot \kvec'\rt) \over E' + m}\rt)
\rt] 
\rt.
\nn \nn \nn
&&{} + \lt. \sin\lt[(\kvec - \kvec') \cdot \rvec'\rt] \lt[\langle \pm | \sgvec | \mp \rangle \cdot
\lt[{\hbar^2 \, \lt(\nabvec_t \PhiG \rt)_{\rm S} \lt(\kvec \times \kvec'\rt) \over (E + m)(E' + m)}
- \hbar \lt({\lt(\lt(\nabvec \PhiG \rt)_{\rm S} \times \kvec\rt) \over E + m} - {\lt(\lt(\nabvec \PhiG \rt)_{\rm S} \times \kvec'\rt) \over E' + m}\rt)\rt]\rt]\rt\},
\label{H-matrix-element-Dirac}
\ee
where the spin flip transition corresponding to the neutrino beam's orientation in space
is represented in terms of a cartesian co-ordinate frame by
\be
\langle \pm | \sgvec | \mp \rangle & = &
\lt[\cos \th \, \cos \vph \pm i \, \sin \vph \rt] \vec{\Cx} +
\lt[\cos \th \, \sin \vph \mp i \, \cos \vph \rt] \vec{\Cy} -
\sin \th \, \vec{\Cz}.
\label{orientation=}
\ee

\subsection{Majorana Neutrinos}

Performing the same set of calculations for the Majorana neutrino requires some preparation beforehand.
This is because the Majorana equivalent of the free-particle spinor (\ref{Uspinor=}) must be {\em self-conjugate} under
the charge conjugation operation
\be
|U^c(\kvec)\rangle_{\pm} & = & C \, \gm^{\hat{0}} \, |U^*(\kvec)\rangle_{\pm}, \qquad C \ = \ i \, \gm^{\hat{2}} \, \gm^{\hat{0}}.
\label{charge-conj}
\ee
For a Majorana spinor $| \chi \rangle$, this implies that \cite{fukugita}
\be
| \chi^c \rangle & = & \pm | \chi \rangle.
\label{self-conj=}
\ee
In order to best satisfy this requirement, we need to reformulate the problem in terms of the Weyl or chiral representation \cite{itzykson}
for the gamma matrices.
The reason follows from knowing that the Dirac spinor (\ref{Uspinor=}) can be decomposed into the
sum of a left- and right-handed projection expressed as more primitive two-component spinors.
Then the Majorana spinor can be constructed as the sum of the Dirac spinor's left-handed projection and its charge conjugate under (\ref{charge-conj}),
which then satisfies (\ref{self-conj=}) automatically.

We can explicitly proceed by first making use of the unitary matrix $\vec{U}$ which relates the Dirac and Weyl representations according to
\be
\gm_{\rm Weyl}^{\Cmu} & = & \vec{U} \, \gm_{\rm Dirac}^{\Cmu} \, \vec{U}^\dag,
\ee
where
\be
\vec{U} & = & {1 \over \sqrt{2}} \lt(1 - \gm^5 \, \gm^{\ze}\rt) \ = \
{1 \over \sqrt{2}} \lt(
\begin{array}{cc}
1 & 1 \\
-1 & 1
\end{array} \rt)
\label{Weyl-trans}
\ee
in the Dirac representation.
With (\ref{Weyl-trans}), the interaction Hamiltonian in the Weyl representation becomes
\be
H_{\PhiG} & = & \lt(\sgvec \cdot \nabvec \PhiG\rt) \otimes
\lt(
\begin{array}{cc}
1 & 0 \\
0 & -1
\end{array} \rt)
+ \lt(\nabvec_t \PhiG\rt) \otimes
\lt(
\begin{array}{cc}
1 & 0 \\
0 & 1
\end{array} \rt).
\label{H-phiG-Weyl=}
\ee
A simultaneous transformation of (\ref{Uspinor=}) then leads to Weyl form of the free-particle Dirac spinor
\be
|U(\vec{k})\rangle_{\pm}^{\rm Weyl} & = & \vec{U} |U(\vec{k})\rangle_{\pm}^{\rm Dirac}
\ = \ {1 \over \sqrt{2}} \, \sqrt{E + m \over 2E}
\lt(
\begin{array}{c}
\lt(1 + {\hbar \, \vec{\sigma} \cdot \vec{k} \over {E + m}}\rt) \\
-\lt(1 - {\hbar \, \vec{\sigma} \cdot \vec{k} \over {E + m}}\rt)
\end{array} \rt) \otimes | \pm \rangle, 
\label{Uspinor-weyl=}
\ee
where the left- and right-handed two-component spinors are
\begin{subequations}
\be
|\nu_L\rangle_{\pm} & = & {1 \over 2} \lt(1 - \gm^5\rt) |U(\vec{k})\rangle_{\pm}^{\rm Weyl}
\ = \ {1 \over \sqrt{2}} \, \sqrt{E + m \over 2E}
\lt(
\begin{array}{c}
0 \\
-\lt(1 - {\hbar \, \vec{\sigma} \cdot \vec{k} \over {E + m}}\rt)
\end{array} \rt) \otimes | \pm \rangle, 
\label{nu-left=}
\nl
|\nu_R\rangle_{\pm} & = & {1 \over 2} \lt(1 + \gm^5\rt) |U(\vec{k})\rangle_{\pm}^{\rm Weyl}
\ = \ {1 \over \sqrt{2}} \, \sqrt{E + m \over 2E}
\lt(
\begin{array}{c}
\lt(1 + {\hbar \, \vec{\sigma} \cdot \vec{k} \over {E + m}}\rt) \\
0
\end{array} \rt) \otimes | \pm \rangle, 
\label{nu-right=}
\ee
\end{subequations}
such that $|U (\kvec) \rangle_{\pm}^{\rm Weyl} = \lt(|\nu_L\rangle + |\nu_R\rangle \rt)_{\pm}$.
By then applying (\ref{charge-conj}) to $|\nu_L\rangle$ and $|\nu_R\rangle$, we obtain the
charge conjugated spinors
\begin{subequations}
\be
|\nu_L^c\rangle_{\pm} & = & {1 \over \sqrt{2}} \, \sqrt{E + m \over 2E}
\lt(
\begin{array}{c}
-i \, \sgvec^{\Cy} \lt(1 - {\hbar \, \sgvec^* \cdot \kvec \over {E + m}} \rt) \\
0
\end{array} \rt) \otimes | \pm \rangle
\ = \
{1 \over \sqrt{2}} \, \sqrt{E + m \over 2E}
\lt(
\begin{array}{c}
-i \lt(1 + {\hbar \, \sgvec \cdot \kvec \over {E + m}} \rt) \sgvec^{\Cy} \\
0
\end{array} \rt) \otimes | \pm \rangle, 
\label{nu-left-c=}
\nl
|\nu_R^c\rangle_{\pm} & = &
{1 \over \sqrt{2}} \, \sqrt{E + m \over 2E}
\lt(
\begin{array}{c}
0 \\
-i \, \sgvec^{\Cy} \lt(1 + {\hbar \, \sgvec^* \cdot \kvec \over {E + m}}\rt)
\end{array} \rt) \otimes | \pm \rangle
\ = \
{1 \over \sqrt{2}} \, \sqrt{E + m \over 2E}
\lt(
\begin{array}{c}
0 \\
-i \lt(1 - {\hbar \, \sgvec \cdot \kvec \over {E + m}}\rt) \sgvec^{\Cy}
\end{array} \rt) \otimes | \pm \rangle, 
\label{nu-right-c=}
\ee
\end{subequations}
where $\lt(\sgvec^{\Cj}\rt)^* = - \sgvec^{\Cy} \, \sgvec^{\Cj} \, \sgvec^{\Cy}$.
Therefore, we can define the free-particle Majorana spinors for the matrix element (\ref{matrix-element}) as
\begin{subequations}
\be
| W_1 (\kvec) \rangle_{\pm}^{\rm Maj.} & \equiv \lt(| \nu_L \rangle + | \nu_L^c \rangle \rt)_{\pm},
\nl
| W_2 (\kvec) \rangle_{\pm}^{\rm Maj.} & \equiv \lt(| \nu_R \rangle - | \nu_R^c \rangle \rt)_{\pm},
\ee
\end{subequations}
which by construction automatically satisfy (\ref{self-conj=}).
That is,
\begin{subequations}
\be
|W_1(\kvec)\rangle_{\pm}^{\rm Maj.} 
& = & {1 \over \sqrt{2}} \, \sqrt{E + m \over 2E}
\lt(
\begin{array}{c}
-i \, \sgvec^{\Cy} \lt(1 - {\hbar \, \sgvec^* \cdot \kvec \over {E + m}} \rt) \\
-\lt(1 - {\hbar \, \sgvec \cdot \kvec \over {E + m}} \rt)
\end{array} \rt) \otimes | \pm \rangle
\ = \
{1 \over \sqrt{2}} \, \sqrt{E + m \over 2E}
\lt(
\begin{array}{c}
-i \lt(1 + {\hbar \, \sgvec \cdot \kvec \over {E + m}} \rt) \sgvec^{\Cy} \\
-\lt(1 - {\hbar \, \sgvec \cdot \kvec \over {E + m}} \rt)
\end{array} \rt) \otimes | \pm \rangle, 
\label{WLspinor=}
\nn
\nl
|W_2(\kvec)\rangle_{\pm}^{\rm Maj.} 
& = & {1 \over \sqrt{2}} \, \sqrt{E + m \over 2E}
\lt(
\begin{array}{c}
\lt(1 + {\hbar \, \sgvec \cdot \kvec \over {E + m}} \rt) \\
i \, \sgvec^{\Cy} \lt(1 + {\hbar \, \sgvec^* \cdot \kvec \over {E + m}}\rt)
\end{array} \rt) \otimes | \pm \rangle
\ = \
{1 \over \sqrt{2}} \, \sqrt{E + m \over 2E}
\lt(
\begin{array}{c}
\lt(1 + {\hbar \, \sgvec \cdot \kvec \over {E + m}} \rt) \\
i \lt(1 - {\hbar \, \sgvec \cdot \kvec \over {E + m}}\rt) \sgvec^{\Cy}
\end{array} \rt) \otimes | \pm \rangle. 
\label{WRspinor=}
\ee
\end{subequations}
%
%
%

%

The most transparent way to evaluate the Majorana matrix element for (\ref{WLspinor=}) and (\ref{WRspinor=}) is
to determine each of the components using (\ref{nu-left=})--(\ref{nu-right-c=}).
For transitions from a left-handed to left-handed (right-handed to right-handed) two-component spinor, we obtain
\be
\lefteqn{
\langle \nu_{L(R)}(\pvec') | H_{\PhiG} | \nu_{L(R)}(\pvec) \rangle \ = \ }
\nn \nn \nn
&& {1 \over 2} \, \sqrt{E + m \over 2E} \sqrt{E' + m \over 2E'} \lt[\lt(\nabvec_t \PhiG \rt)_{\rm S}
\lt(1 + {\lt(\pvec \cdot \pvec'\rt) \over (E + m)(E' + m)} \rt)
+ \lt({\lt(\lt(\nabvec \PhiG \rt)_{\rm S} \cdot \pvec\rt) \over E + m} + {\lt(\lt(\nabvec \PhiG \rt)_{\rm S} \cdot \pvec'\rt) \over E' + m}\rt) \rt.
\nn \nn \nn
&&{} \mp \lt\{\lt(\sgvec \cdot \nabvec \PhiG \rt)_{\rm S}
\lt(1 - {\lt(\pvec \cdot \pvec'\rt) \over (E + m)(E' + m)} \rt) \rt.
\nn \nn \nn
&&{} + \lt. {\sgvec \cdot \pvec \over E + m} \lt(\lt(\nabvec_t \PhiG\rt)_{\rm S} + {\lt(\lt(\nabvec \PhiG \rt)_{\rm S} \cdot \pvec'\rt) \over E' + m}\rt)
+ {\sgvec \cdot \pvec' \over E + m} \lt(\lt(\nabvec_t \PhiG\rt)_{\rm S} + {\lt(\lt(\nabvec \PhiG \rt)_{\rm S} \cdot \pvec\rt) \over E + m}\rt)
\rt\}
\nn \nn \nn
&&{} - \lt. i  \, \sgvec \cdot \lt[ {\lt(\nabvec_t \PhiG \rt)_{\rm S} \lt(\pvec \times \pvec'\rt) \over (E + m)(E' + m)}
- \lt({\lt(\lt(\nabvec \PhiG\rt)_{\rm S} \times \pvec\rt) \over E + m} - {\lt(\lt(\nabvec \PhiG\rt)_{\rm S} \times \pvec'\rt) \over E' + m}\rt)\rt]
\mp i \, {\lt(\nabvec \PhiG\rt)_{\rm S} \cdot \lt(\pvec \times \pvec'\rt) \over (E + m)(E' + m)} \rt],
\label{H-LL-RR=}
\ee
where the upper (lower) sign corresponds to the left- (right)-handed two-component spinor.
We immediately verify from comparison with (\ref{transition-Dirac}) that
\be
\langle \nu_L(\pvec') | H_{\PhiG} | \nu_L(\pvec) \rangle + \langle \nu_R(\pvec') | H_{\PhiG} | \nu_R(\pvec) \rangle
& = & \langle U(\pvec') | H_{\PhiG} | U(\pvec) \rangle^{\rm Dirac},
\ee
as expected.
Similarly, the conjugate left-handed to conjugate left-handed (conjugate right-handed to conjugate right-handed) two-component spinor transition is
\be
\lefteqn{
\langle \nu_{L(R)}^c(\pvec') | H_{\PhiG} | \nu_{L(R)}^c(\pvec) \rangle \ = \ }
\nn \nn \nn
&& {1 \over 2} \, \sqrt{E + m \over 2E} \sqrt{E' + m \over 2E'} \lt[\lt(\nabvec_t \PhiG \rt)_{\rm S}
\lt(1 + {\lt(\pvec \cdot \pvec'\rt) \over (E + m)(E' + m)} \rt)
+ \lt({\lt(\lt(\nabvec \PhiG \rt)_{\rm S} \cdot \pvec\rt) \over E + m} + {\lt(\lt(\nabvec \PhiG \rt)_{\rm S} \cdot \pvec'\rt) \over E' + m}\rt) \rt.
\nn \nn \nn
&&{} \mp \lt\{\lt(\sgvec^* \cdot \nabvec \PhiG \rt)_{\rm S}
\lt(1 - {\lt(\pvec \cdot \pvec'\rt) \over (E + m)(E' + m)} \rt) \rt.
\nn \nn \nn
&&{} + \lt. {\sgvec^* \cdot \pvec \over E + m} \lt(\lt(\nabvec_t \PhiG\rt)_{\rm S} + {\lt(\lt(\nabvec \PhiG \rt)_{\rm S} \cdot \pvec'\rt) \over E' + m}\rt)
+ {\sgvec^* \cdot \pvec' \over E + m} \lt(\lt(\nabvec_t \PhiG\rt)_{\rm S} + {\lt(\lt(\nabvec \PhiG \rt)_{\rm S} \cdot \pvec\rt) \over E + m}\rt)
\rt\}
\nn \nn \nn
&&{} + \lt. i  \, \sgvec^* \cdot \lt[ {\lt(\nabvec_t \PhiG \rt)_{\rm S}  \lt(\pvec \times \pvec'\rt) \over (E + m)(E' + m)}
- \lt({\lt(\lt(\nabvec \PhiG\rt)_{\rm S} \times \pvec\rt) \over E + m} - {\lt(\lt(\nabvec \PhiG\rt)_{\rm S} \times \pvec'\rt) \over E' + m}\rt)\rt]
\pm i \, {\lt(\nabvec \PhiG\rt)_{\rm S} \cdot \lt(\pvec \times \pvec'\rt) \over (E + m)(E' + m)} \rt].
\label{H-LL-RR-c=}
\ee
Therefore, from using (\ref{H-LL-RR=}) and (\ref{H-LL-RR-c=}), it follows that the relevant transition matrix element for the Majorana neutrino is
%
\be
\lefteqn{\langle W_{1(2)}(\pvec') | H_{\PhiG} | W_{1(2)}(\pvec) \rangle^{\rm Maj.} \ = \
\langle U(\pvec') | H_{\PhiG} | U(\pvec) \rangle^{\rm Dirac} }
\nn \nn \nn
&&{} + \sqrt{E + m \over 2E} \sqrt{E' + m \over 2E'} \lt\{i \, \sgvec \cdot
\lt[ {\lt(\nabvec_t \PhiG\rt)_{\rm S} \lt(\pvec \times \pvec'\rt) \over (E + m)(E' + m)}
- \lt({\lt(\lt(\nabvec \PhiG\rt)_{\rm S} \times \pvec\rt) \over E + m}
- {\lt(\lt(\nabvec \PhiG\rt)_{\rm S} \times \pvec'\rt) \over E' + m}\rt) \rt] \rt.
\nn \nn \nn
&&{} - i \, \sgvec^{\Cy} \lt[{\lt(\nabvec_t \PhiG\rt)_{\rm S} \lt(\pvec \times \pvec'\rt)^{\Cy} \over (E + m)(E' + m)}
- \lt({\lt(\lt(\nabvec \PhiG\rt)_{\rm S} \times \pvec\rt)^{\Cy} \over E + m}
- {\lt(\lt(\nabvec \PhiG\rt)_{\rm S} \times \pvec'\rt)^{\Cy} \over E' + m}\rt) \rt]
\nn \nn \nn
&&{} \mp  \lt[
\lt(1 - {\lt(\pvec \cdot \pvec'\rt) \over (E + m)(E' + m)} \rt) \lt[\lt(\sgvec \cdot \nabvec \PhiG \rt)_{\rm S} - \sgvec^{\Cy} \lt(\nabvec_y \PhiG\rt)_{\rm S} \rt] \rt.
\nn \nn \nn
&&{} + \lt. \lt. \lt(\lt(\nabvec_t \PhiG\rt)_{\rm S} + {\lt(\lt(\nabvec \PhiG\rt)_{\rm S} \cdot \pvec'\rt) \over E' + m} \rt)
{\lt(\sgvec \cdot \pvec - \sgvec^{\Cy} \, \pvec^{\Cy}\rt) \over E + m}
+ \lt(\lt(\nabvec_t \PhiG\rt)_{\rm S} + {\lt(\lt(\nabvec \PhiG\rt)_{\rm S} \cdot \pvec\rt) \over E + m} \rt)
{\lt(\sgvec \cdot \pvec' - \sgvec^{\Cy} \, \pvec'^{\Cy}\rt) \over E' + m} \rt] \rt\},
\label{transition-Majorana}
\ee
and when integrated over all phase space, the Majorana matrix element is then
\be
\lefteqn{\langle \psi_{1(2)}(\vec{r}) |H_{\Phi_{\rm G}}| \psi_{1(2)}(\vec{r}) \rangle^{\rm Maj.} \ = \
\langle \psi(\vec{r}) |H_{\Phi_{\rm G}}| \psi(\vec{r}) \rangle^{\rm Dirac} -
{1 \over (2\pi)^3} \int \d^3 r' \ \d^3 k \ \d^3 k' \ \ \xi(\vec{k}) \, \xi(\vec{k}') \, \sqrt{E + m \over 2E} \sqrt{E' + m \over 2E'}}
\nn \nn \nn
&&{} \times \lt\{\sin\lt[(\kvec - \kvec') \cdot \rvec'\rt] \lt[
\langle \pm | \sgvec  | \mp \rangle \cdot \lt[
{\hbar^2 \lt(\nabvec_t \PhiG\rt)_{\rm S} \lt(\kvec \times \kvec'\rt) \over (E + m)(E' + m)}
- \lt({\hbar \lt(\lt(\nabvec \PhiG\rt)_{\rm S} \times \kvec\rt) \over E + m}
- {\hbar \lt(\lt(\nabvec \PhiG\rt)_{\rm S} \times \kvec'\rt) \over E' + m}\rt) \rt] \rt. \rt.
\nn \nn \nn
&&{} - \lt. \lt. \langle \pm | \sgvec | \mp \rangle^{\Cy}  \lt[{\hbar^2\lt(\nabvec_t \PhiG\rt)_{\rm S} \lt(\kvec \times \kvec'\rt)^{\Cy} \over (E + m)(E' + m)}
- \lt({\hbar \lt(\lt(\nabvec \PhiG\rt)_{\rm S} \times \kvec\rt)^{\Cy} \over E + m}
- {\hbar \lt(\lt(\nabvec \PhiG\rt)_{\rm S} \times \kvec'\rt)^{\Cy} \over E' + m}\rt)\rt] \rt]\rt.
\nn \nn \nn
&&{} \pm \, \cos\lt[(\kvec - \kvec') \cdot \rvec'\rt] \langle \pm |
\lt[\lt(1 - {\hbar^2\lt(\kvec \cdot \kvec'\rt) \over (E + m)(E' + m)} \rt) \lt[\lt(\sgvec \cdot \nabvec \PhiG \rt)_{\rm S} - \sgvec^{\Cy} \lt(\nabvec_y \PhiG\rt)_{\rm S} \rt] \rt.
\nn \nn \nn
&&{} + \lt. \lt. \hbar \lt(\lt(\nabvec_t \PhiG\rt)_{\rm S} + {\hbar \lt(\lt(\nabvec \PhiG\rt)_{\rm S} \cdot \kvec'\rt) \over E' + m} \rt)
{\lt(\sgvec \cdot \kvec - \sgvec^{\Cy} \, \kvec^{\Cy}\rt) \over E + m}
+ \hbar \lt(\lt(\nabvec_t \PhiG\rt)_{\rm S} + {\hbar \lt(\lt(\nabvec \PhiG\rt)_{\rm S} \cdot \kvec\rt) \over E + m} \rt)
{\lt(\sgvec \cdot \kvec' - \sgvec^{\Cy} \, \kvec'^{\Cy}\rt) \over E' + m} \rt] | \mp \rangle \rt\}.
\nn
\label{H-matrix-element-Majorana}
\ee

Immediately, we can recognize that the Majorana matrix element (\ref{H-matrix-element-Majorana}) departs significantly in form compared to its
Dirac counterpart (\ref{H-matrix-element-Dirac}).
Although a more detailed analysis follows later in this paper, it is already clear that the outcome of using a self-conjugate spinor is to
induce a frame-dependent expression which favours the $y$-direction.
This comes from the presence of the $\sgvec^{\Cy}$ Pauli matrices in $| \nu_L^c \rangle$ and $| \nu_R^c \rangle$, which contribute
exclusively to (\ref{H-LL-RR-c=}).
Indeed, if we tag $\sgvec^{\Cy}$ in (\ref{nu-left-c=}) and (\ref{nu-right-c=}) with a parameter $\epsilon$,
then (\ref{H-LL-RR-c=}) is of order $\epsilon^2$ and vanishes in the limit $\epsilon \rightarrow 0$,
leaving only (\ref{H-LL-RR=}) to contribute to (\ref{H-matrix-element-Majorana}), as expected.

The Dirac matrix element (\ref{H-matrix-element-Dirac}) and the corresponding Majorana version (\ref{H-matrix-element-Majorana}) for $| W_1 \rangle_{\pm}^{\rm Maj.}$
are the two physically relevant cases that we consider exclusively for the rest of this paper.
However, for the sake of completeness, we also briefly examined the matrix element for a transition from $| W_1 \rangle_{\pm}^{\rm Maj.}$ 
to $| W_2 \rangle_{\pm}^{\rm Maj.}$. 
To do this, we need the transition from the conjugate right-handed two-component spinor to the left-handed (conjugate left-handed to right-handed) expression,
where we obtain
\be
\lefteqn{
\langle \nu_{L(R)}(\pvec') | H_{\PhiG} | \nu_{R(L)}^c(\pvec) \rangle \ = \ }
\nn \nn \nn
&& {1 \over 2} \, \sqrt{E + m \over 2E} \sqrt{E' + m \over 2E'} \lt[
\lt[\lt(\sgvec \times \nabvec \PhiG\rt)_{\rm S}^{\Cy} - i \lt(\nabvec_y \PhiG\rt)_{\rm S}\rt]\lt(1 - {\lt(\pvec \cdot \pvec'\rt) \over (E + m)(E' + m)} \rt)
+ \sgvec^{\Cy} \, {\lt(\nabvec \PhiG\rt)_{\rm S} \cdot \lt(\pvec \times \pvec'\rt) \over (E + m)(E' + m)}  \rt.
\nn \nn \nn
&&{}
+ {\lt[\lt(\sgvec \times \pvec\rt)^{\Cy} - i \, \pvec^{\Cy}\rt] \over E + m}\lt(\lt(\nabvec_t \PhiG\rt)_{\rm S} + {\lt(\lt(\nabvec \PhiG \rt)_{\rm S} \cdot \pvec'\rt) \over E' + m}\rt)
+ {\lt[\lt(\sgvec \times \pvec'\rt)^{\Cy} - i \, \pvec'^{\Cy}\rt] \over E' + m}\lt(\lt(\nabvec_t \PhiG\rt)_{\rm S} + {\lt(\lt(\nabvec \PhiG \rt)_{\rm S} \cdot \pvec\rt) \over E + m}\rt)
\nn \nn \nn
&&{} \pm \lt\{ \lt[ {\lt(\nabvec_t \PhiG \rt)_{\rm S} \lt(\pvec \times \pvec'\rt)^{\Cy} \over (E + m)(E' + m)}
- \lt({\lt(\lt(\nabvec \PhiG\rt)_{\rm S} \times \pvec\rt)^{\Cy} \over E + m} - {\lt(\lt(\nabvec \PhiG\rt)_{\rm S} \times \pvec'\rt)^{\Cy} \over E' + m}\rt)\rt] \rt.
\nn \nn \nn
&&{} +  i  \lt[ {\lt(\nabvec_t \PhiG \rt)_{\rm S} \lt[\sgvec \times \lt(\pvec \times \pvec'\rt)\rt]^{\Cy} \over (E + m)(E' + m)}
- \lt({\lt[\sgvec \times \lt(\lt(\nabvec \PhiG\rt)_{\rm S} \times \pvec\rt)\rt]^{\Cy} \over E + m}
- {\lt[\sgvec \times \lt(\lt(\nabvec \PhiG\rt)_{\rm S} \times \pvec'\rt)\rt]^{\Cy} \over E' + m}\rt)\rt]
\nn \nn \nn
&&{} + \lt. \lt.
i \, \sgvec^{\Cy} \lt[\lt(\nabvec_t \PhiG \rt)_{\rm S}
\lt(1 + {\lt(\pvec \cdot \pvec'\rt) \over (E + m)(E' + m)} \rt)
+ \lt({\lt(\lt(\nabvec \PhiG \rt)_{\rm S} \cdot \pvec\rt) \over E + m} + {\lt(\lt(\nabvec \PhiG \rt)_{\rm S} \cdot \pvec'\rt) \over E' + m}\rt)\rt]
\rt\} \rt].
\label{H-Rc-L=}
\ee
Therefore, it follows that the Majorana transition matrix element from (\ref{WLspinor=}) to (\ref{WRspinor=}) is
\be
\lefteqn{\langle W_2(\pvec') | H_{\PhiG} | W_1(\pvec) \rangle^{\rm Maj.} \ = \ }
\nn \nn \nn
&&{} -i \, \sqrt{E + m \over 2E} \sqrt{E' + m \over 2E'}
\lt[ {\lt(\nabvec_t \PhiG \rt)_{\rm S} \lt[\sgvec \times \lt(\pvec \times \pvec'\rt)\rt]^{\Cy} \over (E + m)(E' + m)}
- \lt({\lt[\sgvec \times \lt(\lt(\nabvec \PhiG\rt)_{\rm S} \times \pvec\rt)\rt]^{\Cy} \over E + m}
- {\lt[\sgvec \times \lt(\lt(\nabvec \PhiG\rt)_{\rm S} \times \pvec'\rt)\rt]^{\Cy} \over E' + m}\rt)\rt]
\nn \nn \nn
&&{} -i \, \sqrt{E + m \over 2E} \sqrt{E' + m \over 2E'} \lt[\lt(\nabvec_y \PhiG\rt)_{\rm S} \lt(1 - {\lt(\pvec \cdot \pvec'\rt) \over (E + m)(E' + m)} \rt)  \rt.
\nn \nn \nn
&&{}
+ \lt. \lt(\lt(\nabvec_t \PhiG\rt)_{\rm S} + {\lt(\lt(\nabvec \PhiG \rt)_{\rm S} \cdot \pvec'\rt) \over E' + m}\rt){\pvec^{\Cy} \over E + m}
+  \lt(\lt(\nabvec_t \PhiG\rt)_{\rm S} + {\lt(\lt(\nabvec \PhiG \rt)_{\rm S} \cdot \pvec\rt) \over E + m}\rt){\pvec'^{\Cy} \over E' + m} \rt]
\nn \nn \nn
&&{}
- \sqrt{E + m \over 2E} \sqrt{E' + m \over 2E'} \lt[ {\lt(\nabvec_t \PhiG \rt)_{\rm S} \lt(\pvec \times \pvec'\rt)^{\Cy} \over (E + m)(E' + m)}
- \lt({\lt(\lt(\nabvec \PhiG\rt)_{\rm S} \times \pvec\rt)^{\Cy} \over E + m} - {\lt(\lt(\nabvec \PhiG\rt)_{\rm S} \times \pvec'\rt)^{\Cy} \over E' + m}\rt)\rt].
\label{transition-Majorana-L-R}
\ee
What is most surprising about this result is that a real and non-trivial contribution survives
due to its first three terms in (\ref{transition-Majorana-L-R}).
This is because they are both antisymmetric under the interchange of $\pvec$ and $\pvec'$ with an overall factor of $i$, which when multiplied by
$\exp\lt[i\lt(\kvec - \kvec'\rt) \cdot \rvec'\rt]$ and integrated over all phase space, yields a real contribution that comes from
$\sin\lt[\lt(\kvec - \kvec'\rt) \cdot \rvec'\rt]$ in the integrand, as demonstrated in (\ref{H-matrix-element-Majorana}).

\section{Evaluation of the gravitational phase matrix element}
\label{section:grav-matrix-element}

\subsection{Formalism}

To evaluate the Dirac and Majorana matrix elements (\ref{H-matrix-element-Dirac}) and (\ref{H-matrix-element-Majorana}), respectively, we
make use of spherical symmetry to integrate over phase space in terms of spherical polar co-ordinates.
Then the wave number $\kvec$ is expressed in the standard form
\be
\kvec^{\Cx} & = & k \, \sin \lm \, \cos \mu, \qquad \kvec^{\Cy} \ = \ k \, \sin \lm \, \sin \mu, \qquad \kvec^{\Cz} \ = \ k \, \cos \lm,
\label{kvec=}
\ee
where $\lm$ and $\mu$ cover the unit sphere and $k$ goes from zero to infinity.
Since we want to determine gravitational corrections to the neutrino oscillation length in terms of neutrino mass,
we proceed to expand (\ref{H-matrix-element-Dirac}) and (\ref{H-matrix-element-Majorana}) by power series in $m$, up to second order.
%
The next step is to integrate over all the position and momentum space angles.
From a first impression, this seems like a formidable challenge considering that the plane wave term $\exp\lt[{i \, \kvec \cdot \rvec'}\rt]$ is angle-dependent.
Fortunately, this problem can be circumvented by employing the Rayleigh plane wave expansion \cite{arfken}
\be
e^{i \, \kvec \cdot \rvec'} & = & 4 \, \pi \sum_{l=0}^\infty \, \sum_{m=-l}^l \, i^l \, j_l\lt(k \, r'\rt) Y_{lm}^*\lt(\lm,\mu\rt) \, Y_{lm}\lt(\th,\vph\rt),
\label{Rayleigh}
\ee
where $j_l\lt(k \, r'\rt)$ are the spherical Bessel functions and $Y_{lm}$ are the spherical harmonics for angles in position and momentum space.
Since the angles in (\ref{H-matrix-element-Dirac}) and (\ref{H-matrix-element-Majorana}) are expressed as polynomials of sinusoidal functions,
they can be expressed as a series expansion of spherical harmonics to take advantage of their orthogonality relations.
It follows that the series expansions truncate by virtue of the orthogonality conditions and the integrations over angles are performed {\em exactly}
for both the Dirac and Majorana matrix elements.

For the final integrations to evaluate, we first proceed to integrate over $k$, in which the integrals take the general form
\be
Q_c(n) & = & \int_0^\infty \exp\lt[-{\lt(k - k_0\rt)^2 \over 4 \, \sg_{\rm p}^2}\rt] \, \cos(k \, r') \, k^n \, \d k,
\label{Qc}
\nl
Q_s(n) & = & \int_0^\infty \exp\lt[-{\lt(k - k_0\rt)^2 \over 4 \, \sg_{\rm p}^2}\rt] \, \sin(k \, r') \, k^n \, \d k,
\label{Qs}
\ee
where $\k \equiv |\kvec_0|$ and $n$ is an integer with $-2 \leq n \leq 2$.
All details related to evaluation of (\ref{Qc}) and (\ref{Qs}) are contained in Appendix~\ref{appendix:momentum-integrals},
while the non-trivial radial integrals which remain are listed in Appendix~\ref{appendix:radial-integrals}.
For future reference, the width of the wave packet is expressed in terms of the dimensionless parameter $q \equiv \k/\sigma_{\rm p}$, and
the neutrino mass is now in units of its mean momentum, such that we have the dimensionless quantity $\bar{m} \equiv m/(\hbar \, \k) \ll 1$.

\subsection{Dirac and Majorana Matrix Elements}

For evaluating (\ref{H-matrix-element-Dirac}), we obtain the expression
\be
\langle \psi(\vec{r}) |H_{\Phi_{\rm G}}| \psi(\vec{r}) \rangle^{\rm Dirac} & = &
(\hbar \, \k)
\lt\{
{M \over r} \lt[C_0 + C_1 \, \bar{m} + C_2 \, \bar{m}^2 \rt]
+ {M \Omega R^2\over r^2} \sin \theta \lt[D_0 + D_1 \, \bar{m} + D_2 \, \bar{m}^2 \rt] \rt\},
\label{amplitude-Dirac}
\ee
for the Dirac matrix element, where
\be
C_j & = & \int_R^\infty \tilde{C}_j(r') \, \d r', \qquad D_j \ = \ \int_R^\infty \tilde{D}_j(r') \, \d r'
\label{C-D-Dirac}
\ee
are dimensionless functions dependent on $q$, $\k$, $R$, and $r$.
The integrands for (\ref{C-D-Dirac}) in terms of (\ref{Qc}) and (\ref{Qs}) are
\begin{subequations}
\be
\tilde{C}_0 (r') & = & \sqrt{2 \over \pi^3} \lt({q \over k_0}\rt)^3
\lt\{-{3 \, r \over k_0 \, r'} \lt[Q_c(1) \, Q_c(2) + Q_s(1) \, Q_s(2)\rt] \rt.
\nn
& &{} + \lt. {r \over k_0 \, r'^2} \lt[3 \, Q_c(2) \, Q_s(0) + 5 \, Q_c(1) \, Q_s(1)\rt] - {5 \, r \over k_0 \, r'^3} \, Q_s(0) \, Q_s(1)\rt\},
\label{C0-Dirac}
\nl
\tilde{C}_1 (r') & = & {1 \over 2} \, \sqrt{2 \over \pi^3} \lt({q \over k_0}\rt)^3
\lt\{ {r \over r'} \lt[Q_c(1)^2 - Q_s(1)^2 + 3 \, Q_c(0) \, Q_c(2) - 3 \, Q_s(0) \, Q_s(2)\rt] \rt.
\nn
& &{} - \lt. {r \over r'^2} \lt[3 \, Q_c(2) \, Q_s(-1) + 5 \, Q_c(0) \, Q_s(1)\rt] + {r \over r'^3} \lt[5 \, Q_s(-1) \, Q_s(1) - Q_s(0)^2 \rt] \rt\},
\label{C1-Dirac}
\nl
\tilde{C}_2 (r') & = & {1 \over 8} \, \sqrt{2 \over \pi^3} \lt({q \over k_0}\rt)^3
\lt\{ {k_0 \, r \over r'} \lt[3 \, Q_c(-1) \, Q_c(2) - 7 \, Q_c(0) \, Q_c(1) + 3 \, Q_s(-1) \, Q_s(2) - 7 \, Q_s(0) \, Q_s(1)\rt] \rt.
\nn
& &{} - \lt. {k_0 \, r \over r'^2} \lt[3 \, Q_c(0) \, Q_s(0) - 5 \, Q_c(-1) \, Q_s(1) - 3 \, Q_c(2) \, Q_s(-2) + 5 \, Q_c(1) \, Q_s(-1)\rt] \rt\}.
\label{C2-Dirac}
\nl
\tilde{D}_0 (r') & = & {1 \over 15} \, \sqrt{2 \over \pi^3} \lt({q \over k_0}\rt)^3
\lt\{{12 \, r^2 \over k_0 \, r'^2} \lt[Q_c(1) \, Q_s(2) + Q_c(2) \, Q_s(1)\rt] \rt.
\nn
& &{} - \lt. {4 \, r^2 \over k_0 \, r'^3} \lt[2 \, Q_c(1)^2 + 3 \, Q_c(0) \, Q_s(2) + 3 \, Q_s(1)^2\rt]
+ {16 \, r^2 \over k_0 \, r'^4} \, Q_c(1) \, Q_s(0) - {8 \, r^2 \over k_0 \, r'^5} \, Q_s(0)^2 \rt\},
\label{D0-Dirac}
\nl
\tilde{D}_1 (r') & = & {1 \over 15} \, \sqrt{2 \over \pi^3} \lt({q \over k_0}\rt)^3
\lt\{-{6 \, r^2 \over r'^2} \lt[Q_c(0) \, Q_s(2) - Q_c(2) \, Q_s(0)\rt] \rt.
\nn
& &{} + {2 \, r^2 \over r'^3} \lt[4 \, Q_c(0) \, Q_c(1) + 3 \, Q_s(-1) \, Q_s(2) - 3 \, Q_s(0) \, Q_s(1)\rt]
\nn
& &{} - \lt. {8 \, r^2 \over r'^4} \lt[Q_c(0) \, Q_s(0) + Q_c(1) \, Q_s(-1) \rt] + {8 \, r^2 \over r'^5} \, Q_s(-1) \, Q_s(0)\rt\},
\label{D1-Dirac}
\nl
\tilde{D}_2 (r') & = & {1 \over 30} \, \sqrt{2 \over \pi^3} \lt({q \over k_0}\rt)^3
\lt\{- {3 \, k_0 \, r^2 \over r'^2} \lt[Q_c(-1) \, Q_s(2) - Q_c(0) \, Q_s(1) - Q_c(1) \, Q_s(0) + Q_c(2) \, Q_s(-1)\rt] \rt.
\nn
& &{} + {k_0 \, r^2 \over r'^3} \lt[4 \, Q_c(-1) \, Q_c(1) - 4 \, Q_c(0)^2 + 3 \, Q_s(-2) \, Q_s(2) - 3 \, Q_s(0)^2 \rt]
\nn
& &{} - {4 \, k_0 \, r^2 \over r'^4} \lt[Q_c(-1) \, Q_s(0) - 2 \, Q_c(0) \, Q_s(-1) + Q_c(1) \, Q_s(-2)\rt]
\nn
& &{} + \lt. {4 \, k_0 \, r^2 \over r'^5} \lt[Q_s(-2) \, Q_s(0) - Q_s(-1)^2 \rt] \rt\}.
\label{D2-Dirac}
\ee
\end{subequations}
Substitution of (\ref{Qc}) and (\ref{Qs}) into (\ref{C0-Dirac})--(\ref{D2-Dirac}) followed by the radial integration
then leads to explicit expressions of $C_j$ and $D_j$, which are found in Appendix~\ref{appendix:dimensionless-functions}.


Similarly for the Majorana matrix element, we obtain from (\ref{H-matrix-element-Majorana}) the expression
%
%
\be
\lefteqn{\langle \psi_{1(2)}(\vec{r}) |H_{\Phi_{\rm G}}| \psi_{1(2)}(\vec{r}) \rangle^{\rm Maj.} \ = \
(\hbar \, \k) \lt\{ {M \over r} \lt[C_0 + C_1 \, \bar{m} + C_2 \, \bar{m}^2 \rt] \rt. }
\nn
&&{} \pm \lt. \sin \th \, \sin \vph
\lt[ {M \over r}
\langle \pm |\sgvec|\mp \rangle^{\Cy} \lt[C_{0 \Cy} + C_{1 \Cy} \, \bar{m} + C_{2 \Cy} \, \bar{m}^2\rt] 
+ {M \Omega R^2\over r^2} \langle \pm |\sgvec|\mp \rangle^{\Cx} \lt[D_{0 \Cx} + D_{1 \Cx} \, \bar{m} + D_{2 \Cx} \, \bar{m}^2\rt] \rt] \rt\},
\nn
\label{amplitude-Maj}
\ee
where
\be
C_{j \Cy} & = & \int_R^\infty \tilde{C}_{j \Cy}(r') \, \d r', \qquad D_{j \Cx} \ = \ \int_R^\infty \tilde{D}_{j \Cx}(r') \, \d r'
\ee
and
\begin{subequations}
\be
\tilde{C}_{0 \Cy} (r') & = & {1 \over 3 \, \pi^2} \, \lt({q \over k_0}\rt)^3
\lt\{{r \over k_0 \, r'} \lt[Q_c(1) \, Q_c(2) + Q_s(1) \, Q_s(2)\rt] \rt.
\nn
& &{} - \lt. {r \over k_0 \, r'^2} \lt[Q_c(2) \, Q_s(0) - Q_c(1) \, Q_s(1)\rt] - {r \over k_0 \, r'^3} \, Q_s(0) \, Q_s(1)\rt\},
\nl
\tilde{C}_{1 \Cy} (r') & = & {1 \over 6 \, \pi^2} \, \lt({q \over k_0}\rt)^3
\lt\{ {r \over r'} \lt[Q_c(1)^2 - Q_s(1)^2 - Q_c(0) \, Q_c(2) + Q_s(0) \, Q_s(2)\rt] \rt.
\nn
& &{} + \lt. {r \over r'^2} \lt[Q_c(2) \, Q_s(-1) - Q_c(0) \, Q_s(1)\rt] + {r \over r'^3} \lt[Q_s(-1) \, Q_s(1) - Q_s(0)^2 \rt] \rt\},
\nl
\tilde{C}_{2 \Cy} (r') & = &  {1 \over 24 \, \pi^2} \, \lt({q \over k_0}\rt)^3
\lt\{ -{k_0 \, r \over r'} \lt[3 \, Q_c(0) \, Q_c(1) + 3 \, Q_s(0) \, Q_s(1) + Q_c(-1) \, Q_c(2) + Q_s(-1) \, Q_s(2)\rt] \rt.
\nn
& &{} + {k_0 \, r \over r'^2} \lt[- Q_c(-1) \, Q_s(1) - Q_c(0) \, Q_s(0) + Q_c(1) \, Q_s(-1) + Q_c(2) \, Q_s(-2)\rt]
\nn
& &{} \lt. + {k_0 \, r \over r'^3} \lt[Q_s(-2) \, Q_s(1) + 3 \, Q_s(-1) \, Q_s(0) \rt] \rt\},
\nl
\tilde{D}_{0 \Cx} (r') & = & {2 \over 15} \, \sqrt{2 \over \pi^3} \lt({q \over k_0}\rt)^3
\lt\{{r^3 \over k_0 \, r'^3} \lt[Q_c(1) \, Q_c(2) - Q_s(1) \, Q_s(2)\rt] \rt.
\nn
& &{} - \lt. {r^3 \over k_0 \, r'^4} \lt[Q_c(2) \, Q_s(0) + 4 \, Q_c(1) \, Q_s(1)\rt]
+ {8 \, r^3 \over k_0 \, r'^5} \, Q_s(0) \, Q_s(1) \rt\},
\nl
\tilde{D}_{1 \Cx} (r') & = & {1 \over 15} \, \sqrt{2 \over \pi^3} \lt({q \over k_0}\rt)^3
\lt\{-{r^3 \over r'^3} \lt[Q_c(0) \, Q_c(2) + 3 \, Q_c(1)^2 + Q_s(0) \, Q_s(2) - Q_s(1)^2 \rt] \rt.
\nn
& &{} + {r^3 \over r'^4} \lt[4 \, Q_c(0) \, Q_s(1) + 3 \, Q_c(1) \, Q_s(0) + Q_c(2) \, Q_s(-1)\rt]
\nn
& &{} - \lt. {4 \, r^3 \over r'^5} \, Q_s(-1) \, Q_s(1)\rt\},
\nl
\tilde{D}_{2 \Cx} (r') & = & {1 \over 60} \, \sqrt{2 \over \pi^3} \lt({q \over k_0}\rt)^3
\lt\{{k_0 \, r^3 \over r'^3} \lt[-Q_c(-1) \, Q_c(2) + 13 \, Q_c(0) \, Q_c(1) + Q_s(-1) \, Q_s(2) + 3 \, Q_s(0) \, Q_s(1)\rt] \rt.
\nn
& &{} - {k_0 \, r^3 \over r'^4} \lt[-4 \, Q_c(-1) \, Q_s(1) + 7 \, Q_c(0) \, Q_s(0) + 10 \, Q_c(1) \, Q_s(-1) - Q_c(2) \, Q_s(-2) \rt]
\nn
& &{} - \lt. {4 \, k_0 \, r^3 \over r'^5} \lt[Q_s(-2) \, Q_s(1) - Q_s(-1) \, Q_s(0)\rt] \rt\}.
\ee
\end{subequations}
The final expressions for $C_{j\hat{y}}$ and $D_{j\hat{x}}$ are also listed in Appendix~\ref{appendix:dimensionless-functions}.


Even at the matrix element level, we notice that the expressions (\ref{amplitude-Dirac}) and (\ref{amplitude-Maj})
for the Dirac and Majorana neutrinos, respectively, have fundamental properties worth noting.
Regarding the Dirac matrix element, we identify the $C_j$ and $D_j$ in (\ref{amplitude-Dirac})
as the spin-diagonal and spin-flip structure functions for $j = 0, 1, 2$, since
$C_j$ is coupled to $M/r$ and $D_j$ is coupled to $M \Om R^2/r^2$.
The fact that $\sin \th$ is also coupled to the spin-flip terms clearly indicates that
only terms with the $z$-component of (\ref{orientation=}) contribute to the matrix element.
This leads to an obvious interpretation that rotational inertia from the Lense-Thirring metric
induces the helicity to flip for any neutrino propagation that is not along the source's axis of symmetry.
The effect is analogous to that of an inhomogeneous magnetic field which forces the particle's spin to flip.
Perhaps most importantly of all, (\ref{amplitude-Dirac}) has terms which are {\em linear} in $\bar{m}$, which
come from the normalization coefficient $\sqrt{(E + m)/(2E)}$ in $| U(\kvec) \rangle^{\rm Dirac}$.
This property becomes most relevant when considering gravitational corrections to neutrino oscillations.

As for the Majorana matrix element (\ref{amplitude-Maj}), it shares the same type of properties
as found in (\ref{amplitude-Dirac}), but with obvious differences.
First, while the spin-diagonal terms $C_j$ are common to both matrix elements,
the remaining terms have an overall factor of $\sin \th \, \sin \vph$,
corresponding to the $y$-component of the neutrino beam.
This is a direct consequence of the Majorana neutrino's self-conjugation condition,
since the charge conjugation operation has $\sgvec^{\Cy}$ present in its definition \cite{fukugita},
and suggests a preferred direction orthogonal to the source's symmetry axis.
However, this is inconsistent with the fact that the Lense-Thirring metric is {\em axisymmetric},
and implies that the $\vph$-dependence on (\ref{amplitude-Maj}) is purely due to how we defined the co-ordinate
axes beforehand.
Since the physics should be unaffected by the choice of co-ordinates to identify points in space,
we need to remove this artificially induced anisotropy by averaging over a complete cycle.
Second, it is unusual to note that the spin-flip parts of (\ref{amplitude-Maj}) are dependent on the $x$- and $y$-components of
(\ref{orientation=}), in contrast to the $z$-component for the Dirac neutrino.
The fact that the two remaining components decouple so cleanly from the integration indicates
a reflection of some fundamental property of the Majorana wave packet when it interacts gravitationally.
Most interestingly is the fact that a spin-flip term still contributes to the Majorana matrix element when $\Om \rightarrow 0$,
which is not true for the Dirac counterpart.
This again follows naturally from the self-conjugation property of Majorana neutrinos,
though it appears counterintuitive for such terms to appear as written.

\section{Gravitational effects on neutrino oscillation length and absolute neutrino masses}
\label{section:oscill-length}

In order to determine the observable consequences of (\ref{amplitude-Dirac}) and (\ref{amplitude-Maj}), we need to
obtain the perturbation energy which follows from these matrix elements.
To demonstrate this explicitly, we make use of the Brillouin-Wigner (BW) method \cite{ballentine} of time-independent perturbation
theory commonly used in the context of condensed matter physics, as opposed to the more well-known
Rayleigh-Schr\"{o}dinger (RS) method \cite{sakurai} we adopted in the previous investigation of this problem \cite{singh2}.
In general form, the perturbed energy eigenvalue following the BW method is
\be
E & = & E_0^{(n)} + \lt\langle n \rt|H_{\rm int.}\lt| n \rt\rangle + \sum_{m \neq n} {\lt|\lt\langle m \rt|H_{\rm int.}\lt| n \rt\rangle \rt|^2 \over E - E_0^{(m)}}
\nn
& &{} + \sum_{m_1, \, m_2 \neq n} {
\lt\langle n \rt|H_{\rm int.}\lt| m_1 \rt\rangle \lt\langle m_1 \rt|H_{\rm int.}\lt| m_2 \rt\rangle \lt\langle m_2 \rt|H_{\rm int.}\lt| n \rt\rangle
\over \lt(E - E_0^{(m_1)}\rt)\lt(E - E_0^{(m_2)}\rt)} + \cdots,
\label{BW-perturbation}
\ee
where $H_{\rm int.}$ is the interaction Hamiltonian, $\lt| n \rt\rangle$ is the unperturbed eigenstate,
and $E_0^{(n)}$ is its associated energy eigenvalue.
Although the perturbed eigenvalues from the BW method are generally less accessible than from the RS method, given the
presence of $E$ in the denominator for the second- and higher-order terms in the perturbation expansion,
for the special case of a second-order expansion for a two-level system, the perturbed energy eigenvalues can be obtained {\em exactly},
which is not possible to achieve with the RS method.
For this investigation, having a way to precisely determine the mass-induced energy difference for a two-neutrino system is extremely useful
for accurately predicting the effects of spin-gravity coupling on the neutrino oscillation length.

To determine the corrections to the known neutrino oscillation length due to gravitation, we need to first obtain the unperturbed
energy eigenvalue $E_0^{(\pm)}$ from the zeroth-order part of the Hamiltonian $H_0$.
Given (\ref{H0=}), it follows that
\be
H_0 \lt| \psi_0(\rvec) \rt\rangle & = & E_0^{(\pm)} \lt| \psi_0(\rvec) \rt\rangle,
\label{E0-equation}
\ee
where
\be
E_0^{(\pm)} & \approx & \sqrt{(\hbar \, \k)^2 + m^2} - {2M \over r} \, (\hbar \, \k)
+ {4 \over 5} \,  {M \Omega R^2\over r^3} \lt(L^{\hat{z}} \pm {\hbar \over 2}\rt)
\nn
& \approx & (\hbar \, \k) \lt[\lt(1 - {2M \over r}\rt) + {1 \over 2} \, \bar{m}^2 \rt]
 + {4 \over 5} \,  {M \Omega R^2\over r^3} \lt(L^{\hat{z}} \pm {\hbar \over 2}\rt),
\label{E0=}
\ee
and $L^{\hat{z}}$ is the orbital angular momentum along the gravitational source's axis of rotation.

It is clear from (\ref{E0=}) that the unperturbed part of the Dirac Hamiltonian cannot generate any gravitational corrections to the neutrino
oscillation length, since the mass-dependent part of $E_0^{(\pm)}$ is not coupled to the gravitational potential.
However, the situation is different with the introduction of the interaction Hamiltonian $H_{\PhiG}$ given by (\ref{H-phiG-Weyl=}).
Following (\ref{BW-perturbation}) and introducing for convenience (with a slight abuse of notation) the
unperturbed mass eigenstates
\be
\lt| \pm \rt\rangle & \equiv & \lt\{
\begin{array}{cl}
\lt(| \nu_L \rangle + | \nu_R \rangle \rt)_{\pm} & \qquad {\rm (Dirac),} \\ \\
\lt(| \nu_L \rangle + | \nu_L^c \rangle \rt)_{\pm} & \qquad {\rm (Majorana),}
\end{array} \rt.
\label{pm-mass-eigenstates}
\ee
it is shown that the second-order perturbed energy $E_{\bar{m}}^{(\pm)}$ is
\be
E_{\bar{m}}^{(\pm)} & = & E_0^{(\pm)} + \lt\langle \pm \rt| H_{\PhiG} \lt| \pm \rt\rangle
+ {\lt|\lt\langle \mp \rt| H_{\PhiG} \lt| \pm \rt\rangle\rt|^2 \over E_{\bar{m}}^{(\pm)} - E_0^{(\mp)}}.
\label{E-perturbed1=}
\ee
As stated earlier, this expression can be solved exactly for $E_{\bar{m}}^{(\pm)}$, leading to
\be
E_{\bar{m}}^{(\pm)} & = & {1 \over 2} \lt\{\lt[\lt(E_0^{(\pm)} + E_0^{(\mp)}\rt) + \lt\langle \pm \rt| H_{\PhiG} \lt| \pm \rt\rangle \rt] \rt.
\nn
& &{} + \lt. \sqrt{\lt[\lt(E_0^{(\pm)} + E_0^{(\mp)}\rt) + \lt\langle \pm \rt| H_{\PhiG} \lt| \pm \rt\rangle \rt]^2
- 4\lt[\lt(E_0^{(\pm)} + \lt\langle \pm \rt| H_{\PhiG} \lt| \pm \rt\rangle\rt)E_0^{(\mp)}
- \lt|\lt\langle \mp \rt| H_{\PhiG} \lt| \pm \rt\rangle\rt|^2 \rt]} \rt\},
\label{E-perturbed2=}
\ee
which reduces to $E_0^{(\pm)}$ in the limit as $H_{\PhiG} \rightarrow 0$.

\subsection{Dirac Neutrinos}

For the case of Dirac neutrinos, the matrix elements to evaluate (\ref{E-perturbed2=}) are
\begin{subequations}
\be
\lt\langle \pm \rt| H_{\PhiG} \lt| \pm \rt\rangle & = & (\hbar \, \k) \, {M \over r} \lt[C_0 + C_1 \, \bar{m} + C_2 \, \bar{m}^2 \rt],
\label{diagonal-term-Dirac-Maj}
\nl
\lt\langle \mp \rt| H_{\PhiG} \lt| \pm \rt\rangle & = & (\hbar \, \k) \, {M \Omega R^2\over r^2} \sin \th \lt[D_0 + D_1 \, \bar{m} + D_2 \, \bar{m}^2 \rt].
\label{spinflip-term-Dirac}
\ee
\end{subequations}
Upon substitution into (\ref{E-perturbed2=}) and after performing a power series expansion with respect to $\bar{m}$, it is shown that
the perturbed energy is
\be
E_{\bar{m}}^{(\pm)} & = & {4 \over 5} \,  {M \Omega R^2\over r^3} \, L^{\hat{z}}
+ (\hbar \, \k) \lt[\lt(1 - {2M \over r}\rt) + {1 \over 2} \, \bar{m}^2\rt]
+ (\hbar \, \k) \lt[\lt(F_0^{\rm Dirac}\rt) + \lt(F_1^{\rm Dirac}\rt) \bar{m} + \lt(F_2^{\rm Dirac}\rt) \bar{m}^2 \rt],
\label{E-perturbed-Dirac=}
\ee
where
\begin{subequations}
\be
F_0^{\rm Dirac} & = & {1 \over 2} \lt[C_0 \lt({M \over r}\rt) + {1 \over 5} \, \lt({\Dl_\pm \over \k \, r}\rt) \rt],
\label{F0-Dirac=}
\nl
F_1^{\rm Dirac} & = & {1 \over 2} \lt[C_1 \lt({M \over r}\rt) +  \lt({\Om_\pm \over \Dl_\pm}\rt)\rt],
\label{F1-Dirac=}
\nl
F_2^{\rm Dirac} & = & {1 \over 2} \lt[C_2 \lt({M \over r}\rt) + {1 \over 2} \lt[\lt({\chi_\pm \over \Dl_\pm}\rt)
 - 5 \, (\k \, r) \lt({\Om_\pm \over \Dl_\pm^{3/2}}\rt)^2\rt]\rt],
\label{F2-Dirac=}
\nl
\Om_\pm^{\rm Dirac} & = & 5 \, (\k \, r) \lt[C_0 \, C_1 \lt({M \over r}\rt)^2 + 4 \, \sin^2 \th \, D_0 \, D_1 \lt({M \Omega R^2\over r^2}\rt)^2 \rt]
\pm 4 \, C_1 \lt({M \over r}\rt)\lt({M \Omega R^2\over r^2}\rt),
\label{Omega-Dirac=}
\nl
\chi_\pm^{\rm Dirac} & = &  5 \, (\k \, r) \lt[\lt(C_1^2 + 2 \, C_0 \, C_2\rt) \lt({M \over r}\rt)^2
 + 4 \, \sin^2 \th \lt(D_1^2 + 2 \, D_0 \, D_2\rt) \lt({M \Omega R^2\over r^2}\rt)^2\rt]
\nn
& &{}
\pm 8 \, C_2 \lt({M \over r}\rt)\lt({M \Omega R^2\over r^2}\rt),
\label{Chi-Dirac=}
\nl
\lt(\Dl_\pm^{\rm Dirac}\rt)^2 & = & 25 \, (\k \, r)^2 C_0^2 \lt({M \over r}\rt)^2 + \lt[16 + 100 \, (\k r)^2 \sin^2 \th \, D_0^2\rt]\lt({M \Omega R^2\over r^2}\rt)^2
\nn
& &{}
\pm 40 \, (\k \, r) \, C_0 \lt({M \over r}\rt)\lt({M \Omega R^2\over r^2}\rt).
\label{Delta^2-Dirac=}
\ee
\end{subequations}
This leads \cite{fukugita,kim} to the final expression for the neutrino oscillation length $L_{\rm osc.} = 2
\pi/\lt(E_{\bar{m}_2}^{(\pm)} - E_{\bar{m}_1}^{(\pm)}\rt)$, where
\be
E_{\bar{m}_2}^{(\pm)} - E_{\bar{m}_1}^{(\pm)} & = & (\hbar \, \k) \lt[ 
\lt(F_1^{\rm Dirac}\rt) \lt(\bar{m}_2 -  \bar{m}_1\rt) + \lt(F_2^{\rm Dirac} + {1 \over 2}\rt) \lt(\bar{m}_2^2 -  \bar{m}_1^2\rt) \rt].
\label{Energy-shift-Dirac=}
\ee

It is worth emphasizing again that (\ref{Energy-shift-Dirac=}) is formally an {\em exact} expression, up to second order in $\bar{m}$,
for the mass-dependent perturbation energy difference,
and all plots presented for the Dirac neutrino case are based on (\ref{E-perturbed-Dirac=})--(\ref{Delta^2-Dirac=}).
Nonetheless, given that both $M/r \ll 1$ and $M \Om R^2/r^2 \ll 1$ for all numerical analyses considered, it is worthwhile to determine
the leading order contributions to the mass-dependent energy difference.
Therefore, it is shown that
\be
E_{\bar{m}}^{(\pm)} & \approx & {4 \over 5} \,  {M \Omega R^2\over r^3} \, L^{\hat{z}}
+ (\hbar \, \k) \lt[\lt(1 - {2M \over r}\rt) + {1 \over 2} \, \bar{m}^2\rt]
+ (\hbar \, \k) \, {M \over r} \lt[\lt(G_{0 (\pm)}^{\rm Dirac}\rt) + \lt(G_{1 (\pm)}^{\rm Dirac}\rt) \bar{m} + \lt(G_{2 (\pm)}^{\rm Dirac}\rt) \bar{m}^2 \rt]
\nn
& &{} + (\hbar \, \k) \, {M \Om R^2 \over r^2} \lt[\lt(K_0^{\rm Dirac}\rt) + \lt(K_1^{\rm Dirac}\rt) \bar{m} + \lt(K_2^{\rm Dirac}\rt) \bar{m}^2 \rt],
\label{E-perturbed-leading-Dirac=}
\nl
E_{\bar{m}_2}^{(\pm)} - E_{\bar{m}_1}^{(\pm)} & \approx &  (\hbar \, \k) \lt[{1 \over 2} \lt(\bar{m}_2^2 -  \bar{m}_1^2\rt) \rt.
\nn
& &{} + \lt. \lt[{M \over r}\lt(G_{1 (\pm)}^{\rm Dirac}\rt) + {M \Om R^2 \over r^2} \lt(K_1^{\rm Dirac}\rt)\rt]\lt(\bar{m}_2 -  \bar{m}_1\rt)
+ \lt[{M \over r}\lt(G_{2 (\pm)}^{\rm Dirac}\rt) + {M \Om R^2 \over r^2} \lt(K_2^{\rm Dirac}\rt)\rt]\lt(\bar{m}_2^2 -  \bar{m}_1^2\rt) \rt],
\nn
\label{Energy-shift-leading-Dirac=}
\ee
where
\begin{subequations}
\be
G_{0 (\pm)}^{\rm Dirac} & = & \lt({1 \over 2} \pm {1 \over \Lm} \rt)C_0,
\label{G0-Dirac=}
\nl
G_{1 (\pm)}^{\rm Dirac} & = & {1 \over 2} \, C_1 \pm {2 \over \Lm} \lt[C_1 - 50 \lt({\k \, r \over \Lm}\rt)^2 \sin^2 \th \, C_0 \, D_0 \, D_1 \rt],
\label{G1-Dirac=}
\nl
G_{2 (\pm)}^{\rm Dirac} & = & {1 \over 2} \, C_2 \pm {1 \over \Lm} \lt[C_2
- {25 \over 2} \lt({\k \, r \over \Lm}\rt)^2 \sin^2 \th \lt[C_0 \lt(D_1^2 + 2 \, D_0 \, D_2\rt) + 2 \, C_1 \, D_0 \, D_1 \rt. \rt.
\nn
& &{} - \lt. \lt. 75 \lt({\k \, r \over \Lm}\rt)^2 \sin^2 \th \, C_0 \, D_0^2 \, D_1^2\rt] \rt],
\label{G2-Dirac=}
\nl
K_0^{\rm Dirac} & = & {1 \over 5} \lt({\Lm \over \k \, r}\rt),
\label{K0-Dirac=}
\nl
K_1^{\rm Dirac} & = & 5 \lt({\k \, r \over \Lm}\rt) \sin^2 \th \, D_0 \, D_1,
\label{K1-Dirac=}
\nl
K_2^{\rm Dirac} & = & {5 \over 2} \lt({\k \, r \over \Lm}\rt) \sin^2 \th \lt[\lt(D_1^2 + D_0 \, D_2\rt)
- 25 \lt({\k \, r \over \Lm}\rt)^2 \sin^2 \th \, D_0^2 \, D_1^2 \rt],
\label{K2-Dirac=}
\nl
\lt(\Lm^{\rm Dirac}\rt)^2 & = & 4 + 25 \, (\k \, r)^2 \, \sin^2 \th \, D_0^2.
\label{Lambda-Dirac=}
\ee
\end{subequations}

The explicit expression for the leading-order gravitational corrections to the mass-induced energy difference (\ref{Energy-shift-leading-Dirac=})
can be grouped into two categories, namely the special case where $\th = 0, \, \pi$, followed by observations off the axis of symmetry where $\th \neq 0, \, \pi$.
For the first case, it is readily shown that
\begin{subequations}
\label{G-K-Dirac-th-1}
\be
G_{1 (\pm)}^{\rm Dirac} & = & \pm \, {1 \over 2} \, C_1,
\label{G1-Dirac-1=}
\nl
G_{2 (\pm)}^{\rm Dirac} & = & {1 \over 2} \lt(1 \pm 1\rt) C_2,
\label{G2-Dirac-1=}
\nl
K_1^{\rm Dirac} & = & K_2^{\rm Dirac} \ = \ 0, \qquad \th = 0, \, \pi
\label{K1-K2-Dirac-1=}
\ee
\end{subequations}
where
\begin{subequations}
\be
C_1 & = & \sqrt{2 \over \pi^3}\, {r \over R} \lt\{ \lt({r \over R} - 1\rt) \lt[\lt({r \over R} + 1\rt)\mu^{-3} - 10 \, \mu^{-2} \rt]  e^{-q^2/2} \, e^{-[\k(r - R)/q]^2} \rt.
\nn
& &{} + \lt[-{1 \over 2} \lt(q^2 - 12 \pi \, q + 8\rt) \ln \lt({r \over R}\rt) e^{-q^2/2} \, e^{-[\k(r - R)/q]^2}
\rt.
\nn
& &{} + \lt. 8 \pi \lt(1 - 2 \, \cos^2\lt(\k R\rt)\rt)e^{-2\krq} + 10 \sqrt{\pi} \, \sin \lt(\k R\rt) e^{-q^2/4} \, e^{-\krq}\rt] \mu^{-1}
\nn
& &{} + \lt[q \lt(11 \, q - 6 \pi\rt)e^{-q^2/2} - 8 \pi \, q \, \cos\lt(\k R\rt) \, \sin\lt(\k R\rt) e^{-2\krq} \rt.
\nn
& &{} - \lt. \lt. \sqrt{\pi} \lt[\lt(11 \, q - 6 \pi\rt) \cos \lt(\k R\rt) + 10 \, \sin \lt(\k R\rt) \rt] e^{-q^2/4} \, e^{-\krq} \rt] \rt\} + O\lt(\mu\rt),
\label{C1-approx}
\nl
C_2 & = & \sqrt{2 \over \pi^3}\, {r \over R} \lt\{ \lt[{3 \pi \over 2} \lt({r \over R} - 1\rt)q \, \mu^{-2}
+ {1 \over 8} \lt(6 \pi \, q^2 - q + 2\pi\rt) \ln \lt({r \over R}\rt) \mu^{-1} \rt]  e^{-q^2/2} \, e^{-[\k(r - R)/q]^2} \rt.
\nn
& &{} - {3 \sqrt{\pi^3} \over 2} \, \sin \lt(\k R\rt) q \, \mu^{-1} \, e^{-q^2/4} \, e^{-\krq} + {1 \over 16} \lt[q\lt(27 \, q^2 - 24 \pi \, q + 2\rt) + 20 \pi \rt]q \, e^{-q^2/2}
\nn
& &{} + \lt.  {3 \sqrt{\pi^3} \over 4} \lt[2 \cos\lt(\k r\rt) - \sin\lt(\k r\rt) \rt]
e^{-q^2/4} \, e^{-\krq} \rt\} + O\lt(\mu\rt).
\label{C2-approx}
\ee
\end{subequations}
For the case when $\th \neq 0, \, \pi$, it can be shown that
\begin{subequations}
\label{G-K-Dirac-th-2}
\be
G_{1 (\pm)}^{\rm Dirac} & \approx & {1 \over 2} \, C_1 \pm {2 \over 5} \, {1 \over \sin \th} \, {X_1 \over \lt(\k r\rt)},
\label{G1-Dirac-2=}
\nl
G_{2 (\pm)}^{\rm Dirac} & \approx & {1 \over 2} \, C_2 \pm {1 \over 5} \, {1 \over \sin \th} \, {X_2 \over \lt(\k r\rt)},
\label{G2-Dirac-2=}
\nl
K_1^{\rm Dirac} & \approx & \sin \th \, D_1,
\label{K1-Dirac-2=}
\nl
K_2^{\rm Dirac} & \approx & {1 \over 2} \, \sin \th \, D_2, \qquad \th \neq 0, \, \pi
\label{K2-Dirac-2=}
\ee
\end{subequations}
where
\begin{subequations}
\be
D_1 & = & \sqrt{2 \over \pi^3}\, {r^2 \over R^2} \lt\{-{8 \over 5} \lt[\ln \lt({r \over R}\rt) q \, \mu^{-2}
- \lt(q + \pi\rt)  \mu^{-1} \rt]  e^{-q^2/2} \, e^{-[\k(r - R)/q]^2} \rt.
\nn
& &{} - {8 \sqrt{\pi} \over 5} \, \cos \lt(\k R\rt) \mu^{-1} \, e^{-q^2/4} \, e^{-\krq}
\nn
& &{} + {1 \over 15} \lt[9 \, q + 20 \pi \, e^{-[\k(r - R)/q]^2} \rt] e^{-q^2/2} - {8 \pi \over 5} \, q \, e^{-2\krq}
\nn
& &{} - \lt.  {\sqrt{\pi} \over 5} \lt[8 \, \cos\lt(\k R\rt) + 4\lt(3 \, q + 2 \pi\rt) \sin\lt(\k R\rt) \rt]
e^{-q^2/4} \, e^{-\krq} \rt\} + O\lt(\mu\rt),
\label{D1-approx}
\nl
D_2 & = & \sqrt{2 \over \pi^3}\, {r^2 \over R^2} \lt\{{1 \over 5} \lt[\ln \lt({r \over R}\rt) \lt(2\pi \, q - 1\rt) q \, \mu^{-2}
- \lt(\pi \, q - 1\rt)  \mu^{-1} \rt]  e^{-q^2/2} \, e^{-[\k(r - R)/q]^2} \rt.
\nn
& &{} + {1 \over 5} \lt[q^2 \, e^{-q^2/2} + 2 \sqrt{\pi^3} \, \cos \lt(\k R\rt) \, e^{-q^2/4} \, e^{-\krq} \rt]q \, \mu^{-1}
\nn
& &{} - {8 \sqrt{\pi} \over 5} \, \cos \lt(\k R\rt) \mu^{-1} \, e^{-q^2/4} \, e^{-\krq}
\nn
& &{} + {1 \over 120} \lt[5 \, q\lt(3 \, q + 4 \pi \rt) -  6 \, e^{-[\k(r - R)/q]^2} \rt] q \, e^{-q^2/2}
\nn
& &{} + \lt.  {\sqrt{\pi} \over 5} \lt[q \, \cos\lt(\k R\rt) - 2 \, \sin\lt(\k R\rt) \rt]q \,
e^{-q^2/4} \, e^{-\krq} \rt\} + O\lt(\mu\rt),
\label{D2-approx}
\ee
\end{subequations}
and
\begin{subequations}
\be
X_1 & = & {1 \over D_0^2} \lt[C_1 \, D_0 - 2 \, C_0 \, D_1\rt] \ \approx \ - {5 \over 32} \lt(1 + {R \over r}\rt)q,
\label{X1}
\nl
X_2 & = & {C_0 \over D_0^3} \lt[D_1^2 - D_0 \, D_2\rt] + {1 \over D_0^2} \lt[C_2 \, D_0 - C_1 \, D_1\rt]
\nn
& \approx & {R \over r}\lt[{5 \over 256} \lt({r/R + 1 \over r/R - 1}\rt)  q^3 \lt(6 \pi \, q - 1\rt) \ln \lt({r \over R}\rt) - {15 \over 64} \, \pi \, q^2 \rt]\mu.
\label{X2}
\ee
\end{subequations}

\subsection{Majorana Neutrinos}

When performing the same calculation for Majorana neutrinos, some care has to be taken for evaluating the spin-flip part
of the matrix element, since it has an explicit dependence on $\vph$.
That is,
\be
\lefteqn{\lt\langle \mp \rt| H_{\PhiG} \lt| \pm \rt\rangle \ = \ }
\nn
&& \pm \, (\hbar \, \k) \, \sin \th \, \sin \vph \lt\{ {M \over r} \langle \pm |\sgvec|\mp \rangle^{\Cy} \lt[C_{0 \Cy} + C_{1 \Cy} \, \bar{m} + C_{2 \Cy} \, \bar{m}^2\rt]
+ {M \Omega R^2\over r^2} \langle \pm |\sgvec|\mp \rangle^{\Cx} \lt[D_{0 \Cx} + D_{1 \Cx} \, \bar{m} + D_{2 \Cx} \, \bar{m}^2\rt] \rt\}
\nn
& = & \pm \, (\hbar \, \k) \, \sin \th \, \sin \vph \lt\{\cos \th
\lt[\sin \vph \lt({M \over r}\rt) \lt[C_{0 \Cy} + C_{1 \Cy} \, \bar{m} + C_{2 \Cy} \, \bar{m}^2\rt]
+ \cos \vph \lt({M \Omega R^2\over r^2}\rt) \lt[D_{0 \Cx} + D_{1 \Cx} \, \bar{m} + D_{2 \Cx} \, \bar{m}^2\rt] \rt] \rt.
\nn
& = & \pm \lt. i \lt[\cos \vph \lt({M \over r}\rt)\lt[C_{0 \Cy} + C_{1 \Cy} \, \bar{m} + C_{2 \Cy} \, \bar{m}^2\rt]
- \sin \vph \lt({M \Omega R^2\over r^2}\rt) \lt[D_{0 \Cx} + D_{1 \Cx} \, \bar{m} + D_{2 \Cx} \, \bar{m}^2\rt] \rt] \rt\}.
\label{spinflip-term-Maj}
\ee
Because the Lense-Thirring metric is axisymmetric, the $\vph$-dependence in (\ref{spinflip-term-Maj}) needs to be averaged over
a complete rotation to remove this unphysical anisotropy in the azimuthal direction.
Therefore, by setting
\be
\lt| \lt\langle \mp \rt| H_{\PhiG} \lt| \pm \rt\rangle \rt|^2 & \rightarrow &
{1 \over 2 \pi} \int_0^{2\pi} \lt| \lt\langle \mp \rt| H_{\PhiG} \lt| \pm \rt\rangle \rt|^2 \, \d \vph,
\label{spinflip-avg-Maj}
\ee
it follows that
\be
\lt| \lt\langle \mp \rt| H_{\PhiG} \lt| \pm \rt\rangle \rt|^2 & = & {1 \over 2} \,  (\hbar \, \k)^2 \, \sin^2 \th
\lt[\lt(1 - {3 \over 16} \, \sin^2 \th\rt) \lt({M \over r}\rt)^2 \lt[C_{0 \Cy} + C_{1 \Cy} \, \bar{m} + C_{2 \Cy} \, \bar{m}^2\rt]^2 \rt.
\nn
& &{} + \lt.
\lt(1 - {1 \over 16} \, \sin^2 \th\rt) \lt({M \Omega R^2\over r^2}\rt)^2 \lt[D_{0 \Cx} + D_{1 \Cx} \, \bar{m} + D_{2 \Cx} \, \bar{m}^2\rt]^2 \rt].
\label{spinflip^2-Maj}
\ee

Following the same procedure as performed for the Dirac neutrino, the perturbed energy eigenvalue for the Majorana neutrino is
\be
E_{\bar{m}}^{(\pm)} & = & {4 \over 5} \,  {M \Omega R^2\over r^3} \, L^{\hat{z}}
+ (\hbar \, \k) \lt[\lt(1 - {2M \over r}\rt) + {1 \over 2} \, \bar{m}^2\rt]
+ (\hbar \, \k) \lt[\lt(F_0^{\rm Maj.}\rt) + \lt(F_1^{\rm Maj.}\rt) \bar{m} + \lt(F_2^{\rm Maj.}\rt) \bar{m}^2 \rt],
\label{E-perturbed-Maj=}
\ee
where
\begin{subequations}
\be
F_0^{\rm Maj.} & = & {1 \over 2} \lt[C_0 \lt({M \over r}\rt) + {1 \over 20} \, \lt({\Dl_\pm \over \k \, r}\rt) \rt],
\label{F0-Maj=}
\nl
F_1^{\rm Maj.} & = & {1 \over 2} \lt[C_1 \lt({M \over r}\rt) +  \lt({\Om_\pm \over \Dl_\pm}\rt)\rt],
\label{F1-Maj=}
\nl
F_2^{\rm Maj.} & = & {1 \over 2} \lt[C_2 \lt({M \over r}\rt) + {1 \over 2} \lt[\lt({\chi_\pm \over \Dl_\pm}\rt)
 - 20 \, (\k \, r) \lt({\Om_\pm \over \Dl_\pm^{3/2}}\rt)^2\rt]\rt],
\label{F2-Maj=}
\nl
\Om_\pm^{\rm Maj.} & = & {5 \over 2} \, (\k \, r) \lt\{\lt[8 \, C_0 \, C_1 + \sin^2 \th \lt(3 \, \cos^2 \th + 13 \rt) C_{0 \Cy} \, C_{1 \Cy}\rt]\lt({M \over r}\rt)^2
 + \sin^2 \th \lt(\cos^2 \th + 15\rt) D_{0 \Cx} \, D_{1 \Cx} \lt({M \Omega R^2\over r^2}\rt)^2 \rt\}
\nn
& &{} \pm 16 \, C_1 \lt({M \over r}\rt)\lt({M \Omega R^2\over r^2}\rt),
\label{Omega-Maj=}
\nl
\chi_\pm^{\rm Maj.} & = & {5 \over 2} \, (\k \, r) \lt\{\lt[8 \lt(C_1^2 + 2 \, C_0 \, C_2\rt)
+ \sin^2 \th \lt(3 \, \cos^2 \th + 13 \rt)\lt(C_{1 \Cy}^2 + 2 \,C_{0 \Cy} \, C_{2 \Cy}\rt)\rt]\lt({M \over r}\rt)^2 \rt.
\nn
& &{} + \lt. \sin^2 \th \lt(\cos^2 \th + 15\rt)\lt(D_{1 \Cx}^2 + 2 \, D_{0 \Cx} \, D_{2 \Cx} \rt) \lt({M \Omega R^2\over r^2}\rt)^2 \rt\}
\pm 32 \, C_2 \lt({M \over r}\rt)\lt({M \Omega R^2\over r^2}\rt),
\label{Chi-Maj=}
\nl
\lt(\Dl_\pm^{\rm Maj.}\rt)^2 & = & 50 \, (\k \, r)^2 \lt[8 \, C_0^2 + \sin^2 \th \lt(3 \, \cos^2 \th + 13 \rt) C_{0 \Cy}^2\rt] \lt({M \over r}\rt)^2
\nn
& &{} + \lt[256 + 50 \, (\k r)^2 \sin^2 \th \lt(\cos^2 \th + 15 \rt) D_{0 \Cx}^2 \rt]\lt({M \Omega R^2\over r^2}\rt)^2
\pm 640 \, (\k \, r) \, C_0 \lt({M \over r}\rt)\lt({M \Omega R^2\over r^2}\rt).
\label{Delta^2-Maj=}
\ee
\end{subequations}
This leads to the corresponding oscillation length expression for Majorana neutrinos, where
\be
E_{\bar{m}_2}^{(\pm)} - E_{\bar{m}_1}^{(\pm)} & = & (\hbar \, \k) \lt[ 
\lt(F_1^{\rm Maj.}\rt) \lt(\bar{m}_2 -  \bar{m}_1\rt) + \lt(F_2^{\rm Maj.} + {1 \over 2}\rt) \lt(\bar{m}_2^2 -  \bar{m}_1^2\rt) \rt].
\label{Energy-shift-Maj=}
\ee
Similarly, the leading order representation of (\ref{E-perturbed-Maj=}) is
\be
E_{\bar{m}}^{(\pm)} & \approx & {4 \over 5} \,  {M \Omega R^2\over r^3} \, L^{\hat{z}}
+ (\hbar \, \k) \lt[\lt(1 - {2M \over r}\rt) + {1 \over 2} \, \bar{m}^2\rt]
+ (\hbar \, \k) \, {M \over r} \lt[\lt(G_{0 (\pm)}^{\rm Maj.}\rt) + \lt(G_{1 (\pm)}^{\rm Maj.}\rt) \bar{m} + \lt(G_{2 (\pm)}^{\rm Maj.}\rt) \bar{m}^2 \rt]
\nn
& &{} + (\hbar \, \k) \, {M \Om R^2 \over r^2} \lt[\lt(K_0^{\rm Maj.}\rt) + \lt(K_1^{\rm Maj.}\rt) \bar{m} + \lt(K_2^{\rm Maj.}\rt) \bar{m}^2 \rt],
\label{E-perturbed-leading-Maj=}
\nl
E_{\bar{m}_2}^{(\pm)} - E_{\bar{m}_1}^{(\pm)} & \approx &  (\hbar \, \k) \lt[{1 \over 2} \lt(\bar{m}_2^2 -  \bar{m}_1^2\rt) \rt.
\nn
& &{} + \lt. \lt[{M \over r}\lt(G_{1 (\pm)}^{\rm Maj.}\rt) + {M \Om R^2 \over r^2} \lt(K_1^{\rm Maj.}\rt)\rt]\lt(\bar{m}_2 -  \bar{m}_1\rt)
+ \lt[{M \over r}\lt(G_{2 (\pm)}^{\rm Maj.}\rt) + {M \Om R^2 \over r^2} \lt(K_2^{\rm Maj.}\rt)\rt]\lt(\bar{m}_2^2 -  \bar{m}_1^2\rt) \rt],
\nn
\label{Energy-shift-leading-Maj=}
\ee
where
\begin{subequations}
\be
G_{0 (\pm)}^{\rm Maj.} & = & \lt({1 \over 2} \pm {8 \over \Lm} \rt)C_0,
\label{G0-Maj=}
\nl
G_{1 (\pm)}^{\rm Maj.} & = & {1 \over 2} \, C_1 \pm {8 \over \Lm} \lt[C_1 - 25 \lt({\k \, r \over \Lm}\rt)^2 \sin^2 \th \lt(\cos^2 \th + 15\rt) C_0 \, D_{0 \Cx} \, D_{1 \Cx} \rt],
\label{G1-Maj=}
\nl
G_{2 (\pm)}^{\rm Maj.} & = & {1 \over 2} \, C_2 \pm {8 \over \Lm} \lt[C_2
- 25 \lt({\k \, r \over \Lm}\rt)^2 \sin^2 \th \lt(\cos^2 \th + 15\rt)\lt[C_0 \lt(D_{1 \Cx}^2 + 2 \, D_{0 \Cx} \, D_{2 \Cx}\rt)
+ 2 \, C_1 \, D_{0 \Cx} \, D_{1 \Cx} \rt. \rt.
\nn
& &{} - \lt. \lt. 150 \lt({\k \, r \over \Lm}\rt)^2 \sin^2 \th \lt(\cos^2 \th + 15\rt) C_0 \, D_{0 \Cx}^2 \, D_{1 \Cx}^2\rt] \rt],
\label{G2-Maj=}
\nl
K_0^{\rm Maj.} & = & {1 \over 40} \lt({\Lm \over \k \, r}\rt),
\label{K0-Maj=}
\nl
K_1^{\rm Maj.} & = & {5 \over 4} \lt({\k \, r \over \Lm}\rt) \sin^2 \th \lt(\cos^2 \th + 15\rt) D_{0 \Cx} \, D_{1 \Cx},
\label{K1-Maj=}
\nl
K_2^{\rm Maj.} & = & {5 \over 8} \lt({\k \, r \over \Lm}\rt) \sin^2 \th \lt(\cos^2 \th + 15\rt) \lt[\lt(D_{1 \Cx}^2 + D_{0 \Cx} \, D_{2 \Cx}\rt)
- 50 \lt({\k \, r \over \Lm}\rt)^2 \sin^2 \th \lt(\cos^2 \th + 15\rt) D_{0 \Cx}^2 \, D_{1 \Cx}^2 \rt],
\label{K2-Maj=}
\nl
\lt(\Lm^{\rm Maj.}\rt)^2 & = & 256 + 50 \, (\k \, r)^2 \, \sin^2 \th \lt(\cos^2 \th + 15\rt) D_{0 \Cx}^2.
\label{Lambda-Maj=}
\ee
\end{subequations}

As shown for the Dirac case, we can obtain the leading order contributions to (\ref{Energy-shift-leading-Maj=}).  Again, we treat as a special case
the calculations where $\th = 0, \, \pi$, which leads to
\begin{subequations}
\label{G-K-Maj-th-1}
\be
G_{1 (\pm)}^{\rm Maj.} & = & {1 \over 2}  \lt(1 \pm 1\rt) C_1,
\label{G1-Maj-1=}
\nl
G_{2 (\pm)}^{\rm Maj.} & = & {1 \over 2} \lt(1 \pm 1\rt) C_2,
\label{G2-Maj-1=}
\nl
K_1^{\rm Maj.} & = & K_2^{\rm Maj.} \ = \ 0, \qquad \th = 0, \, \pi.
\label{K1-K2-Maj-1=}
\ee
\end{subequations}
When $\th \neq 0, \, \pi$, we can show that
\begin{subequations}
\label{G-K-Maj-th-2}
\be
G_{1 (\pm)}^{\rm Maj.} & \approx & {1 \over 2} \, C_1 \pm {8 \over 5 \sqrt{2}} \, {1 \over \sin \th \lt(\cos^2 \th + 15\rt)^{1/2}} \, {Y_1 \over \lt(\k r\rt)},
\label{G1-Maj-2=}
\nl
G_{2 (\pm)}^{\rm Maj.} & \approx & {1 \over 2} \, C_2 \pm {8 \over 5 \sqrt{2}} \, {1 \over \sin \th \lt(\cos^2 \th + 15\rt)^{1/2}} \, {Y_2 \over \lt(\k r\rt)},
\label{G2-Maj-2=}
\nl
K_1^{\rm Maj.} & \approx & {1 \over 4 \sqrt{2}} \, \sin \th \lt(\cos^2 \th + 15\rt)^{1/2} D_{1 \Cx},
\label{K1-Maj-2=}
\nl
K_2^{\rm Maj.} & \approx & {1 \over 8 \sqrt{2}} \, \sin \th \lt(\cos^2 \th + 15\rt)^{1/2} D_{2 \Cx}, \qquad \th \neq 0, \, \pi
\label{K2-Maj-2=}
\ee
\end{subequations}
where
\begin{subequations}
\be
D_{1 \Cx} & = & \sqrt{2 \over \pi^3}\, {r^3 \over R^3} \lt\{-{1 \over 15} \lt[12 \, \ln \lt({r \over R}\rt) \mu^{-3}
- 20 \, \mu^{-2} + \lt(2\pi \, q + 11\rt)\mu^{-1} - 2\rt]  e^{-q^2/2} \, e^{-[\k(r - R)/q]^2} \rt.
\nn
& &{} + \lt[{16 \pi \over 15} \, \cos^2 \lt(\k R\rt) e^{-2\krq}
-{4 \sqrt{\pi} \over 3} \, \sin \lt(\k R\rt) e^{-q^2/4} \, e^{-\krq} - {8 \pi \over 15} \, e^{-2\krq} + {1 \over 30} \, q^2 \, e^{-q^2/2}\rt] \mu^{-1}
\nn
& &{} - {2 \over 45} \lt(5 \, q - 2\pi\rt) q \, e^{-q^2/2} + {8 \pi \over 15} \, \cos \lt(\k R\rt) \, \sin \lt(\k R\rt) \, q \, e^{-2\krq}
\nn
& &{} + \lt.  {2 \sqrt{\pi} \over 15} \lt[10 \, \sin\lt(\k R\rt) + \lt(3 \, q - 2 \pi\rt) \cos\lt(\k R\rt) \rt]
e^{-q^2/4} \, e^{-\krq} \rt\} + O\lt(\mu\rt),
\label{D1x-approx}
\nl
D_{2 \Cx} & = & \sqrt{2 \over \pi^3}\, {r^3 \over R^3} \lt\{{1 \over 15} \lt[\pi \, q \, \mu^{-2} + {1 \over 8}\lt(5 \, q + 24\pi\rt)q \, \mu^{-1}
 - {5 \pi \over 3} \rt]  e^{-q^2/2} \, e^{-[\k(r - R)/q]^2} \rt.
\nn
& &{} - \lt[
{\sqrt{\pi^3} \over 15} \, \sin \lt(\k R\rt) q \, e^{-q^2/4} \, e^{-\krq} + {\pi \over 60} \, q^2 \, e^{-q^2/2}\rt] \mu^{-1}
\nn
& &{} - {1 \over 360} \lt(9 \, q^2 - 44\rt) q^2 \, e^{-q^2/2} - {4 \pi \over 15} \, \cos \lt(\k R\rt) \, \sin \lt(\k R\rt) \, q \, e^{-2\krq}
\nn
& &{} - \lt.  {\sqrt{\pi} \over 30} \lt[q^2 \, \sin\lt(\k R\rt) + 10 \, q \, \cos\lt(\k R\rt) \rt]
e^{-q^2/4} \, e^{-\krq} \rt\} + O\lt(\mu\rt),
\label{D2x-approx}
\ee
\end{subequations}
and
\begin{subequations}
\be
Y_1 & = & {1 \over D_{0 \Cx}^2} \lt[C_1 \, D_{0 \Cx} - {1 \over 2} \, C_0 \, D_{1 \Cx}\rt] \ \approx \ - {105 \over 32} {\lt(1 - {R^2/r^2}\rt) \over \ln \lt({r/R}\rt)},
\label{Y1}
\nl
Y_2 & = & {C_0 \over D_{0 \Cx}^3} \lt[D_{1 \Cx}^2 - D_{0 \Cx} \, D_{2 \Cx}\rt] + {1 \over D_{0 \Cx}^2} \lt[C_2 \, D_{0 \Cx} - C_1 \, D_{1 \Cx}\rt]
\ \approx \ - {45 \over 8} {\lt(1 - {R^2/r^2}\rt) \over \ln \lt({r/R}\rt)}.
\label{Y2}
\ee
\end{subequations}

\subsection{Analytic Comparison of Dirac and Majorana Neutrino Interactions due to Spin-Gravity Coupling}

A complete numerical analysis of the main results presented here is contained in the next section.
However, even a preliminary formal analysis shows clear evidence that Dirac and Majorana neutrinos are distinguishable
due to the presence of spin-gravity coupling.
Focussing on (\ref{G-K-Dirac-th-1}) and (\ref{G-K-Maj-th-1}) for now, we notice that when the neutrino beam is directed along
the axis of symmetry it follows that $K_j^{\rm Dirac}$ and $K_j^{\rm Maj.}$ are identically zero,
implying that the helicity states are unaffected along $\th = 0, \pi$.
For neutrinos with negative initial helicity, both $G_2^{\rm Dirac}$ and $G_2^{\rm Maj.}$ are also identically zero,
while there remains a non-zero contribution of $G_1^{\rm Dirac} = -C_1/2$ which does not exist for $G_1^{\rm Maj.}$.

When considering neutrino beams off the axis of symmetry, it becomes obvious from (\ref{G-K-Dirac-th-2}) and (\ref{G-K-Maj-th-2})
that there are significant differences in $\th$-dependence which appear, along with relatively trivial differences in the coefficients.
Furthermore, it follows that $K_j^{\rm Dirac}$ are proportional to $D_j$, which in turn have an overall dependence of $(r/R)^2$ from (\ref{D1-approx})
and (\ref{D2-approx}).
In contrast, $K_j^{\rm Maj.}$ are proportional to $D_{j\hat{x}}$, which are shown by (\ref{D1x-approx}) and (\ref{D2x-approx}) to be dependent on $(r/R)^3$.


\section{Numerical Analysis}
\label{section:analysis}

Having obtained the final analytic expressions for (\ref{Energy-shift-Dirac=}) and (\ref{Energy-shift-Maj=}) for the Dirac and Majorana neutrinos,
respectively, we proceed to express them numerically for astrophysically relevant situations.
This provides an opportunity to determine the likelihood of observing gravitational corrections to the neutrino oscillation length.
One obvious example is to perform this analysis for neutrinos emitted from the Sun, while another is to use SN1987A as a test case for
comparison involving a known supernova source.
These examples are considered separately below.
Although it is not obvious that these examples suggest any possibility of observing these effects,
they offer interesting theoretical insights about the physical behaviour of Dirac and Majorana neutrinos in a gravitational field.

\subsection{General Properties}

Before going into detail about the examples considered below, it is worthwhile to discuss some of the general properties of the plots common in
both cases.
All the plots presented below are generated based on the expressions found in Appendix~\ref{appendix:dimensionless-functions}.
While the specific examples are, in principle, dependent on the value of the mean momentum, it turns out that the plots are insensitive to any large
variations of the choice of $\k$ considered.
With this realized, we present the plots assuming $\hbar \, \k$ = 1 MeV throughout.
In addition, while there is a formal difference between the functions $F_j^-$ and $F_j^+$ for each of the cases considered due to the initial negative
and positive helicity of the neutrino, respectively, the actual numerical differences do not register on any of these plots,
where at best we have a one part in $10^{40}$ difference due to their initial helicity states.

\subsection{Gravitational Effects on Neutrinos from the Sun}

We begin with an analysis involving the Sun as the gravitational source.
Given that \cite{MTW} in geometric units $M = M_{\odot} = 1.48 \times 10^5$ cm, $R = 6.95 \times 10^{10}$ cm, and $r = 214.8 \, R$ = 1 A.U.,
it follows that $M/r = 9.95 \times 10^{-14}$ for the Sun.
With the sidereal period of the Sun \cite{JPL} equal to about 25.38 days, it also follows that $M \Omega R^2/r^2 = 3.07 \times 10^{-16}$.

\begin{figure*}
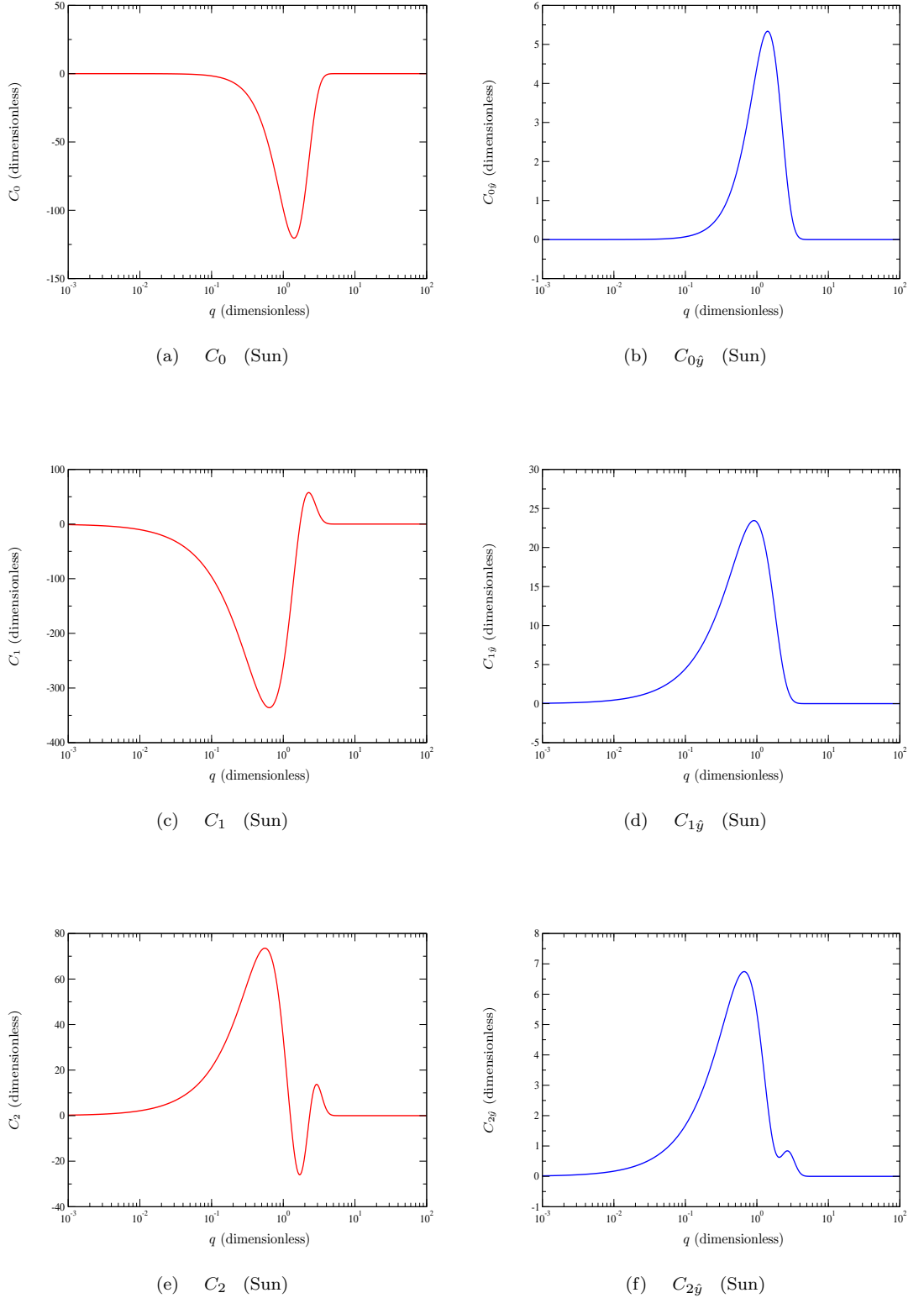

\psfrag{q}[cc][][2.5][0]{\hspace{0.5cm} $q$ (dimensionless)}
\psfrag{C0}[bc][][2.5][0]{$C_0$ (dimensionless)}
\psfrag{C1}[bc][][2.5][0]{$C_1$ (dimensionless)}
\psfrag{C2}[bc][][2.5][0]{$C_2$ (dimensionless)}
\psfrag{C0y}[bc][][2.5][0]{$C_{0\hat{y}}$ (dimensionless)}
\psfrag{C1y}[bc][][2.5][0]{$C_{1\hat{y}}$ (dimensionless)}
\psfrag{C2y}[bc][][2.5][0]{$C_{2\hat{y}}$ (dimensionless)}
\psfrag{D0}[bc][][2.5][0]{$D_0$ (dimensionless)}
\psfrag{D1}[bc][][2.5][0]{$D_1$ (dimensionless)}
\psfrag{D2}[bc][][2.5][0]{$D_2$ (dimensionless)}
\psfrag{D0x}[bc][][2.5][0]{$D_{0\hat{x}}$ (dimensionless)}
\psfrag{D1x}[bc][][2.5][0]{$D_{1\hat{x}}$ (dimensionless)}
\psfrag{D2x}[bc][][2.5][0]{$D_{2\hat{x}}$ (dimensionless)}
\psfrag{F1_Dirac}[bc][][2.5][0]{$F_1^{\rm Dirac}$ (dimensionless)}
\psfrag{F2_Dirac}[bc][][2.5][0]{$F_2^{\rm Dirac}$ (dimensionless)}
\psfrag{F1_Maj}[bc][][2.5][0]{$F_1^{\rm Maj.}$ (dimensionless)}
\psfrag{F2_Maj}[bc][][2.5][0]{$F_2^{\rm Maj.}$ (dimensionless)}
\psfrag{th = 0.5*Pi}[cc][][2.5][0]{\small $\th = 0.5 \pi$}
\psfrag{th = 0.4*Pi}[cc][][2.5][0]{\small $\th = 0.4 \pi$}
\psfrag{th = 0.3*Pi}[cc][][2.5][0]{\small $\th = 0.3 \pi$}
\psfrag{th = 0.2*Pi}[cc][][2.5][0]{\small $\th = 0.2 \pi$}
\psfrag{th = 0.1*Pi}[cc][][2.5][0]{\small $\th = 0.1 \pi$}
\psfrag{th = 0}[cc][][2.5][0]{\small $\th = 0$}
\begin{minipage}[t]{0.3 \textwidth}
\centering
\subfigure[\hspace{0.2cm} $C_0$ \, (Sun)]{
\label{fig:C0-Solar}
\rotatebox{0}{\includegraphics[width = 6.6cm, height = 5.0cm, scale = 1]{2a}}}
\end{minipage}%
\hspace{2.0cm}
\begin{minipage}[t]{0.3 \textwidth}
\centering
\subfigure[\hspace{0.2cm} $C_{0\hat{y}}$ \, (Sun)]{
\label{fig:C0y-Solar}
\rotatebox{0}{\includegraphics[width = 6.6cm, height = 5.0cm, scale = 1]{2b}}}
\end{minipage} \\
\vspace{0.8cm}
\begin{minipage}[t]{0.3 \textwidth}
\centering
\subfigure[\hspace{0.2cm} $C_1$ \, (Sun)]{
\label{fig:C1-Solar}
\rotatebox{0}{\includegraphics[width = 6.6cm, height = 5.0cm, scale = 1]{2c}}}
\end{minipage}%
\hspace{2.0cm}
\begin{minipage}[t]{0.3 \textwidth}
\centering
\subfigure[\hspace{0.2cm} $C_{1\hat{y}}$ \, (Sun)]{
\label{fig:C1y-Solar}
\rotatebox{0}{\includegraphics[width = 6.6cm, height = 5.0cm, scale = 1]{2d}}}
\end{minipage} \\
\vspace{0.8cm}
\begin{minipage}[t]{0.3 \textwidth}
\centering
\subfigure[\hspace{0.2cm} $C_2$ \, (Sun)]{
\label{fig:C2-Solar}
\rotatebox{0}{\includegraphics[width = 6.6cm, height = 5.0cm, scale = 1]{2e}}}
\end{minipage}%
\hspace{2.0cm}
\begin{minipage}[t]{0.3 \textwidth}
\centering
\subfigure[\hspace{0.2cm} $C_{2\hat{y}}$ \, (Sun)]{
\label{fig:C2y-Solar}
\rotatebox{0}{\includegraphics[width = 6.6cm, height = 5.0cm, scale = 1]{2f}}}
\end{minipage}
\caption{\label{fig:Cj-Cjy-Solar} Plots of dimensionless functions $C_j$ and $C_{j\hat{y}}$ with respect to $q$,
where the Sun is the gravitational source.
While both Figures~\ref{fig:C0-Solar} and \ref{fig:C0y-Solar} peak at around $q = 1$, they are of opposite sign
and $C_{0\hat{y}}$ is roughly two orders of magnitude smaller than $C_0$.
Figures~\ref{fig:C1-Solar} and \ref{fig:C1y-Solar} also have opposite sign
and peak at around $q = 1$, though their widths extend from $10^{-2}$ to $1$, and $C_{1\hat{y}}$ is only one order of magnitude smaller
than $C_1$.
Both Figures~\ref{fig:C2-Solar} and \ref{fig:C2y-Solar} have positive-valued peaks around $q = 1$ whose widths range from
$10^{-2}$ to $1$.  However, $C_2$ displays a smaller peak of opposite sign not present in $C_{2\hat{y}}$.
}
\end{figure*}

The expressions for dimensionless functions $C_j$ and $C_{j\hat{y}}$ 
are listed in Figure~\ref{fig:Cj-Cjy-Solar}.
We recall that $C_j$ contribute to the spin diagonal part of the matrix element coupled to $M/r$, while $C_{j\hat{y}}$
for the Majorana neutrino contributes a spin-flip part that is also coupled to $M/r$.
For $C_0$ and $C_{0\hat{y}}$ in Figures~\ref{fig:C0-Solar} and \ref{fig:C0y-Solar}, respectively, they each have a single peak
located around $q = 1$, but are of opposite sign and $C_{0\hat{y}}$ is two orders of magnitude smaller in amplitude
than $C_0$.
The $C_1$ and $C_{1\hat{y}}$ in Figures~\ref{fig:C1-Solar} and \ref{fig:C1y-Solar}
behave in a similar fashion, with two exceptions in that they differ by only one order of magnitude, and that $C_1$ has a smaller
peak that is positive-valued around $q = 2$.
As for $C_2$ and $C_{2\hat{y}}$ in Figures~\ref{fig:C2-Solar} and \ref{fig:C2y-Solar}, they each have some non-trivial structure to them.
Both functions have main peaks which are positive-valued at around $5 \times 10^{-1}$, where $C_2$ is about two orders of magnitude larger
than $C_{2\hat{y}}$.
However, Figure~\ref{fig:C2-Solar} also has a smaller negative peak at $q = 2$ and an even smaller positive peak at around $q = 3$,
while Figure~\ref{fig:C2y-Solar} has a slight shoulder at $q = 3$.
All functions $C_j$ and $C_{j\hat{y}}$ rapidly decay to zero for $q < 10^{-2}$ and $q > 5$.

\begin{figure*}
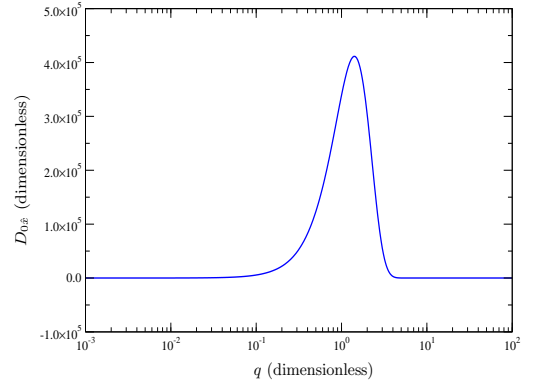
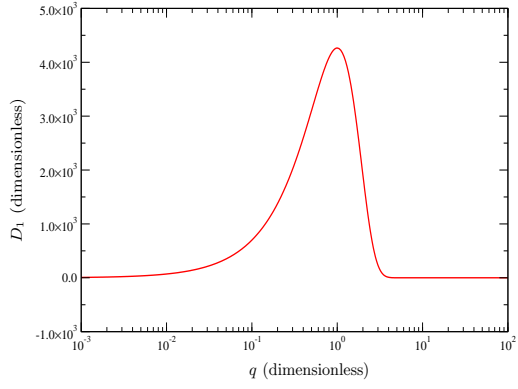
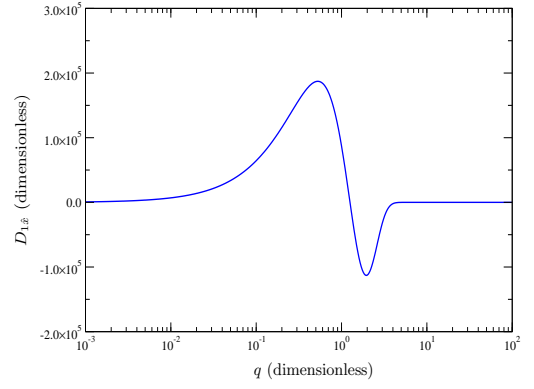
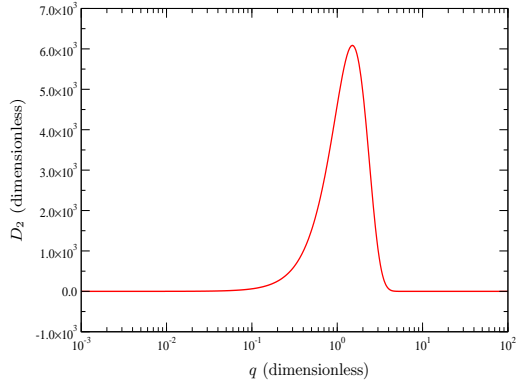
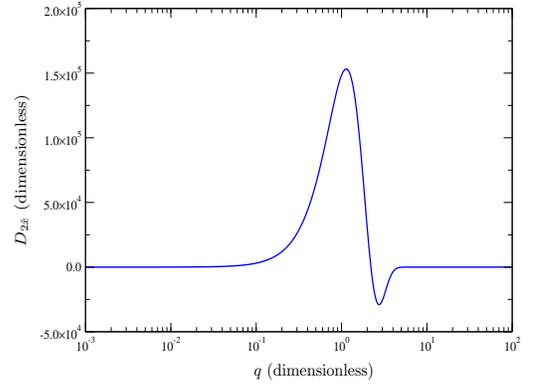

\psfrag{q}[cc][][2.5][0]{\hspace{0.5cm} $q$ (dimensionless)}
\psfrag{C0}[bc][][2.5][0]{$C_0$ (dimensionless)}
\psfrag{C1}[bc][][2.5][0]{$C_1$ (dimensionless)}
\psfrag{C2}[bc][][2.5][0]{$C_2$ (dimensionless)}
\psfrag{C0y}[bc][][2.5][0]{$C_{0\hat{y}}$ (dimensionless)}
\psfrag{C1y}[bc][][2.5][0]{$C_{1\hat{y}}$ (dimensionless)}
\psfrag{C2y}[bc][][2.5][0]{$C_{2\hat{y}}$ (dimensionless)}
\psfrag{D0}[bc][][2.5][0]{$D_0$ (dimensionless)}
\psfrag{D1}[bc][][2.5][0]{$D_1$ (dimensionless)}
\psfrag{D2}[bc][][2.5][0]{$D_2$ (dimensionless)}
\psfrag{D0x}[bc][][2.5][0]{$D_{0\hat{x}}$ (dimensionless)}
\psfrag{D1x}[bc][][2.5][0]{$D_{1\hat{x}}$ (dimensionless)}
\psfrag{D2x}[bc][][2.5][0]{$D_{2\hat{x}}$ (dimensionless)}
\psfrag{F1_Dirac}[bc][][2.5][0]{$F_1^{\rm Dirac}$ (dimensionless)}
\psfrag{F2_Dirac}[bc][][2.5][0]{$F_2^{\rm Dirac}$ (dimensionless)}
\psfrag{F1_Maj}[bc][][2.5][0]{$F_1^{\rm Maj.}$ (dimensionless)}
\psfrag{F2_Maj}[bc][][2.5][0]{$F_2^{\rm Maj.}$ (dimensionless)}
\psfrag{th = 0.5*Pi}[cc][][2.5][0]{\small $\th = 0.5 \pi$}
\psfrag{th = 0.4*Pi}[cc][][2.5][0]{\small $\th = 0.4 \pi$}
\psfrag{th = 0.3*Pi}[cc][][2.5][0]{\small $\th = 0.3 \pi$}
\psfrag{th = 0.2*Pi}[cc][][2.5][0]{\small $\th = 0.2 \pi$}
\psfrag{th = 0.1*Pi}[cc][][2.5][0]{\small $\th = 0.1 \pi$}
\psfrag{th = 0}[cc][][2.5][0]{\small $\th = 0$}
\begin{minipage}[t]{0.3 \textwidth}
\centering
\subfigure[\hspace{0.2cm} $D_0$ \, (Sun)]{
\label{fig:D0-Solar}
\rotatebox{0}{\includegraphics[width = 6.6cm, height = 5.0cm, scale = 1]{3a}}}
\end{minipage}%
\hspace{2.0cm}
\begin{minipage}[t]{0.3 \textwidth}
\centering
\subfigure[\hspace{0.2cm} $D_{0\hat{x}}$ \, (Sun)]{
\label{fig:D0x-Solar}
\rotatebox{0}{\includegraphics[width = 6.6cm, height = 5.0cm, scale = 1]{3b}}}
\end{minipage} \\
\vspace{0.8cm}
\begin{minipage}[t]{0.3 \textwidth}
\centering
\subfigure[\hspace{0.2cm} $D_1$ \, (Sun)]{
\label{fig:D1-Solar}
\rotatebox{0}{\includegraphics[width = 6.6cm, height = 5.0cm, scale = 1]{3c}}}
\end{minipage}%
\hspace{2.0cm}
\begin{minipage}[t]{0.3 \textwidth}
\centering
\subfigure[\hspace{0.2cm} $D_{1\hat{x}}$ \, (Sun)]{
\label{fig:D1x-Solar}
\rotatebox{0}{\includegraphics[width = 6.6cm, height = 5.0cm, scale = 1]{3d}}}
\end{minipage} \\
\vspace{0.8cm}
\begin{minipage}[t]{0.3 \textwidth}
\centering
\subfigure[\hspace{0.2cm} $D_2$ \, (Sun)]{
\label{fig:D2-Solar}
\rotatebox{0}{\includegraphics[width = 6.6cm, height = 5.0cm, scale = 1]{3e}}}
\end{minipage}%
\hspace{2.0cm}
\begin{minipage}[t]{0.3 \textwidth}
\centering
\subfigure[\hspace{0.2cm} $D_{2\hat{x}}$ \, (Sun)]{
\label{fig:D2x-Solar}
\rotatebox{0}{\includegraphics[width = 6.6cm, height = 5.0cm, scale = 1]{3f}}}
\end{minipage}
\caption{\label{fig:Dj-Djx-Solar} Plots of dimensionless functions $D_j$ and $D_{j\hat{x}}$ with respect to $q$
where the Sun is the gravitational source.
In contrast to $C_0$ and $C_{0\hat{y}}$ of Figure~\ref{fig:Cj-Cjy-Solar},
$D_{0\hat{x}}$ is an order of magnitude larger than $D_0$, as shown in Figures~\ref{fig:D0-Solar} and \ref{fig:D0x-Solar}.
Similarly, Figures~\ref{fig:D1-Solar} and \ref{fig:D1x-Solar} show that $D_{1\hat{x}}$ dominates over $D_1$, this time by
two orders of magnitude.
This is also true for Figures~\ref{fig:D2-Solar} and \ref{fig:D2x-Solar}.
}
\end{figure*}

For the dimensionless functions $D_j$ and $D_{j\hat{x}}$ with respect to $q$, they are presented in Figure~\ref{fig:Dj-Djx-Solar}.
The first observation of note is that their magnitudes are about two orders of magnitude larger than their counterparts for
$C_j$ and $C_{j\hat{y}}$.
It is especially interesting to observe that, contrary to the plots in Figure~\ref{fig:Cj-Cjy-Solar}, the $D_{j\hat{x}}$ functions
are one to two orders of magnitude larger than the $D_j$.
As well, Figures~\ref{fig:D1x-Solar} and \ref{fig:D2x-Solar} which describe $D_{1\hat{x}}$ and $D_{2\hat{x}}$, respectively,
have additional structure compared to $D_1$ and $D_2$ with the presence of a negative-valued peak at around $q = 2$.
Like the plots shown Figure~\ref{fig:Cj-Cjy-Solar}, these plots are non-zero only in the range $10^{-2} \lesssim q \lesssim 5$.

\begin{figure*}
\psfrag{q}[cc][][2.5][0]{\hspace{0.5cm} $q$ (dimensionless)}
\psfrag{C0}[bc][][2.5][0]{$C_0$ (dimensionless)}
\psfrag{C1}[bc][][2.5][0]{$C_1$ (dimensionless)}
\psfrag{C2}[bc][][2.5][0]{$C_2$ (dimensionless)}
\psfrag{C0y}[bc][][2.5][0]{$C_{0\hat{y}}$ (dimensionless)}
\psfrag{C1y}[bc][][2.5][0]{$C_{1\hat{y}}$ (dimensionless)}
\psfrag{C2y}[bc][][2.5][0]{$C_{2\hat{y}}$ (dimensionless)}
\psfrag{D0}[bc][][2.5][0]{$D_0$ (dimensionless)}
\psfrag{D1}[bc][][2.5][0]{$D_1$ (dimensionless)}
\psfrag{D2}[bc][][2.5][0]{$D_2$ (dimensionless)}
\psfrag{D0x}[bc][][2.5][0]{$D_{0\hat{x}}$ (dimensionless)}
\psfrag{D1x}[bc][][2.5][0]{$D_{1\hat{x}}$ (dimensionless)}
\psfrag{D2x}[bc][][2.5][0]{$D_{2\hat{x}}$ (dimensionless)}
\psfrag{F1_Dirac}[bc][][2.5][0]{$F_1^{\rm Dirac}$ (dimensionless)}
\psfrag{F2_Dirac}[bc][][2.5][0]{$F_2^{\rm Dirac}$ (dimensionless)}
\psfrag{F1_Maj}[bc][][2.5][0]{$F_1^{\rm Maj.}$ (dimensionless)}
\psfrag{F2_Maj}[bc][][2.5][0]{$F_2^{\rm Maj.}$ (dimensionless)}
\psfrag{th = 0.5*Pi}[cc][][2.5][0]{\small $\th = 0.5 \pi$}
\psfrag{th = 0.4*Pi}[cc][][2.5][0]{\small $\th = 0.4 \pi$}
\psfrag{th = 0.3*Pi}[cc][][2.5][0]{\small $\th = 0.3 \pi$}
\psfrag{th = 0.2*Pi}[cc][][2.5][0]{\small $\th = 0.2 \pi$}
\psfrag{th = 0.1*Pi}[cc][][2.5][0]{\small $\th = 0.1 \pi$}
\psfrag{th = 0}[cc][][2.5][0]{\small $\th = 0$}
\begin{minipage}[t]{0.3 \textwidth}
\centering
\subfigure[\hspace{0.2cm} $F_1^{\rm Dirac}$ \, (Sun)]{
\label{fig:F1-Solar-Dirac}
\rotatebox{0}{\includegraphics[width = 6.6cm, height = 5.0cm, scale = 1]{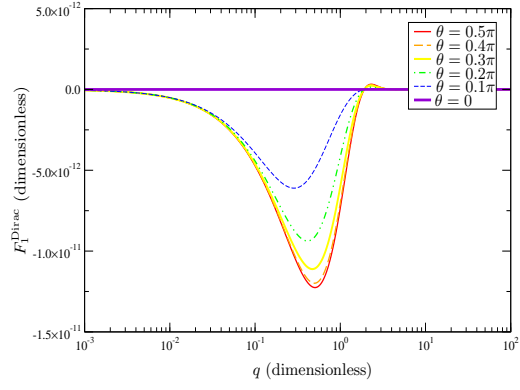}}}
\end{minipage} \\
\vspace{0.8cm}
\begin{minipage}[t]{0.3 \textwidth}
\centering
\subfigure[\hspace{0.2cm} $F_1^{\rm Maj.}$ \, (Sun)]{
\label{fig:F1-Solar-Maj}
\rotatebox{0}{\includegraphics[width = 6.6cm, height = 5.0cm, scale = 1]{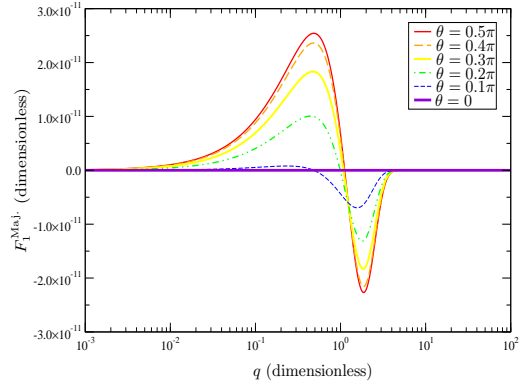}}}
\end{minipage} \\
\vspace{0.8cm}
\begin{minipage}[t]{0.3 \textwidth}
\centering
\subfigure[\hspace{0.2cm} $F_1^{\rm Maj.}$ \, (Sun) \, ($\Om = 0$)]{
\label{fig:F1-Solar-Maj-Omega=0}
\rotatebox{0}{\includegraphics[width = 6.6cm, height = 5.0cm, scale = 1]{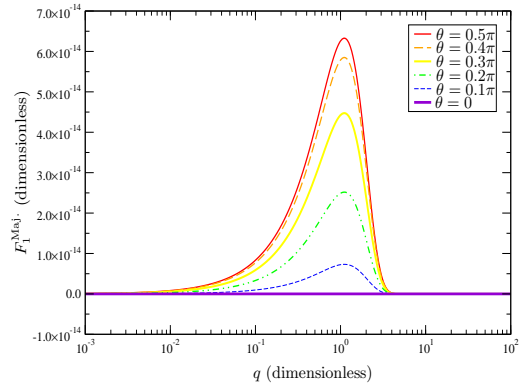}}}
\end{minipage}
\caption{\label{fig:F1-Solar} Comparison of $F_1$ as a function of $q$ for Dirac and Majorana neutrinos
assuming various angles of the neutrino beam off the axis of symmetry, where $\th$ is defined by Figure~\ref{fig:orientation}.
For both Figures~\ref{fig:F1-Solar-Dirac} and \ref{fig:F1-Solar-Maj}, the plot profiles show that magnitude of $F_1$
is most pronounced for $\th = \pi/2$.
For a non-rotating star ($\Om = 0$), Figure~\ref{fig:F1-Solar-Maj-Omega=0} shows that there still exists a non-trivial
contribution to $F_1^{\rm Maj.}$, though reduced by three orders of magnitude compared to Figure~\ref{fig:F1-Solar-Maj}.}
\end{figure*}

Using Figures~\ref{fig:Cj-Cjy-Solar} and \ref{fig:Dj-Djx-Solar}, we can obtain numerical expressions for $F_j^{\rm Dirac}$ and $F_j^{\rm Maj.}$ 
with the Sun as the gravitational source.
We begin with a comparison of $F_1$ for Dirac and Majorana neutrinos presented in Figure~\ref{fig:F1-Solar}, where
Figures~\ref{fig:F1-Solar-Dirac} and \ref{fig:F1-Solar-Maj} describe $F_1^{\rm Dirac}$ and $F_1^{\rm Maj.}$, respectively, for $\Om \neq 0$.
It is evident from these first two plots that the gravitational field can distinguish between Dirac and Majorana neutrinos, given the
distinctive properties of their respective profiles.
From Figure~\ref{fig:F1-Solar-Dirac}, the plot is almost completely negative-valued with a single peak for each choice of $\th$ in the
range of $10^{-1} \lesssim q \lesssim 1$, while Figure~\ref{fig:F1-Solar-Maj} shows both a positive-valued peak in the same range of $q$
and a negative-valued peak around $q = 2$.
The magnitude of the plots for both cases are on the order of $10^{-11}$.
In addition, we have for further comparison Figure~\ref{fig:F1-Solar-Maj-Omega=0}, which describes $F_1^{\rm Maj.}$ for $\Om = 0$.
This has the interesting property of a non-trivial profile with a single peak around $q = 1$ for different choices of $\th$ and
a magnitude on the order of $10^{-14}$.
In contrast, the magnitude of $F_1^{\rm Dirac}$ for $\Om = 0$ only reaches a maximum of $10^{-20}$, which suggests that the rotational part
of the metric makes the dominant contribution to the predicted gravitational effects on the neutrino oscillation length.

\begin{figure*}
\psfrag{q}[cc][][2.5][0]{\hspace{0.5cm} $q$ (dimensionless)}
\psfrag{C0}[bc][][2.5][0]{$C_0$ (dimensionless)}
\psfrag{C1}[bc][][2.5][0]{$C_1$ (dimensionless)}
\psfrag{C2}[bc][][2.5][0]{$C_2$ (dimensionless)}
\psfrag{C0y}[bc][][2.5][0]{$C_{0\hat{y}}$ (dimensionless)}
\psfrag{C1y}[bc][][2.5][0]{$C_{1\hat{y}}$ (dimensionless)}
\psfrag{C2y}[bc][][2.5][0]{$C_{2\hat{y}}$ (dimensionless)}
\psfrag{D0}[bc][][2.5][0]{$D_0$ (dimensionless)}
\psfrag{D1}[bc][][2.5][0]{$D_1$ (dimensionless)}
\psfrag{D2}[bc][][2.5][0]{$D_2$ (dimensionless)}
\psfrag{D0x}[bc][][2.5][0]{$D_{0\hat{x}}$ (dimensionless)}
\psfrag{D1x}[bc][][2.5][0]{$D_{1\hat{x}}$ (dimensionless)}
\psfrag{D2x}[bc][][2.5][0]{$D_{2\hat{x}}$ (dimensionless)}
\psfrag{F1_Dirac}[bc][][2.5][0]{$F_1^{\rm Dirac}$ (dimensionless)}
\psfrag{F2_Dirac}[bc][][2.5][0]{$F_2^{\rm Dirac}$ (dimensionless)}
\psfrag{F1_Maj}[bc][][2.5][0]{$F_1^{\rm Maj.}$ (dimensionless)}
\psfrag{F2_Maj}[bc][][2.5][0]{$F_2^{\rm Maj.}$ (dimensionless)}
\psfrag{th = 0.5*Pi}[cc][][2.5][0]{\small $\th = 0.5 \pi$}
\psfrag{th = 0.4*Pi}[cc][][2.5][0]{\small $\th = 0.4 \pi$}
\psfrag{th = 0.3*Pi}[cc][][2.5][0]{\small $\th = 0.3 \pi$}
\psfrag{th = 0.2*Pi}[cc][][2.5][0]{\small $\th = 0.2 \pi$}
\psfrag{th = 0.1*Pi}[cc][][2.5][0]{\small $\th = 0.1 \pi$}
\psfrag{th = 0}[cc][][2.5][0]{\small $\th = 0$}
\begin{minipage}[t]{0.3 \textwidth}
\centering
\subfigure[\hspace{0.2cm} $F_2^{\rm Dirac}$ \, (Sun)]{
\label{fig:F2-Solar-Dirac}
\rotatebox{0}{\includegraphics[width = 6.6cm, height = 5.0cm, scale = 1]{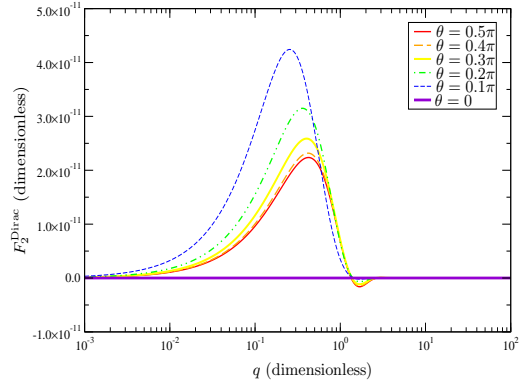}}}
\end{minipage} \\
\vspace{0.8cm}
\begin{minipage}[t]{0.3 \textwidth}
\centering
\subfigure[\hspace{0.2cm} $F_2^{\rm Maj.}$ \, (Sun)]{
\label{fig:F2-Solar-Maj}
\rotatebox{0}{\includegraphics[width = 6.6cm, height = 5.0cm, scale = 1]{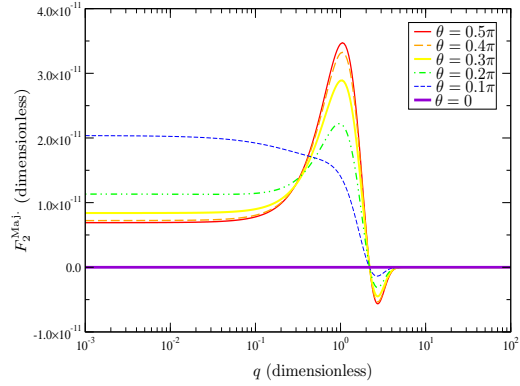}}}
\end{minipage} \\
\vspace{0.8cm}
\begin{minipage}[t]{0.3 \textwidth}
\centering
\subfigure[\hspace{0.2cm} $F_2^{\rm Maj.}$ \, (Sun) \, ($\Om = 0$)]{
\label{fig:F2-Solar-Maj-Omega=0}
\rotatebox{0}{\includegraphics[width = 6.6cm, height = 5.0cm, scale = 1]{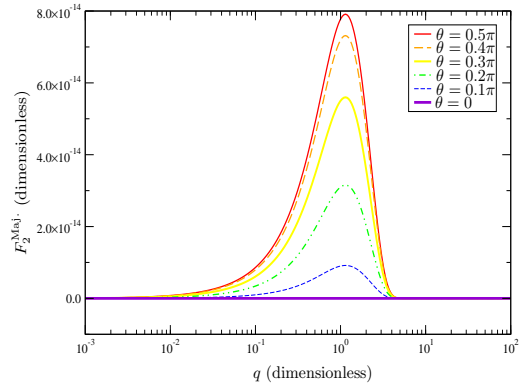}}}
\end{minipage}
\caption{\label{fig:F2-Solar} Comparison of $F_2$ as a function of $q$ for Dirac and Majorana neutrinos
for varying neutrino beam orientation.
Unlike for Figure~\ref{fig:F1-Solar}, an increase of $\th$ on Figure~\ref{fig:F2-Solar-Dirac} leads to a
decrease in the magnitude of $F_2^{\rm Dirac}$, suggesting that rotation of the gravitational source serves to dampen
the quadratic mass correction in the energy difference.
A more intriguing effect appears in the small $q$ behaviour found in Figure~\ref{fig:F2-Solar-Maj},
which does not dampen to zero as $q \rightarrow 0$.  This is somewhat surprising, since the profile of the
matrix element components which contribute to this plot for $F_2^{\rm Maj.}$ all vanish for small $q$.
For the case of a non-rotating star $(\Om = 0)$, Figure~\ref{fig:F2-Solar-Maj-Omega=0} yields a non-zero
plot, again with a reduction of three orders of magnitude compared to Figure~\ref{fig:F2-Solar-Maj}.
}
\end{figure*}

A similar comparison of $F_2$ for Dirac and Majorana neutrinos, as shown in Figure~\ref{fig:F2-Solar}, is also interesting for a different set
of reasons.
From Figure~\ref{fig:F2-Solar-Dirac}, it is evident that the magnitude of $F_2^{\rm Dirac}$ {\em decreases} as $\th$ increases,
which suggests that the rotational part of the metric dampens the gravitational corrections to the quadratic mass term in the energy difference
due to neutrino mass.
This effect does not appear as such for $F_2^{\rm Maj.}$ in Figure~\ref{fig:F2-Solar-Maj}.
However, another anomalous effect is present in the plots for small $q$ behaviour, since the function does not vanish as $q \rightarrow 0$,
but appears to asymptotically approach a finite and non-zero value on the order of $10^{-11}$.
Although it is difficult to precisely determine the cause of this behaviour, it does seem that the presence of rotation in
the metric has a role in this effect.
Comparison with Figure~\ref{fig:F2-Solar-Maj-Omega=0}, which describes $F_2^{\rm Maj.}$ for $\Om = 0$, seems to confirm this
interpretation, as the function rapidly vanishes for $q < 10^{-1}$ with a magnitude on the order of $10^{-14}$.
Again, this is consistent with the results of Figure~\ref{fig:F1-Solar-Dirac}, where the magnitude of $F_2^{\rm Dirac}$ for $\Om = 0$ is also
around $10^{-20}$.

\subsection{Predicted Effects on Neutrinos from SN1987A}

\begin{figure*}
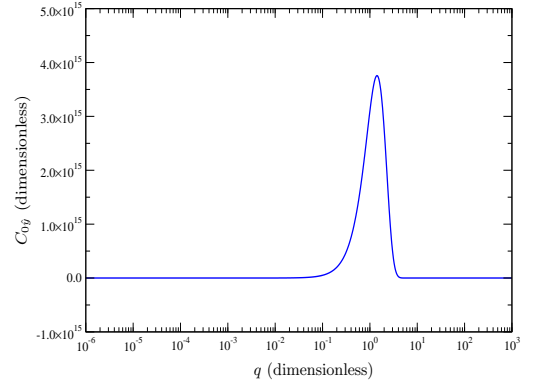
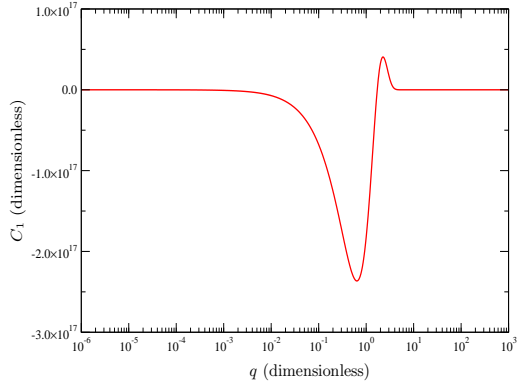
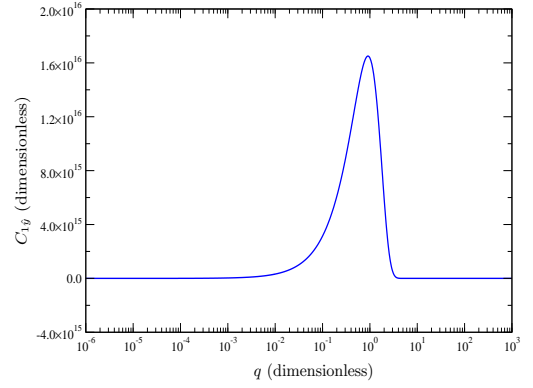
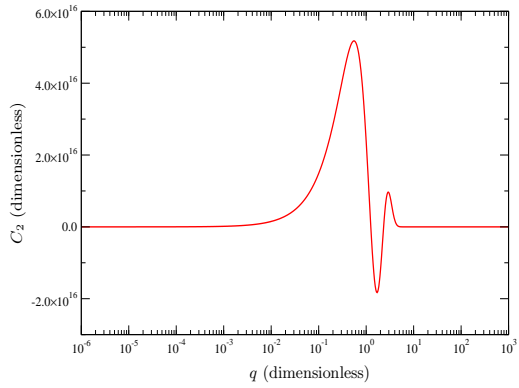
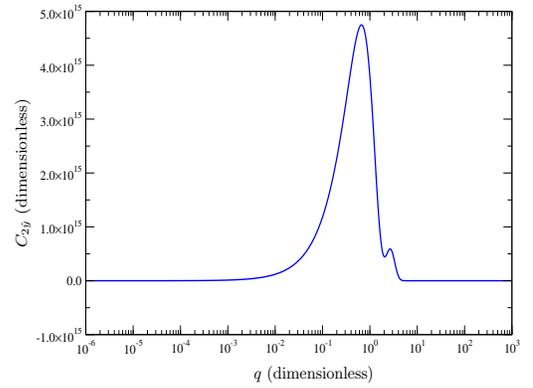

\psfrag{q}[cc][][2.5][0]{\hspace{0.5cm} $q$ (dimensionless)}
\psfrag{C0}[bc][][2.5][0]{$C_0$ (dimensionless)}
\psfrag{C1}[bc][][2.5][0]{$C_1$ (dimensionless)}
\psfrag{C2}[bc][][2.5][0]{$C_2$ (dimensionless)}
\psfrag{C0y}[bc][][2.5][0]{$C_{0\hat{y}}$ (dimensionless)}
\psfrag{C1y}[bc][][2.5][0]{$C_{1\hat{y}}$ (dimensionless)}
\psfrag{C2y}[bc][][2.5][0]{$C_{2\hat{y}}$ (dimensionless)}
\psfrag{D0}[bc][][2.5][0]{$D_0$ (dimensionless)}
\psfrag{D1}[bc][][2.5][0]{$D_1$ (dimensionless)}
\psfrag{D2}[bc][][2.5][0]{$D_2$ (dimensionless)}
\psfrag{D0x}[bc][][2.5][0]{$D_{0\hat{x}}$ (dimensionless)}
\psfrag{D1x}[bc][][2.5][0]{$D_{1\hat{x}}$ (dimensionless)}
\psfrag{D2x}[bc][][2.5][0]{$D_{2\hat{x}}$ (dimensionless)}
\psfrag{F1_Dirac}[bc][][2.5][0]{$F_1^{\rm Dirac}$ (dimensionless)}
\psfrag{F2_Dirac}[bc][][2.5][0]{$F_2^{\rm Dirac}$ (dimensionless)}
\psfrag{F1_Maj}[bc][][2.5][0]{$F_1^{\rm Maj.}$ (dimensionless)}
\psfrag{F2_Maj}[bc][][2.5][0]{$F_2^{\rm Maj.}$ (dimensionless)}
\psfrag{th = 0.5*Pi}[cc][][2.5][0]{\small $\th = 0.5 \pi$}
\psfrag{th = 0.4*Pi}[cc][][2.5][0]{\small $\th = 0.4 \pi$}
\psfrag{th = 0.3*Pi}[cc][][2.5][0]{\small $\th = 0.3 \pi$}
\psfrag{th = 0.2*Pi}[cc][][2.5][0]{\small $\th = 0.2 \pi$}
\psfrag{th = 0.1*Pi}[cc][][2.5][0]{\small $\th = 0.1 \pi$}
\psfrag{th = 0}[cc][][2.5][0]{\small $\th = 0$}
\begin{minipage}[t]{0.3 \textwidth}
\centering
\subfigure[\hspace{0.2cm} $C_0$ \, (SN1987A)]{
\label{fig:C0-SN1987A}
\rotatebox{0}{\includegraphics[width = 6.6cm, height = 5.0cm, scale = 1]{6a}}}
\end{minipage}%
\hspace{2.0cm}
\begin{minipage}[t]{0.3 \textwidth}
\centering
\subfigure[\hspace{0.2cm} $C_{0\hat{y}}$ \, (SN1987A)]{
\label{fig:C0y-SN1987A}
\rotatebox{0}{\includegraphics[width = 6.6cm, height = 5.0cm, scale = 1]{6b}}}
\end{minipage} \\
\vspace{0.8cm}
\begin{minipage}[t]{0.3 \textwidth}
\centering
\subfigure[\hspace{0.2cm} $C_1$ \, (SN1987A)]{
\label{fig:C1-SN1987A}
\rotatebox{0}{\includegraphics[width = 6.6cm, height = 5.0cm, scale = 1]{6c}}}
\end{minipage}%
\hspace{2.0cm}
\begin{minipage}[t]{0.3 \textwidth}
\centering
\subfigure[\hspace{0.2cm} $C_{1\hat{y}}$ \, (SN1987A)]{
\label{fig:C1y-SN1987A}
\rotatebox{0}{\includegraphics[width = 6.6cm, height = 5.0cm, scale = 1]{6d}}}
\end{minipage} \\
\vspace{0.8cm}
\begin{minipage}[t]{0.3 \textwidth}
\centering
\subfigure[\hspace{0.2cm} $C_2$ \, (SN1987A)]{
\label{fig:C2-SN1987A}
\rotatebox{0}{\includegraphics[width = 6.6cm, height = 5.0cm, scale = 1]{6e}}}
\end{minipage}%
\hspace{2.0cm}
\begin{minipage}[t]{0.3 \textwidth}
\centering
\subfigure[\hspace{0.2cm} $C_{2\hat{y}}$ \, (SN1987A)]{
\label{fig:C2y-SN1987A}
\rotatebox{0}{\includegraphics[width = 6.6cm, height = 5.0cm, scale = 1]{6f}}}
\end{minipage}
\caption{\label{fig:Cj-Cjy-SN1987A} Plots of dimensionless functions $C_j$ and $C_{j\hat{y}}$ with respect to $q$, where SN1987A
is the gravitational source.
Apart from their magnitudes, Figures~\ref{fig:C0-SN1987A}--\ref{fig:C2y-SN1987A} behave
in the same general way as their counterparts in Figure~\ref{fig:Cj-Cjy-Solar}.}
\end{figure*}

Using SN1987A as the gravitational source \cite{soida},
$M = 1.4 M_{\odot} = 2.072 \times 10^5$ cm, $R = 10$ km, and $r = 49$ kpc $ = 1.512 \times 10^{23}$~cm, leading to $M/r = 1.3704 \times 10^{-18}$.
Unlike the Sun, SN1987A has a much higher rotational frequency, with an estimated period of 2.14 ms.
This results in $M \Omega R^2/r^2 = 8.8702 \times 10^{-37}$.

\begin{figure*}
\psfrag{q}[cc][][2.5][0]{\hspace{0.5cm} $q$ (dimensionless)}
\psfrag{C0}[bc][][2.5][0]{$C_0$ (dimensionless)}
\psfrag{C1}[bc][][2.5][0]{$C_1$ (dimensionless)}
\psfrag{C2}[bc][][2.5][0]{$C_2$ (dimensionless)}
\psfrag{C0y}[bc][][2.5][0]{$C_{0\hat{y}}$ (dimensionless)}
\psfrag{C1y}[bc][][2.5][0]{$C_{1\hat{y}}$ (dimensionless)}
\psfrag{C2y}[bc][][2.5][0]{$C_{2\hat{y}}$ (dimensionless)}
\psfrag{D0}[bc][][2.5][0]{$D_0$ (dimensionless)}
\psfrag{D1}[bc][][2.5][0]{$D_1$ (dimensionless)}
\psfrag{D2}[bc][][2.5][0]{$D_2$ (dimensionless)}
\psfrag{D0x}[bc][][2.5][0]{$D_{0\hat{x}}$ (dimensionless)}
\psfrag{D1x}[bc][][2.5][0]{$D_{1\hat{x}}$ (dimensionless)}
\psfrag{D2x}[bc][][2.5][0]{$D_{2\hat{x}}$ (dimensionless)}
\psfrag{F1_Dirac}[bc][][2.5][0]{$F_1^{\rm Dirac}$ (dimensionless)}
\psfrag{F2_Dirac}[bc][][2.5][0]{$F_2^{\rm Dirac}$ (dimensionless)}
\psfrag{F1_Maj}[bc][][2.5][0]{$F_1^{\rm Maj.}$ (dimensionless)}
\psfrag{F2_Maj}[bc][][2.5][0]{$F_2^{\rm Maj.}$ (dimensionless)}
\psfrag{th = 0.5*Pi}[cc][][2.5][0]{\small $\th = 0.5 \pi$}
\psfrag{th = 0.4*Pi}[cc][][2.5][0]{\small $\th = 0.4 \pi$}
\psfrag{th = 0.3*Pi}[cc][][2.5][0]{\small $\th = 0.3 \pi$}
\psfrag{th = 0.2*Pi}[cc][][2.5][0]{\small $\th = 0.2 \pi$}
\psfrag{th = 0.1*Pi}[cc][][2.5][0]{\small $\th = 0.1 \pi$}
\psfrag{th = 0}[cc][][2.5][0]{\small $\th = 0$}
\begin{minipage}[t]{0.3 \textwidth}
\centering
\subfigure[\hspace{0.2cm} $D_0$ \, (SN1987A)]{
\label{fig:D0-SN1987A}
\rotatebox{0}{\includegraphics[width = 6.6cm, height = 5.0cm, scale = 1]{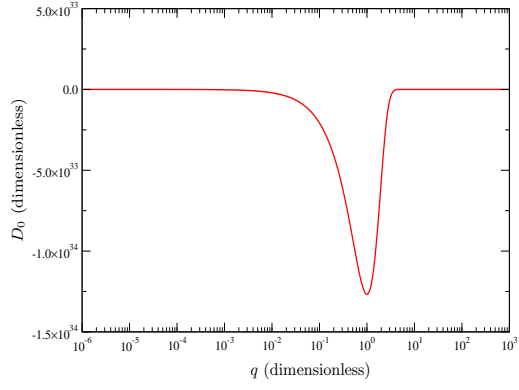}}}
\end{minipage} \\
\vspace{0.8cm}
\begin{minipage}[t]{0.3 \textwidth}
\centering
\subfigure[\hspace{0.2cm} $D_1$ \, (SN1987A)]{
\label{fig:D1-SN1987A}
\rotatebox{0}{\includegraphics[width = 6.6cm, height = 5.0cm, scale = 1]{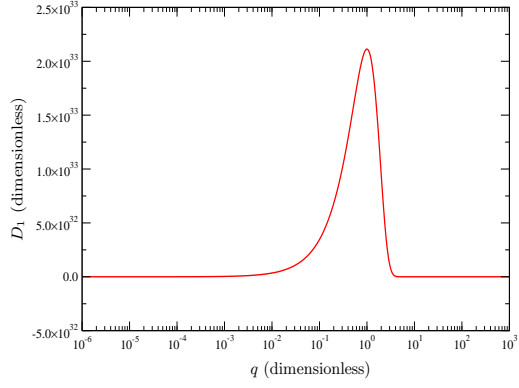}}}
\end{minipage} \\
\vspace{0.8cm}
\begin{minipage}[t]{0.3 \textwidth}
\centering
\subfigure[\hspace{0.2cm} $D_2$ \, (SN1987A)]{
\label{fig:D2-SN1987A}
\rotatebox{0}{\includegraphics[width = 6.6cm, height = 5.0cm, scale = 1]{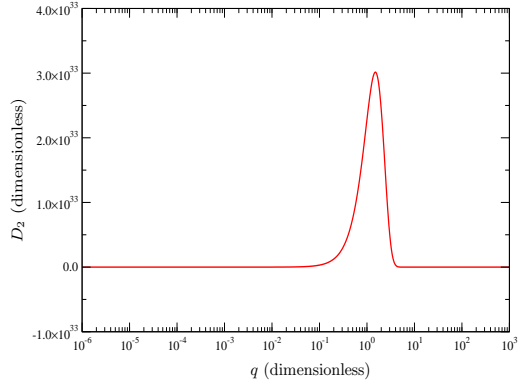}}}
\end{minipage}%
\caption{\label{fig:D0-D1-D2-SN1987A} Dimensionless functions $D_j$ as a function of $q$ for the case of SN1987A.
When the effects of wave packet spreading are taken into account, it happens that the corresponding $D_{j\hat{x}}$
are exponentially damped to zero, and therefore make no contribution to the predicted neutrino energy difference.}
\end{figure*}

Plots of the functions $C_j$ and $C_{j\hat{y}}$ are found in Figure~\ref{fig:Cj-Cjy-SN1987A}.
Essentially, they possess the same properties as found in Figure~\ref{fig:Cj-Cjy-Solar}, except for
an increase in their magnitudes.
The corresponding plots of $D_j$ are found in Figure~\ref{fig:D0-D1-D2-SN1987A}, and they also behave
in the same fashion as found in Figure~\ref{fig:Dj-Djx-Solar}.
As for $D_{j\hat{x}}$, they need to be treated with greater care.
This is because, as stated earlier, $D_{j\hat{x}}$ are proportional to $(r/R)^3$ as opposed to $(r/R)^2$ for $D_j$ (see Appendix~\ref{appendix:dimensionless-functions}),
which implies that the expression $(M\Omega R^2/r^2) D_{j\hat{x}}$ is proportional to $r/R$.
For $r \gg R$, the contribution of $D_{j\hat{x}}$ overwhelms all other contributions to the matrix element.
This appears to be physically unrealistic, since the effect of rotation should decrease with increasing radial
distance from the source.
Therefore, to compensate for this anomaly, we introduce by hand an overall regulator (\ref{regulator}) to account for the effects of wave packet spread, such that
\be
D_{j\hat{x}} & \rightarrow &  \exp\lt[-{\k^2 \lt(r - R\rt)^2 \over q^2}\rt]D_{j\hat{x}},
\label{D-regulated}
\ee
which amounts to exponentially damping these functions to zero for large $r$.

\begin{figure*}
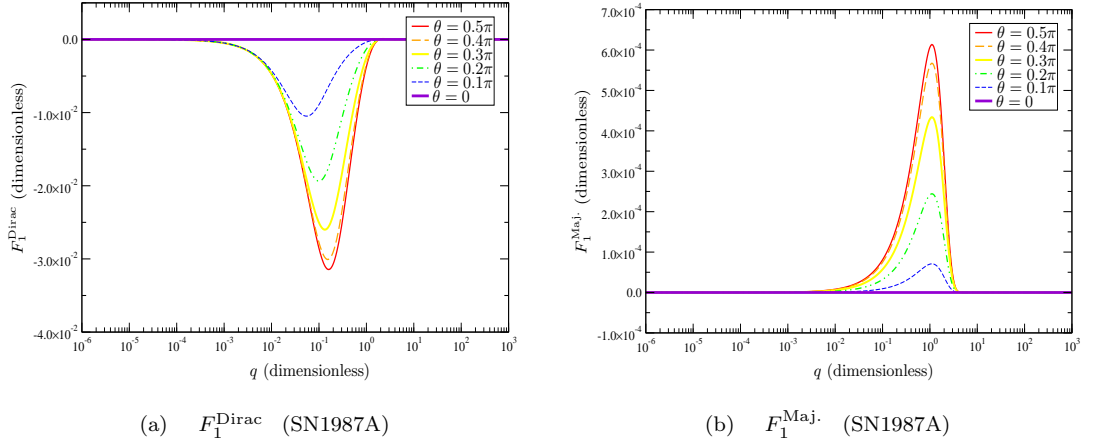

\psfrag{q}[cc][][2.5][0]{\hspace{0.5cm} $q$ (dimensionless)}
\psfrag{C0}[bc][][2.5][0]{$C_0$ (dimensionless)}
\psfrag{C1}[bc][][2.5][0]{$C_1$ (dimensionless)}
\psfrag{C2}[bc][][2.5][0]{$C_2$ (dimensionless)}
\psfrag{C0y}[bc][][2.5][0]{$C_{0\hat{y}}$ (dimensionless)}
\psfrag{C1y}[bc][][2.5][0]{$C_{1\hat{y}}$ (dimensionless)}
\psfrag{C2y}[bc][][2.5][0]{$C_{2\hat{y}}$ (dimensionless)}
\psfrag{D0}[bc][][2.5][0]{$D_0$ (dimensionless)}
\psfrag{D1}[bc][][2.5][0]{$D_1$ (dimensionless)}
\psfrag{D2}[bc][][2.5][0]{$D_2$ (dimensionless)}
\psfrag{D0x}[bc][][2.5][0]{$D_{0\hat{x}}$ (dimensionless)}
\psfrag{D1x}[bc][][2.5][0]{$D_{1\hat{x}}$ (dimensionless)}
\psfrag{D2x}[bc][][2.5][0]{$D_{2\hat{x}}$ (dimensionless)}
\psfrag{F1_Dirac}[bc][][2.5][0]{$F_1^{\rm Dirac}$ (dimensionless)}
\psfrag{F2_Dirac}[bc][][2.5][0]{$F_2^{\rm Dirac}$ (dimensionless)}
\psfrag{F1_Maj}[bc][][2.5][0]{$F_1^{\rm Maj.}$ (dimensionless)}
\psfrag{F2_Maj}[bc][][2.5][0]{$F_2^{\rm Maj.}$ (dimensionless)}
\psfrag{th = 0.5*Pi}[cc][][2.5][0]{\small $\th = 0.5 \pi$}
\psfrag{th = 0.4*Pi}[cc][][2.5][0]{\small $\th = 0.4 \pi$}
\psfrag{th = 0.3*Pi}[cc][][2.5][0]{\small $\th = 0.3 \pi$}
\psfrag{th = 0.2*Pi}[cc][][2.5][0]{\small $\th = 0.2 \pi$}
\psfrag{th = 0.1*Pi}[cc][][2.5][0]{\small $\th = 0.1 \pi$}
\psfrag{th = 0}[cc][][2.5][0]{\small $\th = 0$}
\begin{minipage}[t]{0.3 \textwidth}
\centering
\subfigure[\hspace{0.2cm} $F_1^{\rm Dirac}$ \, (SN1987A)]{
\label{fig:F1-SN1987A-Dirac}
\rotatebox{0}{\includegraphics[width = 6.6cm, height = 5.0cm, scale = 1]{8a}}}
\end{minipage}%
\hspace{2.0cm}
\begin{minipage}[t]{0.3 \textwidth}
\centering
\subfigure[\hspace{0.2cm} $F_1^{\rm Maj.}$ \, (SN1987A)]{
\label{fig:F1-SN1987A-Maj}
\rotatebox{0}{\includegraphics[width = 6.6cm, height = 5.0cm, scale = 1]{8b}}}
\end{minipage}
\caption{\label{fig:F1-SN1987A}  Comparison of $F_1$ as a function of $q$ due to the SN1987A gravitational source,
for varying neutrino beam angle $\th$.  Like the behaviour shown in Figure~\ref{fig:F1-Solar}, the effect of increasing
$\th$ is to increase the magnitude of $F_1^{\rm Dirac}$ and $F_1^{\rm Maj.}$.}
\end{figure*}

\begin{figure*}
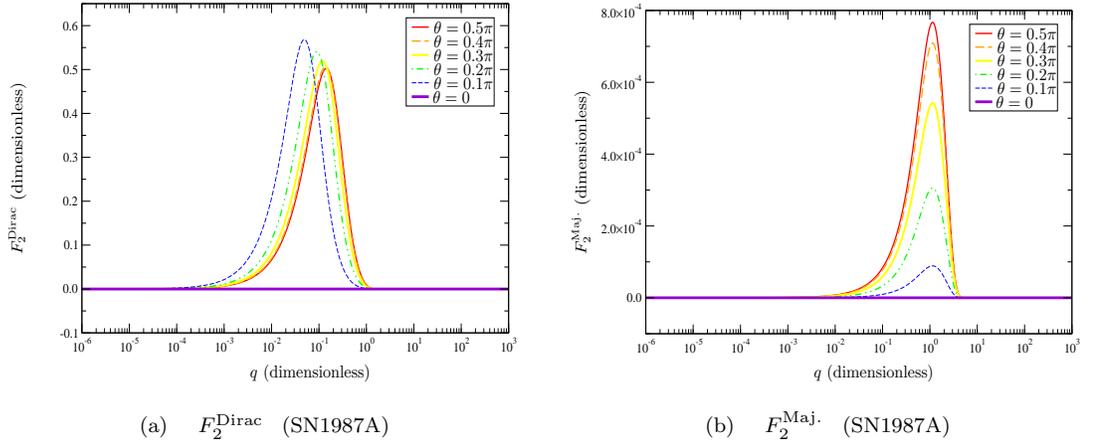

\psfrag{q}[cc][][2.5][0]{\hspace{0.5cm} $q$ (dimensionless)}
\psfrag{C0}[bc][][2.5][0]{$C_0$ (dimensionless)}
\psfrag{C1}[bc][][2.5][0]{$C_1$ (dimensionless)}
\psfrag{C2}[bc][][2.5][0]{$C_2$ (dimensionless)}
\psfrag{C0y}[bc][][2.5][0]{$C_{0\hat{y}}$ (dimensionless)}
\psfrag{C1y}[bc][][2.5][0]{$C_{1\hat{y}}$ (dimensionless)}
\psfrag{C2y}[bc][][2.5][0]{$C_{2\hat{y}}$ (dimensionless)}
\psfrag{D0}[bc][][2.5][0]{$D_0$ (dimensionless)}
\psfrag{D1}[bc][][2.5][0]{$D_1$ (dimensionless)}
\psfrag{D2}[bc][][2.5][0]{$D_2$ (dimensionless)}
\psfrag{D0x}[bc][][2.5][0]{$D_{0\hat{x}}$ (dimensionless)}
\psfrag{D1x}[bc][][2.5][0]{$D_{1\hat{x}}$ (dimensionless)}
\psfrag{D2x}[bc][][2.5][0]{$D_{2\hat{x}}$ (dimensionless)}
\psfrag{F1_Dirac}[bc][][2.5][0]{$F_1^{\rm Dirac}$ (dimensionless)}
\psfrag{F2_Dirac}[bc][][2.5][0]{$F_2^{\rm Dirac}$ (dimensionless)}
\psfrag{F1_Maj}[bc][][2.5][0]{$F_1^{\rm Maj.}$ (dimensionless)}
\psfrag{F2_Maj}[bc][][2.5][0]{$F_2^{\rm Maj.}$ (dimensionless)}
\psfrag{th = 0.5*Pi}[cc][][2.5][0]{\small $\th = 0.5 \pi$}
\psfrag{th = 0.4*Pi}[cc][][2.5][0]{\small $\th = 0.4 \pi$}
\psfrag{th = 0.3*Pi}[cc][][2.5][0]{\small $\th = 0.3 \pi$}
\psfrag{th = 0.2*Pi}[cc][][2.5][0]{\small $\th = 0.2 \pi$}
\psfrag{th = 0.1*Pi}[cc][][2.5][0]{\small $\th = 0.1 \pi$}
\psfrag{th = 0}[cc][][2.5][0]{\small $\th = 0$}
\begin{minipage}[t]{0.3 \textwidth}
\centering
\subfigure[\hspace{0.2cm} $F_2^{\rm Dirac}$ \, (SN1987A)]{
\label{fig:F2-SN1987A-Dirac}
\rotatebox{0}{\includegraphics[width = 6.6cm, height = 5.0cm, scale = 1]{9a}}}
\end{minipage}%
\hspace{2.0cm}
\begin{minipage}[t]{0.3 \textwidth}
\centering
\subfigure[\hspace{0.2cm} $F_2^{\rm Maj.}$ \, (SN1987A)]{
\label{fig:F2-SN1987A-Maj}
\rotatebox{0}{\includegraphics[width = 6.6cm, height = 5.0cm, scale = 1]{9b}}}
\end{minipage}
\caption{\label{fig:F2-SN1987A}  Comparison of $F_2$ as a function of $q$ for Dirac and Majorana neutrinos
with SN1987A as the gravitational source.
While less pronounced as that demonstrated in Figure~\ref{fig:F2-Solar}, the effect of
increasing $\th$ on Figure~\ref{fig:F2-Solar-Dirac} also leads to a decrease in the magnitude of $F_2^{\rm Dirac}$,
with a shift of the peak to lower values of $q$.
Unlike that of Figure~\ref{fig:F2-Solar-Maj}, however, the small $q$ behaviour of Figure~\ref{fig:F2-SN1987A-Maj} is as expected.}
\end{figure*}

The plots of $F_1^{\rm Dirac}$ and $F_1^{\rm Maj.}$ for varying $\th$ are listed in Figure~\ref{fig:F1-SN1987A}.
Like that shown with Figure~\ref{fig:F1-Solar} involving the Sun, Figure~\ref{fig:F1-SN1987A-Dirac} shows
that the magnitude of $F_1^{\rm Dirac}$ grows with increasing $\th$ and is on the order of $10^{-2}$,
while Figure~\ref{fig:F1-SN1987A-Maj} has a magnitude of order $10^{-4}$.
Similarly, the comparison of $F_1^{\rm Dirac}$ and $F_1^{\rm Maj.}$ as a function of $\th$ shown in Figure~\ref{fig:F2-SN1987A}
resembles that of Figures~\ref{fig:F2-Solar-Dirac} and \ref{fig:F2-Solar-Maj-Omega=0}, with magnitudes of $0.5-0.6$ and $8.0 \times 10^{-4}$, respectively.

\subsection{Observational Possibilities}

\begin{figure*}
\psfrag{q}[cc][][2.5][0]{\hspace{0.5cm} $q$ (dimensionless)}
\psfrag{dm}[bc][][2.5][0]{$\Delta \bar{m}_{21}$ (dimensionless)}
\psfrag{e = 10^(-30)}[cc][][2.5][0]{\small $\epsilon = 10^{-30}$}
\psfrag{e = 10^(-20)}[cc][][2.5][0]{\small $\epsilon = 10^{-20}$}
\psfrag{e = 10^(-18)}[cc][][2.5][0]{\small $\epsilon = 10^{-18}$}
\psfrag{e = 10^(-16)}[cc][][2.5][0]{\small $\epsilon = 10^{-16}$}
\psfrag{e = 10^(-14)}[cc][][2.5][0]{\small $\epsilon = 10^{-14}$}
\psfrag{e = 10^(-10)}[cc][][2.5][0]{\small $\epsilon = 10^{-10}$}
\psfrag{e = 10^(-8)}[cc][][2.5][0]{\small $\epsilon = 10^{-8}$}
\psfrag{e = 10^(-6)}[cc][][2.5][0]{\small $\epsilon = 10^{-6}$}
\psfrag{e = 10^(-4)}[cc][][2.5][0]{\small $\epsilon = 10^{-4}$}
\psfrag{e = 10^(-3)}[cc][][2.5][0]{\small $\epsilon = 10^{-3}$}
\psfrag{e = 10^(-2)}[cc][][2.5][0]{\small $\epsilon = 10^{-2}$}
\psfrag{e = 10^(-1)}[cc][][2.5][0]{\small $\epsilon = 10^{-1}$}
\begin{minipage}[t]{0.3 \textwidth}
\centering
\subfigure[\hspace{0.2cm} $\Delta \bar{m}_{21}^{\rm Dirac}$ \, (Sun)]{
\label{fig:delta-m-Solar-Dirac}
\rotatebox{0}{\includegraphics[width = 6.6cm, height = 5.0cm, scale = 1]{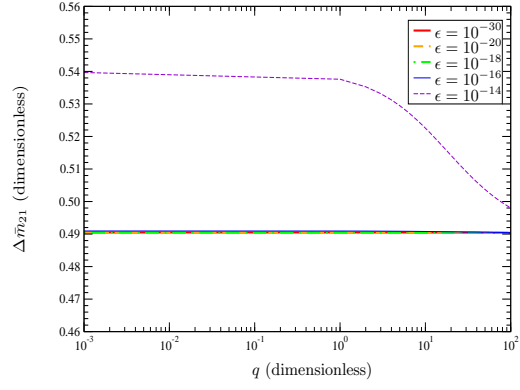}}}
\end{minipage} \\
\vspace{0.8cm}
\begin{minipage}[t]{0.3 \textwidth}
\centering
\subfigure[\hspace{0.2cm} $\Delta \bar{m}_{21}^{\rm Maj.}$ \, (Sun)]{
\label{fig:delta-m-Solar-Maj}
\rotatebox{0}{\includegraphics[width = 6.6cm, height = 5.0cm, scale = 1]{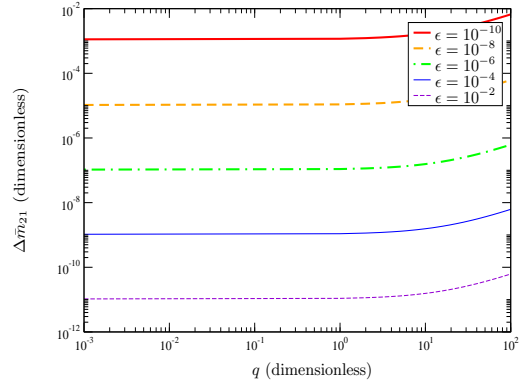}}}
\end{minipage} \\
\vspace{0.8cm}
\begin{minipage}[t]{0.3 \textwidth}
\centering
\subfigure[\hspace{0.2cm} $\Delta \bar{m}_{21}^{\rm Maj.}$ \, (Sun) \, ($\Om = 0$)]{
\label{fig:delta-m-Solar-Maj-Omega=0}
\rotatebox{0}{\includegraphics[width = 6.6cm, height = 5.0cm, scale = 1]{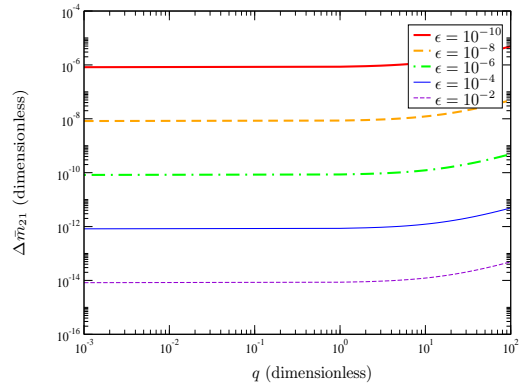}}}
\end{minipage}
\caption{\label{fig:delta-m-Solar} Upper bound of $\Delta \bar{m}_{21}$
for various choices of $\epsilon$, where the Sun is the gravitational source.
It appears that $\Delta \bar{m}_{21}$ is largely a constant in $q$ throughout the whole range available.
Comparison of Figure~\ref{fig:delta-m-Solar-Dirac} with Figure~\ref{fig:delta-m-Solar-Maj}
shows that the mass separation for Dirac neutrinos can be several orders of magnitude larger
compared to that of Majorana neutrinos.
From Figure~\ref{fig:delta-m-Solar-Maj-Omega=0}, the absence of $\Om$ serves to reduce
the upper bound of $\Delta \bar{m}_{21}$ for each choice $\epsilon$ by three orders of magnitude.
}
\end{figure*}

Because (\ref{Energy-shift-Dirac=}) and (\ref{Energy-shift-Maj=}) have both a dependence on $\Delta \bar{m}_{21}^2 \equiv \bar{m}_2^2 - \bar{m}_1^2$
and $\Delta \bar{m}_{21} \equiv \bar{m}_2 - \bar{m}_1$, where by convention we assume $m_2 > m_1$, it follows that we have enough
information to algebraically solve for the absolute masses $m_1$ and $m_2$ by a parameter fit of $q$, $\Delta \bar{m}_{21}$, and $\Delta \bar{m}_{21}^2$
to any data which registers gravitational corrections to the neutrino oscillation length.
This leads to
\be
m_1 & = & {\Delta \bar{m}_{21} \over 2} \lt[{\Delta \bar{m}_{21}^2 \over \lt(\Delta \bar{m}_{21}\rt)^2} - 1\rt] \lt(\hbar \, \k\rt),
\label{m1=}
\nl
m_2 & = & {\Delta \bar{m}_{21} \over 2} \lt[{\Delta \bar{m}_{21}^2 \over \lt(\Delta \bar{m}_{21}\rt)^2} + 1\rt] \lt(\hbar \, \k\rt).
\label{m2=}
\ee
%
It is interesting to note that, while any precise determination of the linear mass difference must come from observation,
it is possible to determine a theoretical upper bound to $\Delta \bar{m}_{21}$ strictly from consistency arguments that follow
from assuming that $m_j \geq 0$.
Since it must be true from (\ref{m1=}) that 
\be
{\Delta \bar{m}_{21} \over \Delta \bar{m}_{21}^2} \leq {1 \over \Delta \bar{m}_{21}},
\label{inequality}
\ee
it follows that we can define an experimental parameter $\epsilon$ as some deviation from the presumed mass-induced energy difference
due to special relativity, such that
\be
\Delta E & = & \lt(\hbar \, \k\rt) \lt[F_1 \, \Delta \bar{m}_{21} + \lt(F_2 + {1 \over 2}\rt) \Delta \bar{m}_{21}^2\rt]
\nn
& \equiv & \lt(\hbar \, \k\rt) \, {1 \over 2} \, \Delta \bar{m}_{21}^2 \lt(1 + 2 \, \epsilon \rt), \qquad |\epsilon| \ll 1
\ee
where $\Delta E \equiv E_{\bar{m}_2}^{(\pm)} - E_{\bar{m}_1}^{(\pm)}$.
Then upon solving for $\epsilon$, leading to
\be
\epsilon & = & {\Delta \bar{E} \over \Delta \bar{m}_{21}^2} - {1 \over 2}
\ = \ F_2 + F_1 \, \lt(\Delta \bar{m}_{21} \over \Delta \bar{m}_{21}^2 \rt),
\label{epsilon}
\ee
it follows that use of (\ref{inequality}) leads to the upper bound for the absolute neutrino mass difference,
\be
\Delta \bar{m}_{21} & \leq & {F_1 \over \epsilon - F_2}.
\label{delta-m}
\ee

\begin{figure*}
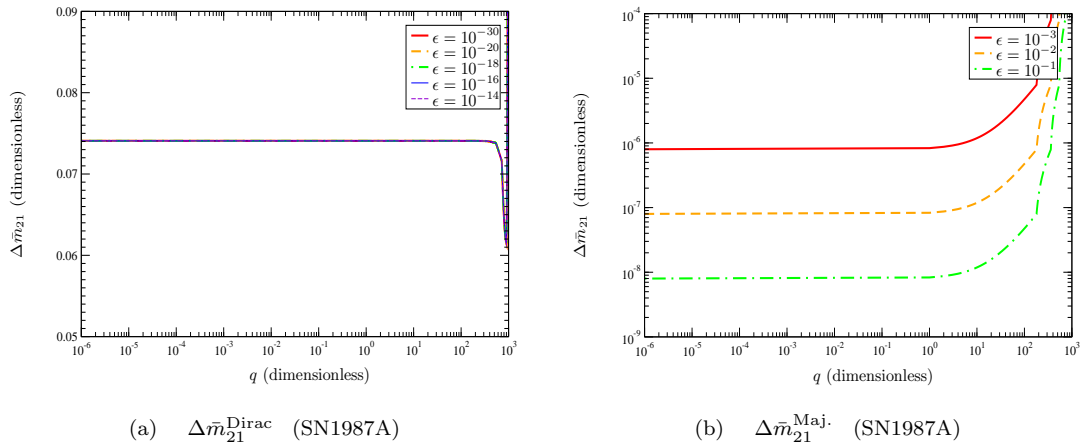

\psfrag{q}[cc][][2.5][0]{\hspace{0.5cm} $q$ (dimensionless)}
\psfrag{dm}[bc][][2.5][0]{$\Delta \bar{m}_{21}$ (dimensionless)}
\psfrag{e = 10^(-30)}[cc][][2.5][0]{\small $\epsilon = 10^{-30}$}
\psfrag{e = 10^(-20)}[cc][][2.5][0]{\small $\epsilon = 10^{-20}$}
\psfrag{e = 10^(-18)}[cc][][2.5][0]{\small $\epsilon = 10^{-18}$}
\psfrag{e = 10^(-16)}[cc][][2.5][0]{\small $\epsilon = 10^{-16}$}
\psfrag{e = 10^(-14)}[cc][][2.5][0]{\small $\epsilon = 10^{-14}$}
\psfrag{e = 10^(-10)}[cc][][2.5][0]{\small $\epsilon = 10^{-10}$}
\psfrag{e = 10^(-8)}[cc][][2.5][0]{\small $\epsilon = 10^{-8}$}
\psfrag{e = 10^(-6)}[cc][][2.5][0]{\small $\epsilon = 10^{-6}$}
\psfrag{e = 10^(-4)}[cc][][2.5][0]{\small $\epsilon = 10^{-4}$}
\psfrag{e = 10^(-3)}[cc][][2.5][0]{\small $\epsilon = 10^{-3}$}
\psfrag{e = 10^(-2)}[cc][][2.5][0]{\small $\epsilon = 10^{-2}$}
\psfrag{e = 10^(-1)}[cc][][2.5][0]{\small $\epsilon = 10^{-1}$}
\begin{minipage}[t]{0.3 \textwidth}
\centering
\subfigure[\hspace{0.2cm} $\Delta \bar{m}_{21}^{\rm Dirac}$ \, (SN1987A)]{
\label{fig:delta-m-SN1987A-Dirac}
\rotatebox{0}{\includegraphics[width = 6.6cm, height = 5.0cm, scale = 1]{11a}}}
\end{minipage}%
\hspace{2.0cm}
\begin{minipage}[t]{0.3 \textwidth}
\centering
\subfigure[\hspace{0.2cm} $\Delta \bar{m}_{21}^{\rm Maj.}$ \, (SN1987A)]{
\label{fig:delta-m-SN1987A-Maj}
\rotatebox{0}{\includegraphics[width = 6.6cm, height = 5.0cm, scale = 1]{11b}}}
\end{minipage}
\caption{\label{fig:delta-m-SN1987A} Upper bound of $\Delta \bar{m}_{21}$ as a function of $q$
for the case of SN1987A.
From Figure~\ref{fig:delta-m-SN1987A-Dirac}, it appears that the maximum mass separation for Dirac
neutrinos is independent of the choice of $\epsilon$, while Figure~\ref{fig:delta-m-SN1987A-Maj}
shows that the upper bound decreases in proportion to each order of magnitude increase of $\epsilon$.
}
\end{figure*}

Figures~\ref{fig:delta-m-Solar} and \ref{fig:delta-m-SN1987A} describe (\ref{delta-m}) for the solar and SN1987A cases, respectively,
as a function of $q$ and for a variety of choices for $\epsilon$.
We notice immediately that the upper bound for $\Delta \bar{m}_{21}$ is essentially constant in $q$
for most choices of $\epsilon$, except when $q \gtrsim 10$.
Direct comparison of $\Delta \bar{m}_{21}^{\rm Dirac}$ with $\Delta \bar{m}_{21}^{\rm Maj.}$
shows that the former allows for a much higher upper bound with even very tiny choices of $\epsilon$, while
the latter is very sensitive to the choice of $\epsilon$ and leads to more restrictive upper bounds for $\Delta \bar{m}_{21}$.
For the case of the Sun, Figure~\ref{fig:delta-m-Solar-Dirac} shows that $\Delta \bar{m}_{21}^{\rm Dirac} \leq 0.491$
for $10^{-30} \leq \epsilon \leq 10^{-16}$, which then increases to $\Delta \bar{m}_{21}^{\rm Dirac} \lesssim 0.54$ for
most choices of $q$ when $\epsilon = 10^{-14}$.
Given that the mean momentum $\hbar \, \k$ for solar neutrinos is on the MeV scale and we assume oscillations between
electron and muon neutrinos, the upper bound for $\Delta m_{21}^{\rm Dirac} = \Delta \bar{m}_{21}^{\rm Dirac}(\hbar \, \k)$ is
largely consistent with the upper bound of $m_{\nu_\mu} \leq $ 0.170 MeV \cite{mohapatra}, as determined by experiment.
In contrast, Figure~\ref{fig:delta-m-Solar-Maj} shows that $\Delta \bar{m}_{21}^{\rm Maj.} \lesssim 10^{-3}$ for
$\epsilon = 10^{-10}$ and $\Delta \bar{m}_{21}^{\rm Maj.} \lesssim 10^{-11}$ for $\epsilon = 10^{-2}$,
while Figure~\ref{fig:delta-m-Solar-Maj-Omega=0} leads to
$\Delta \bar{m}_{21}^{\rm Maj.} \lesssim 10^{-6}$ and $\Delta \bar{m}_{21}^{\rm Maj.} \lesssim 10^{-14}$
for the same respective choices of $\epsilon$.
These results suggest the prediction that Majorana neutrino masses must be closely spaced together
to be self consistent with this model, while Dirac neutrinos have the option of also having a much larger
mass separation.
This property is also reflected in Figures~\ref{fig:delta-m-SN1987A-Dirac} and \ref{fig:delta-m-SN1987A-Maj}, where
$\Delta \bar{m}_{21}^{\rm Dirac} \leq 0.074$ for the full range of $q$ considered, while
$\Delta \bar{m}_{21}^{\rm Maj.} \lesssim 10^{-6}$ for $\epsilon = 10^{-3}$ and reduces to
$\Delta \bar{m}_{21}^{\rm Maj.} \lesssim 10^{-8}$ for $\epsilon = 10^{-1}$.
Furthermore, a much larger $\epsilon$ is required to obtain $\Delta \bar{m}_{21}^{\rm Maj.}$ than for $\Delta \bar{m}_{21}^{\rm Dirac}$,
which immediately follows from the fact that $F_1^{\rm Dirac}$ carries opposite sign compared to $F_2^{\rm Dirac}$,
while $F_1^{\rm Maj.}$ and $F_2^{\rm Maj.}$ have the same sign.

Finally, we consider the likelihood of observing the gravitational corrections in future
neutrino oscillation experiments.
The best possibility appears to be with the efforts of SNO \cite{SNO1} and Borexino \cite{Borexino} to measure
low-energy solar neutrinos on the order of 1~MeV, whose flux rate is many orders of magnitude larger
than for previous measurements.
This can be done in combination with a precision measurement of $\Delta \bar{m}_{21}^2$ from an independent observation, such as from
the T2K \cite{T2K} long baseline neutrino oscillation experiment under development. 
In particular, SNO is focussing attention on measuring the monoenergetic solar neutrinos \cite{chen}
with energy $E_{\nu}$ = 1.44~MeV due to the $pep$ reaction component in the $pp-I$ chain \cite{mohapatra},
with an uncertainty of 3-5\% \cite{chen}.
The advantage with this measurement comes from knowing that this reaction is confined to a very
narrow energy range and suggests a realistic value for $q$ that falls within the non-zero values for
$F_j$.
As for neutrinos from supernovae like SN1987A, while the plots from Figures~\ref{fig:F1-SN1987A} and \ref{fig:F2-SN1987A}
show a significant spin-gravity correction to the mass-dependent energy difference, it seems unlikely \cite{chen}
that we may be able to find suitable candidates which can provide the required precision to identify a
recognizable signal.
However, it remains to be seen whether that is the final answer.

\section{Conclusion}
\label{section:conclusion}

We have shown in this paper that spin-gravity interaction, via gravitational phase and time-independent perturbation theory,
offers the possibility of distinguishing between Dirac and Majorana neutrinos propagating as wave packets.
The distinctions between the two neutrino types are unambiguous at the matrix element level, and lead to meaningful predictions
in gravitational corrections to the neutrino oscillation length.
This has been explicitly demonstrated using both the Sun and SN1987A as test cases for detailed analysis.
The results presented here demonstrate that gravity can be an interesting probe to better understand the fundamental nature of neutrinos
and possibly other subatomic particles.
Because the associated matrix elements for the two neutrino types generates a linear mass term in the perturbed total energy
that is coupled to the background gravitational field, we have the theoretical possibility of determining the absolute
masses for a two-flavour oscillating system.
Furthermore, this model can be used to determine for both Dirac and Majorana neutrinos the allowable upper bound for the
absolute mass difference, and shows that while Dirac neutrinos can have a large mass separation, Majorana neutrinos
must necessarily have a much smaller mass separation under the same conditions.

It is worthwhile to identify details where improvements of the model can be employed.
For instance, we made the assumption that the width of the neutrino wave packet is constant in time.
Since this is strictly not true, it will be necessary to incorporate the spreading of the wave packet in
any future development of this project.
In so doing, however, we change the fundamental nature of the problem such that we then require either
time-dependent perturbation theory or some other compatible technique.
Another technical point to address concerns the use of a regulator to tame some isolated terms in the matrix
elements which otherwise grow infinitely large in the spatial integration.
This issue may actually be resolved if a dynamical wave packet width is accounted for in a more refined model.
Since we know of at least three neutrino flavours in existence, further development of this project
may include a three-flavour oscillation, which introduces considerable complexity to the original problem.
On a more fundamental level, it is possible to consider abandoning the wave packet model altogether in favour of
a quantum field theory approach in curved space-time \cite{birrell}.
It remains to be seen what is the best route to follow for any future developments of this analysis.
Nonetheless, it seems clear that the results presented here merit further investigation.

\section{Acknowledgements}

We wish to thank Mark Chen from Queen's University and the SNO Collaboration for helpful comments about
SNO's upcoming low-energy solar neutrino experiments.
One of us (DS) also thanks Robert Petry from the University of Regina for useful discussions about this project.
This research was supported, in part, by the Natural Sciences and Engineering Research Council of Canada (NSERC).

\begin{appendix}
\section{Evaluation of momentum integrals}
\label{appendix:momentum-integrals}
\renewcommand{\theequation}{A.\arabic{equation}}
\setcounter{subsection}{0} \setcounter{equation}{0}

The required integrations over $k$ are described by the functions
\be
Q_c(n) & = & \int_0^\infty \exp\lt[-{\lt(k - k_0\rt)^2 \over 4 \, \sg_{\rm p}^2}\rt] \, \cos(k \, r) \, k^n \, \d k
\ = \ {\rm Re}\lt[Q(n)\rt],
\label{Qcn}
\nl
Q_s(n) & = & \int_0^\infty \exp\lt[-{\lt(k - k_0\rt)^2 \over 4 \, \sg_{\rm p}^2}\rt] \, \sin(k \, r) \, k^n \, \d k
\ = \ {\rm Im}\lt[Q(n)\rt],
\label{Qsn}
\ee
where
\be
Q(n) & = & \int_0^\infty \exp\lt[-{\lt(k - k_0\rt)^2 \over 4 \, \sg_{\rm p}^2}\rt] \, e^{ikr} \, k^n \, \d k
\label{Qn}
\ee
and $n$ are integers and $-2 \leq n \leq 2$.
For $q = \k/\sg_{\rm p}$, the evaluation of (\ref{Qn}) for non-negative $n$ is relatively straightforward,
the results of which are
\be
Q(0) & = & \sqrt{\pi} \lt({k_0 \over q}\rt) \lt[1 + \erf\lt[{q \over 2} + i \lt({k_0 \, r \over q}\rt)\rt]\rt]e^{-\krq} \, e^{i \k r},
\label{Q0-demo}
\nl
Q(1) & = & 2 \lt({k_0 \over q}\rt)^2 \, e^{-q^2/4}
+ \sqrt{\pi} \lt({k_0 \over q}\rt)^2 \lt[q + 2 \, i \, \lt({k_0 \, r \over q}\rt) \rt]
\lt[1 + \erf\lt[{q \over 2} + i \lt({k_0 \, r \over q}\rt)\rt]\rt]e^{-\krq} \, e^{i \k r},
\label{Q1-demo}
\nl
Q(2) & = & 2 \lt({k_0 \over q}\rt)^3 \lt[q + 2 \, i \, \lt({k_0 \, r \over q}\rt) \rt]e^{-q^2/4}
\nn
& &{} + \sqrt{\pi} \lt({k_0 \over q}\rt)^3 \lt[\lt(q^2 + 2\rt) + 4 \, i \lt(\k r\rt) - 4 \, \lt({k_0 \, r \over q}\rt)^2 \rt]
\lt[1 + \erf\lt[{q \over 2} + i \lt({k_0 \, r \over q}\rt)\rt]\rt]e^{-\krq} \, e^{i \k r}.
\label{Q2-demo}
\ee
In order to allow for the final integration over $r$, the complex error functions present in (\ref{Q0-demo})--(\ref{Q2-demo}) and elsewhere have
to be approximated for suitable limiting cases.
For a complex variable $z = x + i \, y$ with $|z| \gg 1$, it is generally true that
\be
\lt. \erf(z) \rt|_{|z| \gg 1} & = & 1 - {1 \over \sqrt{\pi}} \lt[{1 \over z} - {1 \over 2 \, z^3}
 + {3 \over 4 \, z^5} - {15 \over 8 \, z^7} + {105 \over 16 \, z^9} - \cdots \rt]e^{-z^2}
\nn
& = &  1 - {1 \over \sqrt{\pi}} \lt[{x - i \, y \over x^2 + y^2} - {1 \over 2} \lt({x - i \, y \over x^2 + y^2}\rt)^3
+ {3 \over 4} \lt({x - i \, y \over x^2 + y^2}\rt)^5 - {15 \over 8} \lt({x - i \, y \over x^2 + y^2}\rt)^7 + \cdots
\rt]e^{-z^2}.
\label{erf-z>>1}
\ee
If $x = q/2$ and $y = \k r/q$, then for all the choices of $q$ considered for given $\k$ and $R$, it is always true that
$y \gg 1$, and that $x/y = {1 \over 2} \, (q^2/(\k r)) \ll 1$.
By performing an expansion of (\ref{erf-z>>1}) with respect to $x/y$, it can be shown that to leading order for fixed $\k$,
\be
\erf\lt[{q \over 2} + i \lt({k_0 \, r \over q}\rt)\rt] & \approx & 1 - {1 \over \sqrt{\pi}} \lt({q \over \k r}\rt)
\lt[{q \over 2} \lt({q \over \k r}\rt) - i \rt] e^{-q^2/4} \, e^{\krq} \, e^{-i \k r}.
\label{erf-approx-qsmall}
\ee
%
%
%
Substitution of (\ref{erf-approx-qsmall})
into (\ref{Q0-demo})--(\ref{Q2-demo}) leads to (\ref{Qcn}) and (\ref{Qsn}) once the
real and imaginary parts of the final expressions are evaluated.

The integrals $Q(-1)$ and $Q(-2)$ are more challenging to perform, given that $k$ is in the denominator, suggesting the possibility
of infrared divergences for these expressions.
However, there is a procedure available to obtain these integrals without any troubling behaviour as $k \rightarrow 0$.
To accomplish this, we first define a function $f_n(\k)$ where
\be
f_n(\k) & = & \int_0^\infty \exp\lt[-{\lt(k - k_0\rt)^2 \over 4 \, \sg_{\rm p}^2}\rt] \, {e^{ikr} \over k^n} \, \d k, \qquad n = 1, \, 2
\label{fn}
\ee
such that $Q_c(-n) = {\rm Re}\lt[f_n(\k)\rt]$ and $Q_s(-n) = {\rm Im}\lt[f_n(\k)\rt]$.
By differentiating (\ref{fn}) with respect to $\k$, it follows that the integral must satisfy the first-order differential equation
\be
{df_n(\k) \over d \k} + {\k \over 2 \, \sg_{\rm p}^2} \, f_n(\k) & = & {1 \over 2 \, \sg_{\rm p}^2} \, f_{n-1}(\k),
\label{diff-eq-k0}
\ee
which immediately leads to the solution
\be
f_n(\k) & = & {1 \over 2 \, \sg_{\rm p}^2} \, e^{-\k^2/4\sg_{\rm p}^2} \, \int \, e^{\k^2/4\sg_{\rm p}^2} \, f_{n-1}(\k) \, \d \k + K_n
\label{fn-solution}
\ee
with constant of integration $K_n$.
Since $f_n(\k) \rightarrow 0$ as $\k \rightarrow \infty$, it is clear that $K_n = 0$ from this boundary condition.

By repeated use of (\ref{fn-solution}), $f_n(\k)$ can be explicitly evaluated for all $n \geq 1$, provided that an explicit solution for $f_1(\k)$
is obtainable.
Fortunately, given that $f_0(\k) = Q(0)$ by construction, $f_1(\k)$ can be expressed upon substitution of (\ref{Q0-demo}) into (\ref{fn-solution}).
Utilizing (\ref{erf-approx-qsmall}) 
and the integral
\be
\int \, e^{\k^2/4 \sg_{\rm p}^2} \, e^{i \k r} \, \d \k & = &
i \sqrt{\pi} \lt({k_0 \over q}\rt) e^{\krq} \, \erf\lt[\lt({k_0 \, r\over q}\rt) - i \lt({q \over 2}\rt)\rt],
\ee
where
\be
\erf\lt[\lt({k_0 \, r\over q}\rt) - i \lt({q \over 2}\rt)\rt] & \approx &  1 - {1 \over \sqrt{\pi}} \lt({q \over \k r}\rt)
\lt[1 + i \, {q \over 2} \lt({q \over \k r}\rt)\rt] e^{-\krq} \, e^{q^2/4} \,  e^{i \k r}
\label{erf-approx-qsmall-2}
\ee
by employing the same line of reasoning as used to determine (\ref{erf-approx-qsmall}), it follows that
\be
f_1(\k) & = & Q(-1) \ = \ \lt[-{1 \over 8} \lt({\k^2 \over \sg_{\rm p}^4 \, r^2}\rt) + {i \over 2} \lt({\k \over \sg_{\rm p}^2 \, r}\rt)
+ i \, \pi \, \erf\lt[\lt({k_0 \, r\over q}\rt) - i \lt({q \over 2}\rt)\rt] \rt]e^{-\k^2/4\sg_{\rm p}^2}
\nn
& \approx & \sqrt{\pi} \lt({q \over \k \, r}\rt) \lt[{q \over 2} \lt({q \over \k \, r}\rt) - i\rt] e^{-\krq} \, e^{i\k r}
\nn
& &{}
- \lt[{q^2 \over 8}\lt({q \over \k \, r}\rt)^2 - i \, \pi \lt[1 + {q \over 2 \pi} \lt({q \over \k \, r}\rt) \rt]\rt] e^{-q^2/4}.
\label{f1-qsmall}
\ee
%
%
Finding (\ref{fn-solution}) for $n = 2$ is also a straightforward process, given (\ref{f1-qsmall}).
This leads to
\be
f_2(\k) & = & Q(-2) \ = \ \lt[-{1 \over 48} \lt({\k^3 \over \sg_{\rm p}^6 \, r^2}\rt)
+ {i \, \pi \over 2} \lt({\k \over \sg_{\rm p}^2}\rt) \lt[1 + {1 \over 4 \, \pi} \lt({\k \over \sg_{\rm p}^2 \, r}\rt)\rt]
+ \lt({\pi \over \sg_{\rm p}^2 \, r}\rt) \erf\lt[\lt({k_0 \, r\over q}\rt) - i \lt({q \over 2}\rt)\rt] \rt]e^{-\k^2/4\sg_{\rm p}^2}
\nn
& &{} + {\sqrt{\pi} \over 2} \, {1 \over \sg_{\rm p}^3 \, r^2} \, e^{-(\sg_{\rm p} r)^2} \, e^{i \k r}
\nn
& \approx & -{\sqrt{\pi} \over 2} \lt({q \over \k}\rt) \lt[ \lt({q \over \k \, r}\rt)^2 + i \, q  \lt({q \over \k \, r}\rt)^3 \rt] e^{-\krq} \, e^{i\k r}
\nn
& &{} + \pi \lt({q \over \k}\rt)\lt[\lt({q \over \k \, r}\rt) - {q^3 \over 48 \, \pi} \lt({q \over \k \, r}\rt)^2
+ i \, {q  \over 2} \lt[1 + {2 \, q \over \pi} \lt({q \over \k \, r}\rt) \rt] \rt]e^{-q^2/4}.
\label{f2-qsmall}
\ee
%

Therefore, the final expressions for $Q_c(n)$ and $Q_s(n)$ are the following:
\be
Q_c(-2) & = &
- {\sqrt{\pi} \over 2} \lt({q \over \k}\rt)\lt({q \over \k r}\rt)^2 \lt[\cos(\k r) - q \lt({q \over \k r}\rt) \sin(\k r) \rt]e^{-\krq}
\nn
& &{} + \pi \lt({q \over \k}\rt)\lt({q \over \k r}\rt) \lt[1 - {q^3 \over 48 \pi} \lt({q \over \k r}\rt)\rt]e^{-q^2/4},
\label{Qc(-2)-qsmall}
\nl
Q_s(-2) & = &
- {\sqrt{\pi} \over 2} \lt({q \over \k}\rt)\lt({q \over \k r}\rt)^2 \lt[\sin(\k r) + q \lt({q \over \k r}\rt) \cos(\k r) \rt]e^{-\krq}
\nn
& &{} + {\pi \, q \over 2} \lt({q \over \k}\rt) \lt[1 + {2 \, q \over \pi} \lt({q \over \k r}\rt)\rt]e^{-q^2/4},
\label{Qs(-2)-qsmall}
\nl
Q_c(-1) & = &
\sqrt{\pi} \lt({q \over \k r}\rt) \lt[{q \over 2} \lt({q \over \k r}\rt) \cos(\k r) + \sin(\k r) \rt]e^{-\krq} - {q^2 \over 8}\lt({q \over \k r}\rt)^2 e^{-q^2/4},
\label{Qc(-1)-qsmall}
\nl
Q_s(-1) & = &
\sqrt{\pi} \lt({q \over \k r}\rt) \lt[{q \over 2} \lt({q \over \k r}\rt) \sin(\k r) - \cos(\k r) \rt]e^{-\krq}
+ \pi \lt[1 + {q \over 2\pi} \lt({q \over \k r}\rt)\rt] e^{-q^2/4},
\label{Qs(-1)-qsmall}
\nl
Q_c(0) & = &
2\sqrt{\pi} \lt({\k \over q}\rt)\lt[\cos(\k r) \, e^{-\krq} - {q \over 4 \sqrt{\pi}} \lt({q \over \k r}\rt)^2 e^{-q^2/4} \rt],
\label{Qc(0)-qsmall}
\nl
Q_s(0) & = &
2\sqrt{\pi} \lt({\k \over q}\rt)\lt[\sin(\k r) \, e^{-\krq} + {1 \over 2 \sqrt{\pi}} \, e^{-q^2/4} \rt],
\label{Qs(0)-qsmall}
\nl
Q_c(1) & = &
2\sqrt{\pi} \lt({\k \over q}\rt)^2 \lt[q \, \cos(\k r) - 2 \lt({\k r \over q}\rt) \sin(\k r)\rt] e^{-\krq}
\nn
& &{}
+ 2 \lt({\k \over q}\rt)^2 \lt[1 - {q^2 \over 4} \lt({q \over \k r}\rt)^2 - \lt({\k r \over q}\rt)\rt] e^{-q^2/4},
\label{Qc(1)-qsmall}
\nl
Q_s(1) & = &
2\sqrt{\pi} \lt({\k \over q}\rt)^2\lt\{\lt[q \, \sin(\k r) + 2 \lt({\k r \over q}\rt) \cos(\k r)\rt] e^{-\krq}
+ {q \over 2 \sqrt{\pi}} \lt[1 - \lt({q \over \k r}\rt)\rt] e^{-q^2/4}\rt\},
\label{Qs(1)-qsmall}
\nl
Q_c(2) & = &
2\sqrt{\pi} \lt({\k \over q}\rt)^3 \lt[\lt[2 + q^2 - 4 \lt({\k r \over q}\rt)^2 \rt] \cos(\k r) - 4 \lt(\k r\rt) \sin(\k r)\rt] e^{-\krq}
\nn
& &{}
+ \lt({\k \over q}\rt)^3 \lt[2 \, q  - {q \over 2} \lt[2 + q^2 - 4 \lt({\k r \over q}\rt)^2 \rt] \lt({q \over \k r}\rt)^2 - 4 \lt(\k r\rt)\rt] e^{-q^2/4},
\label{Qc(2)-qsmall}
\nl
Q_s(2) & = &
2\sqrt{\pi} \lt({\k \over q}\rt)^3 \lt[\lt[2 + q^2 - 4 \lt({\k r \over q}\rt)^2 \rt] \sin(\k r) + 4 \lt(\k r\rt) \cos(\k r)\rt] e^{-\krq}
\nn
& &{}
+ \lt({\k \over q}\rt)^3 \lt[4 \lt({\k r \over q}\rt) - 2 \, q \lt(\k r\rt) \lt({q \over \k r}\rt)^2
- \lt[2 + q^2 - 4 \lt({\k r \over q}\rt)^2 \rt] \lt({q \over \k r}\rt)^2 \rt] e^{-q^2/4}.
\label{Qs(2)-qsmall}
\ee

\section{Evaluation of radial integrals}
\label{appendix:radial-integrals}
\renewcommand{\theequation}{B.\arabic{equation}}
\setcounter{subsection}{0} \setcounter{equation}{0}

For this paper, most of the radial integrals have the form
\be
S(\al, \bt, n) & = & \int_R^\infty e^{-\al \krq} \, e^{i \bt \k r} \, r^n \, \d r, \qquad n = -9 \ldots 2
\label{S}
\ee
where $\al$ is a positive integer and $\bt$ is an integer in the range $-\al \leq \bt \leq \al$.
For nonnegative $n$, (\ref{S}) can be immediately integrated, leading to
\be
S(\al, \bt, 0) & = & \lt. {1 \over 2} \, \sqrt{\pi \over \al} \lt({q \over \k}\rt) \, e^{-\bt^2 q^2/(4\al)} \,
\erf \lt[\sqrt{\al} \lt({\k \, r \over q}\rt) - i \, {\bt \over \sqrt{\al}} \lt({q \over 2}\rt) \rt] \rt|_R^\infty,
\label{S0a}
\nl
S(\al, \bt, 1) & = & \lt.  -\lt({q \over \k}\rt)^2 \lt\{ {1 \over 2 \al} \, e^{i \bt \k r} \, e^{-\al \krq} - i \, {\bt \, q \over 4} \, \sqrt{\pi \over \al^3} \,
e^{-\bt^2 q^2/(4\al)} \, \erf \lt[\sqrt{\al} \lt({\k \, r \over q}\rt) - i \, {\bt \over \sqrt{\al}} \lt({q \over 2}\rt) \rt] \rt\}\rt|_R^\infty,
\nn
\label{S1a}
\nl
S(\al, \bt, 2) & = &  -\lt({q \over \k}\rt)^3 \lt\{ \lt[{1 \over 2 \al} \lt({\k \, r \over q}\rt) + i \, {\bt \, q \over 4\al^2}\rt] e^{i \bt \k r} \, e^{-\al \krq} \rt.
\nn
& &{}- \lt. \lt. {i \over 8} \, \sqrt{\pi \over \al^5} \lt(2 \al - \bt^2 \, q^2 \rt)
e^{-\bt^2 q^2/(4\al)} \, \erf \lt[\sqrt{\al} \lt({\k \, r \over q}\rt) - i \, {\bt \over \sqrt{\al}} \lt({q \over 2}\rt) \rt] \rt\}\rt|_R^\infty,
\label{S2a}
\ee
where
\be
\erf \lt[\sqrt{\al} \lt({\k \, r \over q}\rt) - i \, {\bt \over \sqrt{\al}} \lt({q \over 2}\rt) \rt]
& \approx & 1 - {1 \over \sqrt{\al \, \pi}} \lt({q \over \k \, r}\rt) \lt[1 + i \, {\bt \, q \over 2 \al} \lt({q \over \k \, r}\rt) \rt]
e^{-\bt^2 q^2/(4\al)} \, e^{-\al \krq} \, e^{i \bt \k r}.
\label{erf}
\ee
It then follows that
\be
S(\al, \bt, 0) & = & {1 \over 2\al} \lt({q \over \k R}\rt)^2 \lt\{ \lt[\cos(\bt \, \k R)
 - {\bt \, q \over 2 \al} \lt({q \over \k R}\rt) \sin(\bt \, \k R)\rt] \rt.
\nn
& &{} + \lt. i \lt[ {\bt \, q \over 2 \al} \lt({q \over \k R}\rt) \cos(\bt \, \k \, R)
+ \sin(\bt \, \k R) \rt] \rt\} e^{-\al \lt(\k R/q\rt)^2} \, R,
\label{S0}
\nl \nn
S(\al, \bt, 1) & = & {1 \over 2\al} \lt({q \over \k R}\rt)^2 \lt\{\lt[ \lt[1 - {\bt^2 \, q^2 \over 4\al^2} \lt({q \over \k R}\rt)^2 \rt] \cos(\bt \, \k R)
 - {\bt \, q \over 2 \al} \lt({q \over \k R}\rt) \sin(\bt \, \k R) \rt] \rt.
\nn
& &{} + \lt. i \lt[ {\bt \, q \over 2 \al} \lt({q \over \k R}\rt) \cos(\bt \, \k \, R)
+ \lt[1 - {\bt^2 \, q^2 \over 4\al^2} \lt({q \over \k R}\rt)^2 \rt] \sin(\bt \, \k R) \rt] \rt\} e^{-\al \lt(\k R/q\rt)^2} \, R^2,
\label{S1}
\nl \nn
S(\al, \bt, 2) & = & {1 \over 2\al} \lt({q \over \k R}\rt)^2 \lt[1 + {1 \over 4\al^2}\lt(2\al - \bt^2 \, q^2\rt)\lt({q \over \k R}\rt)^2\rt]
\lt\{ \lt[\cos(\bt \, \k R) - {\bt \, q \over 2 \al} \lt({q \over \k R}\rt) \sin(\bt \, \k R)\rt] \rt.
\nn
& &{} + \lt. i \lt[ {\bt \, q \over 2 \al} \lt({q \over \k R}\rt) \cos(\bt \, \k \, R) + \sin(\bt \, \k R) \rt] \rt\} e^{-\al \lt(\k R/q\rt)^2} \, R^3.
\label{S2}
\ee

However, for positive-valued $n$, the evaluation of (\ref{S}) is less straightforward.
Fortunately, by treating $\bt$ as a continuous parameter and making use of the identity
\be
1 & = & \int \d \bt \lt({\d \over \d \bt}\rt),
\label{identity}
\ee
it is possible to solve for $S(\al, \bt, -1)$ using (\ref{S0}).
This leads to the iteration solution
\be
S(\al, \bt, -c) & = & i \, \k \int \, S\lt(\al, \bt, -(c - 1)\rt) \, \d \bt, \qquad c = 1 \ldots 9
\nn
& = & {1 \over 2\al} \lt({q \over \k R}\rt)^2 \lt\{ \lt[\lt[1 - {c \over 2\al} \lt({q \over \k R}\rt)^2 \rt] \cos(\bt \, \k R)
 - {\bt \, q \over 2 \al} \lt({q \over \k R}\rt) \sin(\bt \, \k R)\rt] \rt.
\nn
& &{} + \lt. i \lt[{\bt \, q \over 2 \al} \lt({q \over \k R}\rt) \cos(\bt \, \k \, R)
+ \lt[1 - {c \over 2\al} \lt({q \over \k R}\rt)^2 \rt] \sin(\bt \, \k R) \rt] \rt\} {e^{-\al \lt(\k R/q\rt)^2} \over R^{(c-1)}}. \qquad
\label{S-neg}
\ee
As expected, when $\bt \rightarrow - \bt$ the real parts of (\ref{S0})--(\ref{S2}) and (\ref{S-neg}) remain unaffected, while the imaginary parts
change sign under this transformation.

\section{Evaluation of the dimensionless functions $C_j$ and $D_j$}
\label{appendix:dimensionless-functions}
\renewcommand{\theequation}{C.\arabic{equation}}
\setcounter{subsection}{0} \setcounter{equation}{0}

In this Appendix, we list in analytical form the dimensionless functions $C_j$ and $D_j$ for the
Dirac and Majorana matrix elements, due to the gravitational phase $\PhiG$.
Up to the approximation of the complex error functions, as described in
Appendices~\ref{appendix:momentum-integrals} and~\ref{appendix:radial-integrals},
these expressions for $C_j$ and $D_j$ are {\em exact}.
While almost all the terms in the final expressions below are well behaved for the full range of $q$,
there are some isolated terms which diverge when $q \rightarrow 0$ and/or when
$r \rightarrow \infty$.
For these terms, we insert by hand an exponential decay regulator of the form
\be
\xi\lt(q, \k, r, R\rt) & \equiv & \exp\lt[-{\k^2 \lt(r - R\rt)^2 \over q^2}\rt],
\label{regulator}
\ee
which mimics the effects of wave packet spreading with time to provide a physically plausible cutoff of the
respective integrals.
The choice of (\ref{regulator}) results from evaluating the time evolution of a
one-dimensional wave packet with ultrarelativistic energy, following standard techniques \cite{bransden}.

The $C_j$ and $D_j$ can be written in the general form
\begin{subequations}
\label{C-D-generic}
\be
C_j\lt(q, \k, r, R\rt) & = & \lt(C_j\rt)_{ab} \, \cos^a \lt(\k \, R\rt) \, \sin^b \lt(\k \, R\rt),
\label{C-generic}
\nl \nn
D_j\lt(q, \k, r, R\rt) & = & \lt(D_j\rt)_{ab} \, \cos^a \lt(\k \, R\rt) \, \sin^b \lt(\k \, R\rt),
\label{D-generic}
\ee
\end{subequations}
where $\lt(C_j\rt)_{ab}$ and $\lt(D_j\rt)_{ab}$ are expressed in terms of $q$, the dimensionless parameter $\mu \equiv q/\lt(\k \, R\rt) \ll 1$,
and (\ref{regulator}).
We proceed to list the components of (\ref{C-D-generic}) for the Dirac and Majorana matrix elements, respectively.

\subsection{Dirac Neutrinos}

For the $C_j$ functions, we have the respective components $\lt(C_j\rt)_{ab}$, where
\begin{subequations}
\be
\lt(C_0\rt)_{00} & = &
-2 \sqrt{2 \over \pi^3} \lt\{{6 \, r^2 \over R^2} - \lt[19 \, \mu^3 - 23 \, \ln \lt({r \over R}\rt)\mu^2 + 18 \lt({r \over R} - 1\rt)\mu
+ 6\rt]\rt\} \mu^{-3} \, e^{-q^2/2} \, e^{-\lt[\k(r - R)/q\rt]^2} \, {r \over R}
\nn
& &{} - \sqrt{2 \over \pi^3} \lt\{ {1 \over 48} \lt[3 \, q^2\lt(3 \, q^2 - 4\rt)\mu^3 + 16\lt(7 \, q^2 + 4\rt)\mu^2
- 72\lt(2 \, q^2 - 1\rt)\mu + 144 \, q^2 \rt]e^{-q^2/2} \rt.
\nn
& &{} + \lt. {\pi \over 4} \lt[-15 \, \mu^5 + \lt(5 \, q^2 - 12\rt)\mu^3 + 4\lt(3 \, q^2 + 20\rt)\mu + {48 \over \mu}\rt]e^{-2\krq} \rt\} \, {r \over R} \, ,
\nl
\lt(C_0\rt)_{10} & = & - {1 \over \pi \sqrt{2}}
\lt[q^2\lt(13 \, q^2 - 14\rt)\mu^5 - \lt(3 \, q^4 - 32 \, q^2 - 16\rt)\mu^4 - 2\lt(3 \, q^4 + 12 \, q^2 - 4\rt)\mu^3 \rt.
\nn
& &{} + \lt. \lt(3 \, q^4 + 34 \, q^2 - 16\rt)\mu^2 - 8\lt(3 \, q^2 + 2\rt)\mu + 4\lt(3 \, q^2 + 28\rt) + {48 \over \mu^2}\rt] q^{-1} \, e^{-q^2/4} \, e^{-\krq} \, {r \over R} \, ,
\nl
\lt(C_0\rt)_{01} & = & -{1 \over \pi \sqrt{2}}
\lt[-8\lt(2 \, q^2 - 1\rt)\mu^6 + 3\lt(3 \, q^2 - 4\rt)\mu^5 + \lt(3 \, q^4 + 4 \, q^2 + 60\rt)\mu^4 + \lt(5 \, q^2 - 46\rt)\mu^3 \rt.
\nn
& &{} - \lt. 56 \, \mu^2 + 68 \, \mu - 48 + {48 \over \mu}\rt]e^{-q^2/4} \, e^{-\krq} \, {r \over R} \, ,
\nl
\lt(C_0\rt)_{11} & = & - {1 \over \sqrt{2 \, \pi}}
\lt[\lt(13 \, q^2 - 4\rt)\mu^4 + 8\lt(2 \, q^2 + 1\rt)\mu^2 + 64\rt] q^{-1} \, e^{-2\krq} \, {r \over R} \, ,
\nl
\lt(C_0\rt)_{20} & = & -{1 \over 4} \, \sqrt{2 \over \pi}
\lt[-15 \, \mu^4 + 4\lt(4 \, q^2 - 5\rt)\mu^2 + 64\rt] \mu \, e^{-2\krq} \, {r \over R} \, ,
\ee
\end{subequations}
\begin{subequations}
\be
\lt(C_1\rt)_{00} & = &
\sqrt{2 \over \pi^3} \lt\{ {r^2 \over R^2} - \lt[{1 \over 2} \lt(q^2 - 12 \pi \, q + 8\rt) \ln \lt({r \over R}\rt)\mu^2 + 10 \lt({r \over R} - 1\rt)\mu
+ 1\rt]\rt\} \mu^{-3} \, e^{-q^2/2} \, e^{-\lt[\k(r - R)/q\rt]^2} \, {r \over R}
\nn
& &{} + \sqrt{2 \over \pi^3} \lt\{ {1 \over 96} \lt[3 \, q^2\lt(7 \, q^2 - 8\rt)\mu^3 + 8 \, q\lt(3 \pi \, q^2 + 10 \, q - 4 \pi \rt)\mu^2
- 24 \lt(12 \, q^2 - 5 \pi \, q + 5\rt)\mu \rt. \rt.
\nn
& &{} + \lt.  96 \, q\lt(11 \, q^2 - 6 \pi\rt)\rt]e^{-q^2/2}
\nn
& &{} + \lt. {\pi \over 32} \lt[-25 \, q^2 \, \mu^7 + \lt(3 \, q^4 - 10 \, q^2 + 12\rt) \mu^5 + 8\lt(12 \, q^2 + 1\rt)\mu^3
 -96 \, \mu + {256 \over \mu}\rt]e^{-2\krq} \rt\} \, {r \over R} \, ,
\nl
\lt(C_1\rt)_{10} & = & {\sqrt{2} \over 32 \pi}
\lt[q\lt(q^2 + 10\rt)\lt(3 \, q^2 - 4\rt)\mu^7 + 10 \, q \lt(q^2 + 8\rt)\mu^6 + 136 \, q \lt(q^2 - 2\rt)\mu^5
+ 8\lt(6 \, q^3 + 11 \pi \, q^2 + 21 \, q - 8 \pi\rt)\mu^4 \rt.
\nn
& &{} + \lt. 8 \, q \lt(11 \, q^2 - 8\rt) \mu^3
- 8\lt(5 \, q^3 - 6 \pi \, q^2 + 40 \, q - 8 \pi \rt)\mu^2 - 352 \, q \, \mu - 32 \lt(11 \, q^2 - 6 \pi\rt)\rt]e^{-q^2/4} \, e^{-\krq} \, {r \over R} \, ,
\nn \nl
\lt(C_1\rt)_{01} & = & -{8 \over \pi \sqrt{2}}
\lt[3 \, q^2\lt(3 \, q^2 - 4\rt)\mu^8 + 25 \, q^2 \, \mu^7 - 2 \, q^2 \lt(3 \, q^2 - 20\rt) \mu^6
+ 4\lt(13 \, q^2 + 15 \pi \, q + 12\rt)\mu^5 \rt.
\nn
& &{} - 2 \, q^2 \lt(11 \, q^2 - 104\rt)\mu^4 - 4 \lt(3 \pi \, q^3 + 28 \, q^2 - 6 \pi \, q + 26\rt)\mu^3
- 48 \, q^2 \, \mu^2 - 48 \lt(q^2 + \pi \, q - 3\rt) \mu
\nn
& &{} + \lt. 160 - {160 \over \mu}\rt]e^{-q^2/4} \, e^{-\krq} \, {r \over R} \, ,
\nl
\lt(C_1\rt)_{11} & = & {1 \over 16} \, \sqrt{2 \over \pi}
\lt[\lt(11 \, q^2 + 12\rt)\mu^6 + 24 \, \mu^4 - 16\lt(2 \, q^2 + 3\rt)\mu^2 - 128\rt] q \, e^{-2\krq} \, {r \over R} \, ,
\nl
\lt(C_1\rt)_{20} & = & -{1 \over 32} \, \sqrt{2 \over \pi}
\lt[-25 \, q^2 \, \mu^7 + 2 \lt(3 \, q^4 + 7 \, q^2 - 6\rt) \mu^5 + 120 \, q^2 \, \mu^3 + 64 \, \mu + {512 \over \mu}\rt] e^{-2\krq} \, {r \over R} \, ,
\ee
\end{subequations}
\begin{subequations}
\be
\lt(C_2\rt)_{00} & = &
\, {1 \over \sqrt{2\pi^3}} \lt[{1 \over 4} \lt(6 \pi \, q^2 - q + 2 \pi\rt) \ln \lt({r \over R}\rt)\mu + 3 \pi \, q \lt({r \over R} - 1\rt)\rt] \mu^{-2} \,
e^{-q^2/2} \, e^{-\lt[\k(r - R)/q\rt]^2} \, {r \over R}
\nn
& &{} + \sqrt{2 \over \pi^3} \lt\{ {1 \over 1536} \lt[9 \, q^3\lt(9 \, q^2 - 28\rt)\mu^3 + 8 \, q\lt(6 \pi \, q^3 + 33 \, q^2 - 28 \pi \, q - 40 \rt)\mu^2
- 48 \lt(33 \, q^3 + \pi \, q^2 - 24 \, q + 14 \pi\rt)\mu \rt. \rt.
\nn
& &{} + \lt.  96 \, q\lt(27 \, q^2 - 24 \pi \, q + 2\rt) + 1920 \pi\rt]q \, e^{-q^2/2}
\nn
& &{} + \lt.  {\pi \over 128} \lt[\lt(5 \, q^2 + 30\rt) \mu^6 + 56 \, \mu^4 + 4\lt(3 \, q^2 - 10 \rt)\mu^2 - 256 \rt] q^2 \, \mu \, e^{-2\krq} \rt\} \, {r \over R} \, ,
\nl
\lt(C_2\rt)_{10} & = & {\sqrt{2} \over 128 \pi}
\lt[7 \, q^2\lt(3 \, q^2 - 4\rt) \mu^9 + 60 \, q^2 \, \mu^8 + q^2 \lt(q^2 - 178\rt)\mu^7
- 2\lt(3 \, q^4 - 22 \, q^2 + 56 \rt)\mu^6 + 2 \lt(51 \, q^4 - 184 \, q^2 + 60\rt)\mu^5 \rt.
\nn
& &{} + 4 \lt(11 \pi \, q^3 - 9 \, q^2 - 30 \, \pi \, q + 6\rt)\mu^4 + 2\lt(33 \, q^4 + 110 \, q^2 + 40\rt)\mu^3
+ 8\lt(6 \pi \, q^3 - 26 \, q^2 + 10 \pi \, q + 4\rt)\mu^2
\nn
& &{} + \lt. 8 \lt(33 \, q^2 - 40\rt)\mu + 192 \pi \, q \rt] q \, e^{-q^2/4} \, e^{-\krq} \, {r \over R} \, ,
\nl
\lt(C_2\rt)_{01} & = & {\sqrt{2} \over 128 \pi}
\lt[q\lt(3 \, q^4 + 44 \, q^2 - 24\rt) \mu^8 + 40 \, q \lt(q^2 + 3\rt) \mu^7 + q \lt(3 \, q^4 - 162 \, q^2 - 136\rt)\mu^6 \rt.
\nn
& &{} + 2\lt(5 \, q^3 - 30 \pi \, q^2 + 22 \, q + 12 \pi\rt)\mu^5 + q^3 \lt(27 \, q^2 - 346\rt)\mu^4
\nn
& &{} + 4 \lt(3 \pi \, q^4 + 6 \, q^3 - 22 \pi \, q^2 + 16 \, q + 4 \pi\rt)\mu^3 + 4 \, q\lt(21 \, q^2 + 68\rt)\mu^2
+ 16\lt(3 \pi \, q^2 - 20 \, q - 4 \pi\rt)\mu
\nn
& &{} - \lt. 96 \, q - {192 \pi \over \mu}\rt]q \, \, e^{-q^2/4} \, e^{-\krq} \, {r \over R} \, ,
\nl
\lt(C_2\rt)_{11} & = & -{1 \over 64} \, \sqrt{2 \over \pi}
\lt[- 15 \, q^2 \, \mu^4 + 16\lt(3 \, q^4 - 6 \, q^2 -4\rt)\mu^2 + 2 \lt(9 \, q^2 + 2\rt)\rt] q \, \mu^4 \, e^{-2\krq} \, {r \over R} \, ,
\nl
\lt(C_2\rt)_{20} & = & -{1 \over 128} \, \sqrt{2 \over \pi}
\lt[\lt(25 \, q^2 + 10\rt) \mu^2 + 2 \lt(3 \, q^2 - 2\rt)\rt] q^2 \, \mu^5 \, e^{-2\krq} \, {r \over R} \, .
\ee
\end{subequations}
For the $D_j$ functions, the components $\lt(D_j\rt)_{ab}$ are
\begin{subequations}
\be
\lt(D_0\rt)_{00} & = &
-{16 \over 45} \, \sqrt{2 \over \pi^3} \lt\{ 3 \lt[\lt(3 \, q^2 + 2\rt) \ln \lt({r \over R}\rt)\mu + 6 \lt({r \over R} - 1\rt)\rt]
+ \lt[\mu^2 + 6 \, \mu - 3 \lt(9 \, q^2 + 7\rt)\rt]\mu^2\rt\}q^{-1} \, \mu^{-3}
\nn
& &{} \times  e^{-q^2/2} \, e^{-\lt[\k(r - R)/q\rt]^2} \, {r^2 \over R^2}
\nn
& &{} - \sqrt{2 \over \pi^3} \lt\{ {1 \over 450}\lt[10 \, q^3 \, \mu^4 - 12 \, q \lt(3 \, q^2 + 2\rt)\mu^3
- 15 \lt(9 \, q^4 + 14 \, q^2 + 8\rt) q^{-1} \, \mu^2 + 20 \, q \lt(3 \, q^2 + 14 \rt)\mu + 1620 \, q \rt]e^{-q^2/2} \rt.
\nn
& &{} + \lt. {2\pi \over 15} \lt[-5 \, \mu^6 - \lt(7 \, q^2 + 17\rt) \mu^4 + \lt(3 \, q^4 + q^2 + 24\rt)\mu^2 + 4 \lt(3 \, q^2 + 4\rt)
 - {48 \over \mu^2}\rt] q^{-1} \, e^{-2\krq} \rt\} \, {r^2 \over R^2} \, ,
\nl
\lt(D_0\rt)_{10} & = & -{2\sqrt{2} \over 15 \pi}
\lt[12 \, q^2 \, \mu^6 - 5 \lt(3 \, q^2 + 2\rt)\mu^5 + \lt(3 \, q^4 - 60 \, q^2 - 32 \rt)\mu^4  + 15 \lt(q^2 + 2\rt) \mu^3 \rt.
\nn
& &{} - \lt.
12 \lt(q^2 - 6\rt)\mu^2 + 12 \lt(2 \, q^2 + 3\rt) \mu - 96 + {96 \over \mu}\rt]e^{-q^2/4} \, e^{-\krq} \, {r^2 \over R^2} \, ,
\nl
\lt(D_0\rt)_{01} & = & {2\sqrt{2} \over 15 \pi}
\lt[12 \, q^2 \, \mu^7 - 5 \lt(3 \, q^2 + 2\rt)\mu^6 + 2\lt(5 \, q^4 - 6 \, q^2 - 16 \rt)\mu^5  + \lt(3 \, q^4 + 17 \, q^2 + 70\rt) \mu^4
+ 4 \lt(3 \, q^4 - 21 \, q^2 - 4\rt)\mu^3 \rt.
\nn
& &{} - \lt. \lt(3 \, q^4 - 22 \, q^2 + 4\rt) \mu^2 + 8\lt(3 \, q^2 + 4\rt) \mu - 4 \lt(3 \, q^2 + 20\rt)
- {96 \over \mu} + {48 \over \mu^2}\rt] q^{-1} \, e^{-q^2/4} \, e^{-\krq} \, {r^2 \over R^2} \, ,
\nl
\lt(D_0\rt)_{11} & = & -{4 \over 15} \, \sqrt{2 \over \pi}
\lt[5 \, \mu^5 + \lt(5 \, q^2 + 1\rt)\mu^3 + \lt(12 \, q^2 + 8\rt)\mu + {48 \over \mu}\rt] e^{-2\krq} \, {r^2 \over R^2} \, ,
\nl
\lt(D_0\rt)_{20} & = & {2 \over 15} \, \sqrt{2 \over \pi}
\lt[-5 \, \mu^6 - \lt(2 \, q^2 + 17\rt) \mu^4 + 2\lt(3 \, q^4 - 7 \, q^2 + 12\rt) \mu^2 + 8\lt(3 \, q^2 + 2\rt) - {96 \over \mu^2}\rt] q^{-1} \, e^{-2\krq} \, {r^2 \over R^2} \, ,
\ee
\end{subequations}
\begin{subequations}
\be
\lt(D_1\rt)_{00} & = &
-{4 \over 15} \, \sqrt{2 \over \pi^3} \lt\{ 6 \, q \, \ln \lt({r \over R}\rt) - \lt[5 \pi \, \mu + 6 \lt(q + \pi\rt)\rt]\mu\rt\}
\mu^{-2}e^{-q^2/2} \, e^{-\lt[\k(r - R)/q\rt]^2} \, {r^2 \over R^2}
\nn
& &{} + \sqrt{2 \over \pi^3} \lt\{ {1 \over 450}\lt[20 \, q^3 \, \mu^4 - 12 \, q \lt(3 \, q^2 - 2 \pi \, q + 2\rt)\mu^3
- 15 \lt(6 \, q^3 + 3 \pi \, q^2 + 8 \, q + 2 \, \pi\rt) \mu^2 \rt. \rt.
\nn
& &{} - \lt. 10 \lt(3 \, q^3 + 12 \pi \, q^2 + 40 \, q - 16 \, \pi\rt)\mu + 270 \, q \rt]e^{-q^2/2}
\nn
& &{} - \lt. {\pi \over 120} \lt[28 \, \mu^8 + \lt(11 \, q^2 + 46\rt) \mu^6 - 2\lt(3 \, q^2 + 86\rt) \mu^4 + 16 \, \mu^2
+ 192\rt] q \, e^{-2\krq} \rt\} \, {r^2 \over R^2} \, ,
\nl
\lt(D_1\rt)_{10} & = & {\sqrt{2} \over 60 \pi}
\lt[2 \, q^2\lt(q^2 + 14\rt)\mu^8 - \lt(3 \, q^4 + 38 \, q^2 + 24\rt)\mu^7 - 2\lt(3 \, q^4 - 20 \, q^2 + 40\rt)\mu^6
-8 \lt(2 \, q^2 - 10 \pi \, q - 17\rt)\mu^5 \rt.
\nn
& &{} + 2\lt(3 \, q^4 - 86 \, q^2 + 4\rt)\mu^4  + 12 \lt(\pi \, q^3 - 2 \, q^2 - 6 \pi \, q + 8\rt) \mu^3
+ 16 \lt(9 \, q^2 + 4\rt)\mu^2 - 4 \lt(18 \, q^2 - 12 \pi \, q + 40\rt)\mu
\nn
& &{} - \lt. 96 - {96 \over \mu}\rt]e^{-q^2/4} \, e^{-\krq} \, {r^2 \over R^2} \, ,
\nl
\lt(D_1\rt)_{01} & = & -{\sqrt{2} \over 60 \pi}
\lt[16 \, q^3 \, \mu^9 - 7 \, q\lt(3 \, q^2 + 2\rt)\mu^8 - 4 \, q\lt(11 \, q^2 - 12\rt)\mu^7 + 4\lt(3 \, q^3 + 27 \, q + 20 \, \pi\rt)\mu^6
+ 8 \, q \lt(3 \, q^2 + 34\rt)\mu^5 \rt.
\nn
& &{} + 4\lt(3 \, q^4 + 5 \pi \, q^2 - 36 \, q + 34 \pi\rt)\mu^4 + 4 \, q\lt(15 \, q^2 -  32\rt) \mu^3
+ 4 \lt(3 \, q^3 + 6 \, q^2 - 32 \, q - 40 \pi\rt)\mu^2 - 48 \, q \, \mu
\nn
& &{} + \lt. 12 \lt(12 \, q + 8 \pi\rt)\rt]e^{-q^2/4} \, e^{-\krq} \, {r^2 \over R^2} \, ,
\nl
\lt(D_1\rt)_{11} & = & {1 \over 60} \, \sqrt{2 \over \pi}
\lt[8\lt(2 \, q^2 + 3\rt)\mu^6 + \lt(3 \, q^4 - 14 \, q^2 + 72\rt)\mu^4 - 4\lt(q^2 + 14\rt)\mu^2 - 64\rt]\mu \, e^{-2\krq} \, {r^2 \over R^2} \, ,
\nl
\lt(D_1\rt)_{20} & = & {1 \over 120} \, \sqrt{2 \over \pi}
\lt[28 \, \mu^4 + \lt(7 \, q^2 - 2\rt) \mu^2 - 160\rt] q \, \mu^4 \, e^{-2\krq} \, {r^2 \over R^2} \, ,
\ee
\end{subequations}
\begin{subequations}
\be
\lt(D_2\rt)_{00} & = &
{1 \over 20} \, \sqrt{2 \over \pi^3} \lt\{ 4 \, q \lt(2 \pi \, q - 1\rt) \ln \lt({r \over R}\rt) + \lt[\mu - 4 \lt(q^2 - \pi \, q + 1\rt)\rt]\mu\rt\}
\mu^{-2} \, e^{-q^2/2} \, e^{-\lt[\k(r - R)/q\rt]^2} \, {r^2 \over R^2}
\nn
& &{} - \sqrt{2 \over \pi^3} \lt\{ {1 \over 3600}\lt[-5 \, q^2 \lt(9 \, q^2 - 16\rt) \mu^4
+ 3 \, q \lt(27 \, q^3 - 8 \pi \, q^2 + 18 \, q + 64 \pi\rt)\mu^3 \rt. \rt.
\nn
& &{} + 5 \lt(36 \, q^4 + 9 \pi \, q^3 + 36 \, q^2 + 6 \pi \, q + 24 \, \pi^2\rt) \mu^2
+ 20 \, q \lt(3 \pi \, q^2 - 39 \, q + 2 \pi\rt)\mu
\nn
& &{} - \lt. 150 \, q \lt(3 \, q + 4 \pi\rt) \rt] e^{-q^2/2}
\nn
& &{} - {\pi \over 480} \lt[9 \, q^2 \, \mu^8 - 4 \lt(q^2 - 7\rt) \mu^6 + \lt(3 \, q^4 + 21 \, q^2 + 94\rt) \mu^4
+ 2 \lt(3 \, q^4 - 9 \, q^2 - 2\rt) \mu^2 \rt.
\nn
& &{} + \lt. \lt. 8\lt(3 \, q^2 - 14\rt)\rt] \mu^2 \, e^{-2\krq} \rt\} q \, {r^2 \over R^2} \, ,
\nl
\lt(D_2\rt)_{10} & = & {\sqrt{2} \over 480 \pi}
\lt[36 \, q^3 \, \mu^{10} - 16 \, q \lt(3 \, q^2 + 2\rt)\mu^9 - 84 \, q\lt(q^2 + 4\rt)\mu^8 - 2\lt(9 \, q^3 + 8 \, q^2 - 114 \, q + 96 \pi\rt)\mu^7 \rt.
\nn
& &{} + q \lt(3 \, q^4 + 212 \, q^2 - 108\rt)\mu^6 + 8 \lt(3 \, q^3 + 10 \pi \, q^2 + 7 \, q - 24 \pi\rt)\mu^5
+ 3 \, q \lt(9 \, q^2 - 10\rt)\lt(q^2 - 8\rt)\mu^4
\nn
& &{} + 4 \lt(3 \pi \, q^4 + 6 \, q^3 - 6 \pi \, q^2 - 52 \, q + 56 \pi\rt) \mu^3
+ 12 \, q\lt(11 \, q^2 - 12\rt)\mu^2 + 96 \lt(\pi \, q^2 - q  - \pi\rt)\mu
\nn
& &{} + \lt. 96 \, q + {192 \pi \over \mu}\rt]q \, e^{-q^2/4} \, e^{-\krq} \, {r^2 \over R^2} \, ,
\nl
\lt(D_2\rt)_{01} & = & -{\sqrt{2} \over 240 \pi}
\lt[-2 \, q^2 \lt(q^2 + 40\rt) \mu^9 + \lt(3 \, q^4 + 23 \, q^2 - 56 \pi \, q + 14\rt)\mu^8 + \lt(13 \, q^4 + 182 \, q^2 + 48\rt)\mu^7 \rt.
\nn
& &{} + 3\lt(q^4 + 2 \, q^2 + 24 \pi \, q - 28\rt)\mu^6 + 36 \, q^2 \lt(q^2 + 5\rt)\mu^5
+ 2 \lt(5 \pi \, q^3 + 2 \, q^2 + 50 \pi \, q - 80\rt)\mu^4
\nn
& &{} + \lt. \lt(21 \, q^4 - 186 \, q^2 - 8\rt) \mu^3 - 8\lt(3 \, q^2 + 7 \pi \, q - 10 \rt)\mu^2 + 84 \, q^2 \, \mu + 96
\rt] q \, e^{-q^2/4} \, e^{-\krq} \, {r^2 \over R^2} \, ,
\nl
\lt(D_2\rt)_{11} & = & -{1 \over 240} \, \sqrt{2 \over \pi}
\lt[q^2 \, \mu^6 - \lt(5 \, q^2 + 18\rt)\mu^4 - \lt(11 \, q^2 + 6\rt)\mu^2 + 48\rt]q^2 \, \mu^3 \, e^{-2\krq} \, {r^2 \over R^2} \, ,
\nl
\lt(D_2\rt)_{20} & = & -{1 \over 480} \, \sqrt{2 \over \pi}
\lt[9 \, q^2 \, \mu^8 + 24 \, q^2 \, \mu^6 + \lt(6 \, q^4 - 29 \, q^2 - 30\rt) \mu^4
+ 4\lt(3 \, q^4 + 2 \, q^2 + 12\rt) \mu^2 \rt.
\nn
& &{} + \lt. 48 \lt(q^2 - 2\rt)\rt] q \, \mu^2 \, e^{-2\krq} \, {r^2 \over R^2} \, .
\ee
\end{subequations}

\subsection{Majorana Neutrinos}

The dimensionless functions $C_{j \Cy}$ and $D_{j \Cx}$ which define the Majorana matrix element can be expressed as shown in (\ref{C-D-generic}).
Therefore, it follows that the components for $C_{j \Cy}$ are
\begin{subequations}
\be
\lt(C_{0 \Cy}\rt)_{00} & = &
{1 \over 3 \pi^2} \lt\{ {4 \, r^2 \over R^2} - \lt[2 \, \mu^3 - 10 \, \ln \lt({r \over R}\rt)\mu^2 + 12 \lt({r \over R} - 1\rt)\mu
+ 4\rt]\rt\} \mu^{-3} \, e^{-q^2/2} \, e^{-\lt[\k(r - R)/q\rt]^2} \, {r \over R}
\nn
& &{} + {1 \over 144 \pi^2} \lt\{ \lt[3 \, q^2\lt(3 \, q^2 + 4\rt)\mu^3 + 16\lt(q^2 + 4\rt)\mu^2
- 24\lt(2 \, q^2 + 7\rt)\mu + 48 \, q^2 \rt]e^{-q^2/2} \rt.
\nn
& &{} - \lt. 12 \pi \lt[-3 \, \mu^5 + \lt(q^2 + 4\rt)\mu^3 - 4\lt(q^2 - 4\rt)\mu - {16 \over \mu}\rt]e^{-2\krq} \rt\} \, {r \over R} \, ,
\nl
\lt(C_{0 \Cy}\rt)_{10} & = & {1 \over 6 \sqrt{\pi^3}}
\lt[q^2\lt(3 \, q^2 + 14\rt)\mu^5 - \lt(q^4 - 16\rt)\mu^4 - 2\lt(q^2 + 2\rt)^2\mu^3 \rt.
\nn
& &{} + \lt. \lt(q^2 + 8\rt)\lt(q^2 - 2\rt)\mu^2 - 8\lt(q^2 - 2\rt)\mu + 4\lt(q^2 + 4\rt) + {16 \over \mu^2}\rt] q^{-1} \, e^{-q^2/4} \, e^{-\krq} \, {r \over R} \, ,
\nl
\lt(C_{0 \Cy}\rt)_{01} & = & {1 \over 6 \sqrt{\pi^3}}
\lt[-8 \, \mu^6 + 3\lt(q^2 + 4\rt)\mu^5 + \lt(q^2 + 2\rt)^2\mu^4 - \lt(q^2 + 10\rt)\mu^3 \rt.
\nn
& &{} - \lt. 8 \, \mu^2 + 12 \, \mu - 16 + {16 \over \mu}\rt]e^{-q^2/4} \, e^{-\krq} \, {r \over R} \, ,
\nl
\lt(C_{0 \Cy}\rt)_{11} & = & - {1 \over 6\pi} \lt[\lt(q^2 - 4\rt)\mu^2 + 8\rt] q^{-1} \, \mu^2 \, e^{-2\krq} \, {r \over R} \, ,
\nl
\lt(C_{0 \Cy}\rt)_{20} & = & -{1 \over 12 \pi} \lt(3 \, \mu^2 + 4\rt) \mu^3 \, e^{-2\krq} \, {r \over R} \, ,
\ee
\end{subequations}
\begin{subequations}
\be
\lt(C_{1 \Cy}\rt)_{00} & = &
{1 \over 6 \pi^2}  \lt\{ {2r^2 \over R^2} - \lt[\lt(q^2 + 4 \pi \, q - 8\rt) \ln \lt({r \over R}\rt)\mu^2 + 4 \lt({r \over R} - 1\rt)\mu
+ 2\rt]\rt\} \mu^{-3} \, e^{-q^2/2} \, e^{-\lt[\k(r - R)/q\rt]^2} \, {r \over R}
\nn
& &{} - {1 \over 288 \pi^2} \lt\{ \lt[3 \, q^2\lt(q^2 + 8\rt)\mu^3 + 8 \, q\lt(\pi \, q^2 - 2 \, q + 4 \pi \rt)\mu^2
- 24 \lt(\pi \, q - 3\rt)\mu \rt. \rt.
\nn
& &{} + \lt.  96 \, q \lt(q - 2 \pi\rt)\rt]e^{-q^2/2}
\nn
& &{} + \lt. 3 \pi \lt[5 \, q^2 \, \mu^6 + \lt(q^4 - 14 \, q^2 - 12\rt) \mu^4 + 8\lt(2 \, q^2 + 3\rt)\mu^2 - 32\rt] \mu \, e^{-2\krq} \rt\} \, {r \over R} \, ,
\nl
\lt(C_{1 \Cy}\rt)_{10} & = & -{1 \over 96 \sqrt{\pi^3}}
\lt[q\lt(q^2 + 10\rt)\lt(q^2 + 4\rt)\mu^7 - 2 \, q \lt(q^2 + 8\rt)\mu^6 - 8 \, q \lt(q^2 - 10\rt)\mu^5
+ 8\lt(2 \, q^3 + \pi \, q^2 + 7 \, q + 8 \pi\rt)\mu^4 \rt.
\nn
& &{} + \lt. 8 \, q \lt(q^2 - 8\rt) \mu^3
+ 8\lt(q^3 + 2 \pi \, q^2 - 8 \, q - 8 \pi \rt)\mu^2 + 32 \, q \, \mu - 32 \lt(q - 2 \pi\rt)\rt]e^{-q^2/4} \, e^{-\krq} \, {r \over R} \, ,
\nn \nl
\lt(C_{1 \Cy}\rt)_{01} & = & {1 \over 48 \sqrt{\pi^3}}
\lt[3 \, q^2\lt(q^2 + 4\rt)\mu^8 - 5 \, q^2 \, \mu^7 - 2 \, q^2 \lt(q^2 + 20\rt) \mu^6
+ 4\lt(7 \, q^2 - 3 \pi \, q + 12\rt)\mu^5 \rt.
\nn
& &{} - 2 \, q^2 \lt(q^2 - 8\rt)\mu^4 - 4 \lt(\pi \, q^3 + 4 \, q^2 - 2 \pi \, q + 14\rt)\mu^3
- 16 \, q^2 \, \mu^2 - 16 \lt(q^2 + \pi \, q - 3\rt) \mu
\nn
& &{} - \lt. 32 + {32 \over \mu}\rt]e^{-q^2/4} \, e^{-\krq} \, {r \over R} \, ,
\nl
\lt(C_{1 \Cy}\rt)_{11} & = & -{1 \over 48 \pi}
\lt[\lt(q^2 + 4\rt)\mu^4 - 8 \, \mu^2 - 16\rt] q \, \mu^2 \, e^{-2\krq} \, {r \over R} \, ,
\nl
\lt(C_{1 \Cy}\rt)_{20} & = & {1 \over 96 \, \pi}
\lt[5 \, q^2 \, \mu^6 + 2 \lt(q^4 - 3 \, q^2 + 6\rt) \mu^4 + 8 \, q^2 \, \mu^2 - 64\rt] e^{-2\krq} \, {r \over R} \, ,
\ee
\end{subequations}
\begin{subequations}
\be
\lt(C_{2 \Cy}\rt)_{00} & = &
{1 \over 24 \pi^2} \lt[\lt(2 \pi \, q^2 + q - 2 \pi\rt) \ln \lt({r \over R}\rt)\mu + 4 \pi \, q \lt({r \over R} - 1\rt)\rt] \mu^{-2} \,
e^{-q^2/2} \, e^{-\lt[\k(r - R)/q\rt]^2} \, {r \over R}
\nn
& &{} + {1 \over 4608 \pi^2} \lt\{  \lt[3 \, q^3\lt(9 \, q^2 + 20\rt)\mu^3 + 8 \, q\lt(2 \pi \, q^3 - 13 \, q^2 - 4 \pi \, q + 8 \rt)\mu^2
- 48 \lt(11 \, q^3 + 3 \, \pi \, q^2 - 8 \, q + 2 \pi\rt)\mu \rt. \rt.
\nn
& &{} + \lt.  96 \, q\lt(9 \, q^2 - 8 \pi \, q - 2\rt) + 1152 \pi\rt]q \, e^{-q^2/2}
\nn
& &{} - \lt.  12 \pi \lt[\lt(q^2 + 10\rt) \mu^6 + 4 \lt(q^2 - 6\rt) \mu^4 - 4\lt(q^2 - 2 \rt)\mu^2 +64 \rt] q^2 \, \mu \, e^{-2\krq} \rt\} \, {r \over R} \, ,
\nl
\lt(C_{2 \Cy}\rt)_{10} & = & {1 \over 384 \sqrt{\pi^3}}
\lt[7 \, q^2\lt(q^2 + 4\rt) \mu^9 - 12 \, q^2 \, \mu^8 - q^2 \lt(q^2 + 62\rt)\mu^7
- 2\lt(q^4 - 6 \, q^2 + 8 \rt)\mu^6 + 2 \lt(5 \, q^4 + 24 \, q^2 + 36\rt)\mu^5 \rt.
\nn
& &{} + 4 \lt(\pi \, q^3 + q^2 - 2 \, \pi \, q + 10\rt)\mu^4 + 2\lt(11 \, q^4 - 22 \, q^2 - 8\rt)\mu^3
+ 8\lt(2 \pi \, q^3 - 6 \, q^2 - 2 \pi \, q - 4\rt)\mu^2
\nn
& &{} + \lt. 8 \lt(11 \, q^2 - 8\rt)\mu + 64 \pi \, q \rt] q \, e^{-q^2/4} \, e^{-\krq} \, {r \over R} \, ,
\nl
\lt(C_{2 \Cy}\rt)_{01} & = & {1 \over 384 \sqrt{\pi^3}}
\lt[q\lt(q^4 + 28 \, q^2 + 24\rt) \mu^8 + 8 \, q^3 \, \mu^7 + q \lt(q^4 + 42 \, q^2 - 120\rt)\mu^6 \rt.
\nn
& &{} - 2\lt(q^3 - 6 \pi \, q^2 - 10 \, q + 12 \pi\rt)\mu^5 + q \lt(9 \, q^4 - 110 \, q^2 + 64\rt)\mu^4
\nn
& &{} + 4 \lt(\pi \, q^4 + 2 \, q^3 - 10 \pi \, q^2 + 8 \, q + 12 \pi\rt)\mu^3 + 4 \, q\lt(7 \, q^2 + 12\rt)\mu^2
+ 16\lt(\pi \, q^2 - 4 \, q - 4 \pi\rt)\mu
\nn
& &{} - \lt. 32 \, q - {64 \pi \over \mu} \rt]q \, e^{-q^2/4} \, e^{-\krq} \, {r \over R} \, ,
\nl
\lt(C_{2 \Cy}\rt)_{11} & = & -{1 \over 192 \pi}
\lt[3 \, q^2 \, \mu^4 + \lt(q^4 + 2 \, q^2 + 4\rt)\mu^2 - 2 \lt(q^2 + 2\rt)\rt] q \, \mu^4 \, e^{-2\krq} \, {r \over R} \, ,
\nl
\lt(C_{2 \Cy}\rt)_{20} & = & -{1 \over 384 \pi}
\lt[\lt(3 \, q^2 + 10\rt) \mu^2 - 6 \lt(q^2 + 2\rt)\rt] q^2 \, \mu^5 \, e^{-2\krq} \, {r \over R} \, .
\ee
\end{subequations}
The components for $D_{j \Cx}$ are
\begin{subequations}
\be
\lt(D_{0 \Cx}\rt)_{00} & = &
{4 \over 45} \, \sqrt{2 \over \pi^3} \lt\{ 12 \, \ln \lt({r \over R}\rt) - \lt[7 \, \mu^2 - 15 \, \mu + 30 \rt]\mu\rt\} \mu^{-3} \,
e^{-q^2/2} \, e^{-\lt[\k(r - R)/q\rt]^2} \, {r^3 \over R^3}
\nn
& &{} + \sqrt{2 \over \pi^3} \lt\{ {1 \over 900}\lt[5 \, q^2 \lt(q^2 - 6\rt) \mu^3 + 12 \lt(3 \, q^2 - 10\rt)\mu^2
- 120 \lt(q^2 - 3\rt)\mu + 200 \, q^2 \rt]e^{-q^2/2} \rt.
\nn
& &{} + \lt. {2\pi \over 15} \lt[-5 \, \mu^5 + \lt(2 \, q^2 - 5\rt) \mu^3 + \lt(2 \, q^2 + 4\rt)\mu + {8 \over \mu}\rt] e^{-2\krq} \rt\} \, {r^3 \over R^3} \, ,
\nl
\lt(D_{0 \Cx}\rt)_{10} & = & {\sqrt{2} \over 30 \pi}
\lt[3 \, q^2 \lt(5 \, q^2 - 22\rt)\mu^5 + 2 \lt(q^4 + 22 \, q^2 - 40 \rt)\mu^4  + 8 \lt(q^4 - 13 \, q^2 + 9\rt) \mu^3 \rt.
\nn
& &{} - \lt.
2 \lt(q^4 - 32 \, q^2 + 4\rt)\mu^2 + 16 \lt(q^2 - 1\rt) \mu + 32 - {64 \over \mu} + {32 \over \mu^2}\rt]e^{-q^2/4} \, e^{-\krq} \, {r^3 \over R^3} \, ,
\nl
\lt(D_{0 \Cx}\rt)_{01} & = & {\sqrt{2} \over 15 \pi}
\lt[- 3 \lt(5 \, q^2 - 6\rt)\mu^6 - 5\lt(q^2 + 10\rt)\mu^5  + \lt(q^4 - 25 \, q^2 + 58\rt) \mu^4
+ 4 \lt(2 \, q^2 - 11\rt)\mu^3 \rt.
\nn
& &{} - \lt. 4\lt(q^2 - 7\rt) \mu^2 + 8\lt(q^2 + 1\rt) \mu - 32 + {32 \over \mu}\rt]  e^{-q^2/4} \, e^{-\krq} \, {r^3 \over R^3} \, ,
\nl
\lt(D_{0 \Cx}\rt)_{11} & = & -{2 \over 15} \, \sqrt{2 \over \pi}
\lt[-\lt(7 \, q^2 - 6\rt)\mu^4 + q^2 \lt(q^2 - 9\rt)\mu^2 - 12 - {16 \over \mu^2}\rt] q^{-1} \, e^{-2\krq} \, {r^3 \over R^3} \, ,
\nl
\lt(D_{0 \Cx}\rt)_{20} & = & -{2 \over 15} \, \sqrt{2 \over \pi}
\lt[-5 \, \mu^5 + \lt(4 \, q^2 - 11\rt) \mu^3 + 4\lt(q^2 + 1\rt) \mu + {16 \over \mu}\rt] e^{-2\krq} \, {r^3 \over R^3} \, ,
\ee
\end{subequations}
\begin{subequations}
\be
\lt(D_{1 \Cx}\rt)_{00} & = &
-{1 \over 30} \, \sqrt{2 \over \pi^3} \lt\{ 24 \, \ln \lt({r \over R}\rt) - \lt[4 \, \mu^2 + \lt(q^2 - 4 \pi \, q - 22\rt) \mu + 40 \rt]\mu\rt\} \mu^{-3} \,
e^{-q^2/2} \, e^{-\lt[\k(r - R)/q\rt]^2} \, {r^3 \over R^3}
\nn
& &{} - \sqrt{2 \over \pi^3} \lt\{ {1 \over 1800}\lt[5 \, q^2 \lt(5 \, q^2 - 12\rt) \mu^3 + 12 \, q\lt(\pi \, q^2 + 11 \, q - 6\pi\rt)\mu^2
- 60 \lt(6\, q^2 - 2 \pi \, q + 1\rt)\mu \rt. \rt.
\nn
& &{} + \lt. 80 \, q \lt(5 \, q - 2\pi\rt) \rt]e^{-q^2/2}
\nn
& &{} + \lt. {\pi \over 240} \lt[-28 \, q^2 \, \mu^7 + q^2 \lt(q^2 - 2\rt)\mu^5 + 48\lt(q^2 - 2\rt) \mu^3 + 32 \lt(2 \, q^2 + 3\rt)\mu
+ {128 \over \mu}\rt] e^{-2\krq} \rt\} \, {r^3 \over R^3} \, ,
\nn
\nl
\lt(D_{1 \Cx}\rt)_{10} & = & -{\sqrt{2} \over 240 \pi}
\lt[q \lt(q^2 - 6\rt)\lt(q^2 + 14\rt)\mu^7 + 8 \, q\lt(q^2 + 12\rt)\mu^6 + 4 \, q \lt(53 \, q^2 - 104\rt) \mu^5 \rt.
\nn
& &{} + 16 \lt(4 \pi \, q^2 + 15 \, q - 12 \pi\rt)\mu^4 + 8 \, q\lt(q^2 - 56\rt) \mu^3 + -8\lt(q^3 -2 \pi \, q^2 - 32 \, q - 4\pi\rt) \mu^2
\nn
& &{} + \lt. 224 \, q \, \mu - 96 \, q + 64 \pi \rt]e^{-q^2/4} \, e^{-\krq} \, {r^3 \over R^3} \, ,
\nl
\lt(D_{1 \Cx}\rt)_{01} & = & {\sqrt{2} \over 60 \pi}
\lt[2 \, q^2 \lt(q^2 - 6\rt)\mu^8 + 14 \, q^2 \, \mu^7 - q^2 \lt(q^2 - 60\rt)\mu^6 - 4\lt(2 \, q^2 - 10 \pi \, q  + 5\rt)\mu^5
- 3 \lt(3 \, q^4 - 28 \, q^2 + 16\rt) \mu^4 \rt.
\nn
& &{} - 2 \lt(\pi \, q^3 + 12 \, q^2 - 10 \pi \, q - 58\rt)\mu^3 + 8\lt(q^2 - 7\rt) \mu^2 - 4\lt(7 \, q^2 + 2 \pi \, q + 8\rt) \mu
\nn
& &{} + \lt. 80 - {80 \over \mu}\rt]  e^{-q^2/4} \, e^{-\krq} \, {r^3 \over R^3} \, ,
\nl
\lt(D_{1 \Cx}\rt)_{11} & = & -{1 \over 120} \, \sqrt{2 \over \pi}
\lt[\lt(7 \, q^2 + 6\rt)\mu^6 + 72 \, \mu^4 - 16 \lt(q^2 - 4\rt)\mu^2 - 64\rt] q \, e^{-2\krq} \, {r^3 \over R^3} \, ,
\nl
\lt(D_{1 \Cx}\rt)_{20} & = & {1 \over 120} \, \sqrt{2 \over \pi}
\lt[-14 \, q^2 \, \mu^7 +  \lt(q^4 + 5 \, q^2 - 30\rt) \mu^5 + 12\lt(5 \, q^2 - 6\rt) \mu^3 + 32\lt(q^2 + 3\rt) \mu + {128 \over \mu}\rt] e^{-2\krq} \, {r^3 \over R^3} \, ,
\nn
\ee
\end{subequations}
\begin{subequations}
\be
\lt(D_{2 \Cx}\rt)_{00} & = &
-{1 \over 360} \, \sqrt{2 \over \pi^3} \lt[40 \pi \, \mu^2  + 3 \lt(2 \pi \, q^2 - 5 \, q - 24 \pi\rt)\mu - 24\rt]
q \, \mu^{-2} \, e^{-q^2/2} \, e^{-\lt[\k(r - R)/q\rt]^2} \, {r^3 \over R^3}
\nn
& &{} - \sqrt{2 \over \pi^3} \lt\{ {1 \over 28800}\lt[5 \, q^3 \lt(9 \, q^2 - 146\rt) \mu^3
+ 24 \, q \lt(\pi \, q^3 + 20 \, q^2 - 26 \pi \, q - 18 \rt)\mu^2 \rt. \rt.
\nn
& &{} - \lt. 20 \lt(21 \, q^3 - 18 \pi \, q^2 - 138 \, q + 12 \pi \rt) \mu
+ 80 \, q \lt(9 \, q^2 - 44\rt) \rt] q \, e^{-q^2/2}
\nn
& &{} + \lt. {\pi \over 480} \lt[2\lt(q^2 + 14\rt) \mu^6 + \lt(q^2 + 68\rt) \mu^4 - 72 \, \mu^2 - 64\rt] q^2 \, \mu \, e^{-2\krq} \rt\} \, {r^3 \over R^3} \, ,
\nl
\lt(D_{2 \Cx}\rt)_{10} & = & -{\sqrt{2} \over 960 \pi}
\lt[9 \, q^2 \lt(q^2 - 6\rt)\mu^9 + 64 \, q^2 \, \mu^8 - 2 \, q^2\lt(3 \, q^2 + 110 \rt)\mu^7 \rt.
\nn
& &{} + 2 \lt(q^4 + 46 \, q^2 - 24\rt)\mu^6 + \lt(81 \, q^4 - 906 \, q^2 + 400\rt)\mu^5
+ 8 \lt(4 \pi \, q^3 - 10 \, q^2 - 56 \pi \, q - 74\rt)\mu^4
\nn
& &{} + 2 \lt(7 \, q^4 + 88 \, q^2 + 208 \rt) \mu^3 - 16 \lt(5 \, q^2 - 4 \pi \, q + 1\rt)\mu^2 + 8 \lt(7 \, q^2 - 48 \rt)\mu
\nn
& &{} + \lt. 320\rt]q \, e^{-q^2/4} \, e^{-\krq} \, {r^3 \over R^3} \, ,
\nl
\lt(D_{2 \Cx}\rt)_{01} & = & -{\sqrt{2} \over 960 \pi}
\lt[\lt(q^4 + 82 \, q^2 - 48 \pi \, q + 14\rt)\mu^8 - 6 \, q\lt(q^2 - 14\rt)\mu^7 \rt.
\nn
& &{} + q\lt(q^4 - 284 \, q^2- 84\rt)\mu^6 + 8 \lt(3 \, q^3 - 10 \pi \, q^2 + 41 \, q + 20 \pi\rt)\mu^5
+ q \lt(9 \, q^4 - 250 \, q^2 + 752\rt)\mu^4
\nn
& &{} + 4 \lt(\pi \, q^4 + 2 \, q^3 - 24 \pi \, q^2 - 26 \, q  + 116 \pi\rt) \mu^3 + 4 \, q \lt(11 \, q^2 - 40\rt)\mu^2
+ 32 \lt(\pi \, q^2 - 9 \, q - 12 \pi \rt) \mu
\nn
& &{} + \lt. 32 \, q + {64 \pi \over \mu} \rt] q \, e^{-q^2/4} \, e^{-\krq} \, {r^3 \over R^3} \, ,
\nl
\lt(D_{2 \Cx}\rt)_{11} & = & {1 \over 480} \, \sqrt{2 \over \pi}
\lt[-16 \, q^2 \, \mu^8 + \lt(q^4 + 19 \, q^2 + 6\rt)\mu^6 + 2 \lt(q^4 - 5 \, q^2 + 54\rt)\mu^4 \rt.
\nn
& &{} - \lt. 24\lt(q^2 + 2\rt)\mu^2 - 128\rt]q \, e^{-2\krq} \, {r^3 \over R^3} \, ,
\nl
\lt(D_{2 \Cx}\rt)_{20} & = & {1 \over 960} \, \sqrt{2 \over \pi}
\lt[\lt(15 \, q^2 + 14\rt) \mu^4 + 8 \, \mu^2 - 224 \rt] q^2 \, \mu^3 \, e^{-2\krq} \, {r^3 \over R^3} \, .
\ee
\end{subequations}

\end{appendix}

\end{document}